\documentclass[11pt, a4paper]{article}
\RequirePackage{amsthm,amsmath,amsfonts,amssymb}
\RequirePackage[authoryear]{natbib} 
\RequirePackage[colorlinks,citecolor=blue,urlcolor=blue]{hyperref}
\usepackage{bm}
\usepackage{mathrsfs}  
\usepackage{appendix}
\usepackage{commath}
\usepackage{dsfont}
\usepackage{subfig}
\usepackage{xcolor}
\usepackage{mathtools}
\usepackage{enumitem}
\usepackage{fancyhdr}
\usepackage{mathrsfs}

\usepackage[hmargin=1in,vmargin=1in]{geometry}
\usepackage{graphicx}

\usepackage{setspace}

\newcommand{\N}{\mathbb{N}}
\newcommand{\R}{\mathbb{R}}

\newcommand{\eps}{\varepsilon}

\newcommand\E{\mathbb{E}}
\DeclareMathOperator*{\argmax}{arg\,max}
\newcommand{\cp}{\xrightarrow{\mathbb{P}}}
\newcommand{\cps}{\xrightarrow{\mathbb{P}^*}}

\theoremstyle{definition}
\newtheorem{theorem}{Theorem}[section]
\newtheorem{lemma}[theorem]{Lemma}

\newtheorem{proposition}[theorem]{Proposition}
\newtheorem{remark}{Remark}[section]
\newtheorem{example}{Example}[section]
\newtheorem{assumption}{Assumption}[section]

\begin{document}

\title{Regularizing Fairness in Optimal Policy Learning with Distributional Targets \footnote{We thank Michael Lechner, Liyang Sun, Alex Tetenov, Alonso Zuniga, and several colleagues and seminar/workshop participants at WU Vienna, TU Graz, University of Vienna, University of Geneva, UPF Barcelona, and Erasmus University Rotterdam for helpful feedback and suggestions. We are grateful to an Associate Editor and two referees for their comments, which helped us to improve the paper.}}

\author{
\begin{tabular}{c}
Anders Bredahl Kock \\ 
\small Department of Economics \\
\small	University of Oxford \\
\small	10 Manor Rd, Oxford OX1 3UQ
\\
\small	{\small	\href{mailto:anders.kock@economics.ox.ac.uk}{anders.kock@economics.ox.ac.uk}} 
\end{tabular}
\and
\begin{tabular}{c}
David Preinerstorfer\footnote{Corresponding author.} \\ 
{\small	Insitute for Statistics and Mathematics} \\ 
{\small	WU Vienna University of Economics and Business} \\
{\small Welthandelsplatz 1, 1020 Vienna } \\ 
{\small	 \href{mailto:david.preinerstorfer@wu.ac.at}{david.preinerstorfer@wu.ac.at}}
\end{tabular}
}

\onehalfspacing
\date{First version: September 2023 \\ Second version: January 2024 \\ This version: May 2025}

\maketitle	

\begin{abstract}
A decision maker typically (i) incorporates training data to learn about the relative effectiveness of treatments, and (ii) chooses an implementation mechanism that implies an ``optimal'' predicted outcome distribution according to some target functional. Nevertheless, a fairness-aware decision maker may not be satisfied achieving said optimality at the cost of being ``unfair" against a subgroup of the population, in the sense that the outcome distribution in that subgroup deviates too strongly from the overall optimal outcome distribution. We study a framework that allows the decision maker to regularize such deviations, while allowing for a wide range of target functionals and fairness measures to be employed. We establish regret and consistency guarantees for empirical success policies with (possibly) data-driven preference parameters, and provide numerical results. Furthermore, we briefly illustrate the methods in two empirical settings.



\end{abstract}
\newpage

\section{Introduction}

Optimal policy choice, i.e., inferring good answers to the question ``whom to assign to which treatment'' from data, is of fundamental importance to decision makers. Much progress has been achieved in the recent literature: seminal contributions (emphasizing the nonparametric nature of the problem) were made in~\cite{manski2004statistical}, \cite{dehejia2005program}, \cite{stoye2009minimax}, \cite{stoye2012minimax}, \cite{bhattacharya2012inferring}, \cite{Manski10518}, a study of asymptotic efficiency properties was provided in~\cite{hirano2009asymptotics}, whereas more recent contributions focused on optimal policy choice from observational data and include~\cite{kitagawa2018should}, \cite{kitagawa2019equality}, \cite{athey2017efficient}, or \cite{mbakop}; see also the recent overview given in~\cite{hphandb} for further references. 

A number of recent advances in that literature, in particular \cite{kleinberg2018}, cf.~also the recent review of~\cite{cowgill2019economics}, stress that in addition to assigning individuals with the goal to maximize expected welfare, decision makers should incorporate ``fairness'' considerations into optimal policy choice. The problem of a fairness-aware decision maker that we consider in the present article is the following: although the decision rule inferred, e.g., by applying one of the approaches suggested in the literature discussed above, may be optimal on an \emph{aggregate} level, there may be \emph{subgroups} of the population defined by protected characteristics, e.g., gender, ethnicity, or religion, in which the effect of the decision rule deviates markedly from that on the aggregate level. In other words, although the decision maker may have succeeded in choosing a policy that leads to optimal results on the population level, this global optimality is not resembled within all subgroups and hence the policy may be considered unfair. 

This concern fits into the broader context of fair Machine Learning or fair Artificial Intelligence, where the general goal is to avoid that algorithms inferred from data discriminate against certain groups; cf.~\cite{corbett2018measure} and~\cite{pessach} for recent overviews on the subject. They identify the following key features: (i) incorporating fairness is complicated by the lack of a single accepted definition of fairness (cf., e.g.,~Section 3 and Appendix B.1 in~\cite{pessach}), (ii) impossibility results show that different types of fairness cannot be achieved at the same time, and (iii) achieving (a particular type of) fairness typically reduces the efficiency of the policy.

Fairness concerns can be incorporated into the optimal policy choice algorithms studied in \cite{kitagawa2018should} or \cite{athey2017efficient} by adjusting subpopulation weights in the objective function. Nevertheless, such weights need to be chosen by the decision maker upfront, and are typically difficult to defend. Motivated by the lack of methods allowing decision makers to incorporate more general forms of fairness considerations, \cite{vivbra} studied methods that combine fairness and optimal policy choice in a more distinguished way: instead of trying to infer a policy that maximizes expected population welfare, they formulate a multiobjective optimization problem, in which each objective corresponds to the welfare in the subpopulation obtained by considering a given value of a protected characteristic. After characterizing the Pareto frontier, the authors suggest to chose an element from the frontier which maximizes a fairness criterion, which is allowed to be very general, and can be chosen by the user in dependence on the notion of fairness that fits the specific problem at hand. Statistical performance guarantees are provided. Another recent contribution is \cite{fang23}, who do not consider fairness between subgroups defined by a sensitive characteristic, but aim at lifting the lower tail of the outcome distribution implied by a policy to a given level, therefore leading to ``fairness'' between rich and poor individuals. To achieve this, in addition to maximizing expected welfare, they introduce a lower-bound constraint on a quantile chosen by the user, and suggest an algorithm to solve the resulting optimization problem, for which asymptotic guarantees are established. Both articles discuss observational as well as experimental data.

Although decision makers can draw on these methods in situations where expected utility is the main target, there is currently no theory available that shows how one could incorporate fairness considerations in settings with more flexible target measures. Both \cite{vivbra} and \cite{fang23} target expected outcomes, and exploit  linearity properties of that target in their constructions and proofs. For example, the characterization of the Pareto frontier given in Lemma 2.1 of \cite{vivbra}, on which their approach crucially rests, exploits linearity -- or at least convexity/concavity -- of the target measure.\footnote{It needs to be emphasized again that (in contrast to their mean target, which is fixed) the \emph{fairness} measure used by \cite{vivbra} is very general and allows for great flexibility.} Important alternative targets in economic decision making are welfare measures other than expected utility, or measures incorporating poverty concerns (cf.~\cite{cowell} or~\cite{chakravarty2009} for book-length treatments of measures that have been developed and studied in the economics and statistics literature). The importance of flexibility in choosing the target measure was emphasized in, e.g.,~\cite{bitler2006mean} and~\cite{rostek2010quantile}. 

The goal of the present article is to develop a flexible theory of optimal treatment assignment that allows the decision maker to incorporate fairness considerations in a setting where the target is not necessarily the expected outcome, but, e.g., a welfare measure such as the Gini-welfare or a quantile such as the median. More specifically, we develop a theory that allows the decision maker to work with the rich class of targets introduced in~\cite{kpv1}, while incorporating covariate information and monitoring fairness properties of the policy recommendation. To arrive at a widely applicable theory concerning the target functional, we do not take a Pareto frontier based approach as in \cite{vivbra} or a constraint approach as in \cite{fang23} (which requires the somewhat inconvenient assumption that the constraint is actually feasible, which is not necessary in our approach), but we proceed as discussed informally in Section~\ref{sec:briefoutline} below, and refer the reader to later sections for more details. 

In Appendix~\ref{sec:reltokpv3} we discuss in more detail how the present article relates to \cite{kitagawa2019equality} and \cite{kpv3}, where optimal policy choice problems with general targets are treated (without fairness considerations).

\subsection{Informal summary of our methods and results, and practical guidelines}\label{sec:briefoutline}

\subsubsection{Objective}

In Section~\ref{sec:setting}, we first decompose~$\langle \bm{\delta}, \mathcal{F} \rangle$, the (unknown) population distribution implied by a decision rule~$\bm{\delta}$, as
\begin{equation*}
\langle \bm{\delta}, \mathcal{F} \rangle = \sum_{z \in \mathcal{Z}} p_Z(z) \langle \bm{\delta}, \mathcal{F} \rangle_z,
\end{equation*}
where~$\langle \bm{\delta}, \mathcal{F} \rangle_z$ denotes the (unknown) distribution implied by the decision rule~$\bm{\delta}$ in the subpopulation of subjects with protected characteristic equal to~$z \in \mathcal{Z}$, and where~$p_Z(z)$ denotes the (unknown) proportion of such subjects. While the primary goal of the decision maker is to choose~$\bm{\delta}$ such as to maximize~$\mathsf{T}(\langle \bm{\delta}, \mathcal{F} \rangle)$, where~$\mathsf{T}$ is some (potentially nonlinear and nonconvex) functional, the secondary goal of the decision maker is to monitor fairness properties of the policy.\footnote{Because the decision maker does not know~$\langle \bm{\delta}, \mathcal{F} \rangle$ or~$\langle \bm{\delta}, \mathcal{F} \rangle_z$, the decision maker has to estimate these quantities, which introduces a nontrivial statistical aspect that needs to be addressed.} Specifically, we consider a situation where, as the secondary goal, the decision maker finds \emph{parity} across sub-populations (aka groups) defined by a protected characteristic desirable. That is, the decision maker wants to choose~$\bm{\delta}$ such that $$\langle \bm{\delta}, \mathcal{F} \rangle_z \approx \langle \bm{\delta}, \mathcal{F} \rangle \quad \text{ for every } \quad z \in \mathcal{Z},$$ in the sense that $$\max_{z \in \mathcal{Z}} \mathsf{S}\left(\langle \bm{\delta}, \mathcal{F} \rangle_z, \langle \bm{\delta}, \mathcal{F} \rangle\right) \approx 0,$$ where~$\mathsf{S} \geq 0$ is a semi-metric on the set of distribution functions, e.g., the Kolmogorov-Smirnov distance, allowing for some flexibility. If~$\bm{\delta}$ satisfies such a parity condition, this means that the outcome distributions implied by the decision rule~$\bm{\delta}$ do not differ much across the groups defined by the protected characteristic (with respect to the properties encoded in the fairness measure~$\mathsf{S}$).

To combine the two goals outlined above, i.e., maximization of~$\mathsf{T}$ evaluated at the implied outcome distribution~$\langle \bm{\delta}, \mathcal{F} \rangle$ and achieving some degree of parity formalized via~$\mathsf{S}$, we adapt a penalty-based approach. This idea has repeatedly been applied to incorporate fairness considerations into  regression and classification problems, cf., e.g.,~\cite{kamishima2011}, \cite{Kamishima2012}, \cite{berk2017convex}, or~\cite{bechavod2017penalizing}. Specifically, we aim at studying statistical methods to choose a decision rule~$\bm{\delta}$ that maximizes the penalized objective
\begin{equation}\label{eqn:introtarget}
(1-\lambda) \mathsf{T}(\langle \bm{\delta}, \mathcal{F} \rangle) - \lambda \max_{z \in \mathcal{Z}} \mathsf{S}\left(\langle \bm{\delta}, \mathcal{F} \rangle_z, \langle \bm{\delta}, \mathcal{F} \rangle\right),
\end{equation}
where~$\lambda \in [0, 1]$ is a preference parameter, which has to be chosen by the decision maker (possibly in a data-driven way). Note that the larger~$\lambda$, the more emphasis is put on fairness.

\subsubsection{Policies}

The policies we study are all of the \emph{empirical success} type: In the case of discrete covariates, which is our main focus, we study a plug-in approach, where we first estimate~$\mathcal{F}$ through~$\hat{\mathcal{F}}$, say, and then solve (numerically) the empirical version of the penalized objective in~\eqref{eqn:introtarget}, cf.~Section~\ref{sec:esr} for a detailed description of the policy and its regret and consistency properties. We also discuss variants of our policies when the preference parameter~$\lambda$ in~\eqref{eqn:introtarget} is chosen in a data-driven way. To this end, the decision maker (DM) may obtain optimal policies for a set of preference parameters~$\Lambda \subseteq [0, 1]$ first (making use of the just-mentioned empirical success type policy), and then select one of these policies by balancing the fairness/efficiency trade-off of the policies. This can be done informally, e.g., by investigating plots of the estimated fairness/efficiency properties of the policies, or more formally, e.g., by fixing a ``budget'' the DM can afford spending on fairness (cf.~\cite{berk2017convex}), and then selecting the policy which realizes the maximal degree of fairness affordable. The latter approach is described and studied in detail in Section~\ref{sec:tune}. The former, more graphical approach, is illustrated in the applications in Section~\ref{sec:empirics} in the context of the Pennsylvania reemployment bonus experiment (cf.~\cite{bilias}) and the entrepreneurship program data from \cite{lyonszhang}, which was also investigated in \cite{vivbra}. 

The case of non-discrete covariates is studied in Section~\ref{sec:contX}, cf.~Section~\ref{sec:contXes} for a detailed description of the policies and a regret upper bound. In this context, we do not work with a plug-in for~$\hat{\mathcal{F}}$ but estimate the objective function (introduced earlier) more directly, an approach that may be viewed as an extension of \cite{kitagawa2019equality} to our setting. 

\subsubsection{Practical guideline}

In a concrete application, one can proceed as follows:
\begin{enumerate}
\item First, one has to fix a target functional~$\mathsf{T}$ and a measure of unfairness~$\mathsf{S}$. Canonical choices are, e.g., the Gini-welfare measure or the expectation for the former and the Kolmogorov-Smirnov distance (possibly one-sided) for the latter. Many relevant target functionals are described in~\cite{chakravarty2009} or~\cite{cowell}, to which we refer the practitioner who is unsure about which target functional fits the concrete application best. Once this choice has been made, the next step depends on the nature of the covariates.
\item If the covariates are discrete, one can use the policy in Section~\ref{sec:esr}, otherwise one uses the policy in Section~\ref{sec:contXes}. How this is precisely carried out further depends on whether a concrete value for~$\lambda$ is available or not:
\begin{itemize}
\item If a concrete value for the preference parameter~$\lambda$ is known, approximately solve the objective in Equation~\eqref{eqn:approxmax} (for the discrete case) or in Equation~\eqref{eqn:approxmax2} (otherwise) with a numerical optimization procedure, and roll out the policy obtained by random assignment; e.g., by making use of the concrete assignment mechanism described in Footnote~\ref{foot:concrass}.
\item If~$\lambda$ is not specified, we recommend to solve the respective objective for a grid of~$\lambda$ values, and use graphical inspection and/or the procedure in Section~\ref{sec:tune} to select a suitable value of~$\lambda$; cf.~the empirical examples in Section~\ref{sec:empirics} for some guidance. Then, roll out the policy by random assignment (e.g., by making use of the concrete assignment mechanism described in Footnote~\ref{foot:concrass}).
\end{itemize}
\end{enumerate}    

\section{Setting}\label{sec:setting}

\subsection{Observational structure and two assumptions}

We consider a DM, who must decide at which proportions~$K\geq 2$ available treatments should be rolled out to a heterogeneous population. Before the ``roll-out phase" begins, the DM needs to find out which proportions lead to the ``optimal'' (this will be made precise below) outcome distribution in the population. To make this decision, we assume that the DM is in possession of a training sample of~$n$ subjects that:
\begin{enumerate}
\item were independently drawn from the target population,
\item were assigned to one of the~$K$ treatments, possibly by making use of covariate information in the assignment decision,
\item and were subjected to an outcome measurement.
\end{enumerate}

We adopt the potential outcomes framework, where to each subject~$j = 1, \hdots, n$, there corresponds a potential outcomes vector
\begin{equation*}
\bm{Y}_j = \left(Y_{1,j}, \hdots, Y_{K,j}\right)' \in \R^K.
\end{equation*}
That is,~$Y_{i,j}$ is the potential outcome of subject~$j$ under treatment~$i$. As usual, for each subject~$j = 1, \hdots, n$ in the training sample, the decision maker does not observe the whole potential outcome vector~$\bm{Y}_j$, but only observes the single coordinate~$$\tilde{Y}_j := Y_{D_j, j},$$ where~$D_j \in \{1, \hdots, K\}$ denotes the treatment subject~$j$ was assigned to, which the DM observes as well. In addition to~$\tilde{Y}_j $ and~$D_j$, the DM observes two \emph{categorical} covariates $$X_j \in \mathcal{X},~~0 < |\mathcal{X}|< \infty, \quad \text{ and } \quad Z_j \in \mathcal{Z},~~0 < |\mathcal{Z}|< \infty.$$ 

For now we focus on the case where both covariates~$X_j$ and~$Z_j$ can take on only finitely many values, because this simplifies the analysis and exposition considerably, and because this is an important case for practice. In Section~\ref{sec:contX} we consider the case where~$\mathcal{X}$ is not finite.

\begin{remark}\label{rem:covintpt}
At this stage, the reader may think of~$X_j$ as a covariate that can be used to assign individuals to treatments during the roll-out phase. That is, the proportions used during the roll-out phase can depend on these covariates, in the sense that different proportions are rolled out for different values of~$x \in \mathcal{X}$. The covariate~$Z_j$, on the other hand, should be thought of as a ``protected characteristic'' that \emph{cannot} be used for assigning individuals in the roll-out phase, e.g., because it is simply not possible to measure this information during the roll-out phase or because it is not allowed to make different assignment decisions for different values of the protected characteristic.\footnote{Note that the variables~$X_j$ and~$Z_j$ may be dependent, and that a decision depending on the former characteristic only may therefore also depend on~$Z_j$. In this sense, the decision rules we allow for are certainly not ``independent" of the information inherent in the protected characteristic in general. If such an ``independence'' condition is required, a different approach that tries to eliminate this dependence has to be taken, which is not the goal in the present article. In the terminology of \cite{frauen2023fair}, who consider optimal policy choice with fairness considerations for an expected utility target, our focus is oriented towards ``value fairness'' whereas the fairness notion discussed in this footnote corresponds to ``action fairness''.} Nevertheless, this covariate is assumed to be available for all subjects in the training sample of the DM. \cite{tan2022rise}, who consider robust decision learning targeting the expected outcome, use an analogous setting concerning the treatment of sensitive characteristics.
\end{remark}

Summarizing, each subject is characterized by an underlying (only partly observed) random vector
\begin{equation*}
\bm{W}_j := \left(\bm{Y}_j', X_j, Z_j, D_j\right)' \in \R^{K} \times \mathcal{X} \times \mathcal{Z} \times \{1, \hdots, K\}.
\end{equation*}
The \emph{training sample} available to the DM is (a realization of)
\begin{equation}\label{eqn:obs}
\tilde{\bm{W}}_j := \left(\tilde{Y}_j, X_j, Z_j, D_j\right)' \in \R \times  \mathcal{X} \times \mathcal{Z} \times \{1, \hdots, K\} \quad \text{ for } j = 1, \hdots, n.
\end{equation}
All random variables are defined on an underlying probability space~$(\Omega, \mathcal{A}, \mathbb{P})$ with corresponding expectation operator~$\mathbb{E}$; outer probability and outer expectation are denoted by~$\mathbb{P}^*$ and~$\mathbb{E}^*$, respectively, which will help us to avoid imposing measurability conditions.

Without further mentioning in the theoretical results, we shall \emph{throughout this article impose} an independent sampling condition:
\begin{assumption}\label{as:iid}
The random vectors~$\bm W_j$,~$j = 1, \hdots, n$, are i.i.d..
\end{assumption}
The independent sampling assumption is standard in the treatment assignment literature, and weakening it is not an objective of the present article. Also without further mentioning in the theoretical results, we \emph{impose throughout this article} that the assignment mechanism is unconfounded conditional on the covariates observed, that is:
\begin{assumption}\label{as:cia}
Conditional on~$X_j$ and~$Z_j$ the potential outcome vector~$\bm Y_j$ is independent of~$D_j$; i.e.,~$$\bm Y_j  \perp\!\!\!\perp D_j ~|~ (X_j, Z_j).$$
\end{assumption}

\subsection{Decision rules and policies}\label{sec:recfun}

Based on the training sample in~\eqref{eqn:obs}, the DM needs to specify how to proceed in the roll-out phase. The assignment decision can only be based on the observed value of the unprotected characteristic~$x \in \mathcal{X}$, cf.~Remark~\ref{rem:covintpt}. Formally, and allowing for randomized decision rules, for each~$x \in \mathcal{X}$ the DM therefore has to decide (based on the training sample) on the proportions at which to roll out the treatments~$i = 1, \hdots, K$. A \emph{decision rule} thus specifies, for each covariate value~$x \in \mathcal{X}$, a ``probability vector''
\begin{equation}
\bm{\delta}(x) = \left(\delta_1(x), \hdots, \delta_K(x)\right)' \in \mathscr{S}_K,
\end{equation} where~$\mathscr{S}_K = \{\alpha = (\alpha_1, \hdots, \alpha_K)' \in [0, 1]^K: \alpha_1 + \hdots + \alpha_K = 1\}$ denotes the standard simplex in~$\R^K$. This class of decision rules allows for (but does not impose) randomization.

The~$i$th coordinate~$\delta_i(x)$ of~$\bm{\delta}(x)$ corresponds to the probability of assigning a subject with covariate~$x$ to treatment~$i$ during the roll-out phase. The dependence of the proportions on~$x$ permits the DM to take advantage of the population's heterogeneity. Note again (cf.~Remark~\ref{rem:covintpt}) that~$\bm{\delta}$ does not have~$z$ as an argument.

Every data-driven method that turns the training sample defined in Equation~\eqref{eqn:obs} into a decision rule as just described shall be referred to as a \emph{policy}; i.e., a policy is a function
\begin{equation}\label{eqn:poly}
\bm{\pi}_n: \R^n \times \mathcal{X}^n \times \mathcal{Z}^n \times \{1, \hdots, K\}^n \to \mathbb{S}_K,
\end{equation}
where~$\mathbb{S}_K = \mathscr{S}_K^{|\mathcal{X}|}$ denotes the set of decision rules, i.e., all \emph{functions} from~$\mathcal{X}$ to the simplex~$\mathscr{S}_K$. Throughout, we shall equip~$\mathbb{S}_K$ with the distance~$d_1$, which, for~$\bm{\delta}$ and~$\tilde{\bm{\delta}}$ in~$\mathbb{S}_K$, is defined as 
\begin{equation}\label{eqn:dstS}
d_1(\bm{\delta},  \tilde{\bm{\delta}}) := \sum_{x \in \mathcal{X}} \|\bm{\delta}(x) - \tilde{\bm{\delta}}(x)\|_1,
\end{equation}
where, for~$p \geq 1$, the symbol~$\|\cdot\|_p$ denotes the~$p$-norm on a Euclidean space.
\begin{remark}\label{rem:constr}
As in~\cite{kpv3}, we could also consider policies that are restricted to lie in a non-empty and closed subset of~$\mathbb{S}_K$ (corresponding to budget, similarity, or compatibility constraints). The theory and algorithms developed in Sections~\ref{sec:setting} and \ref{sec:esr} would go through upon fairly straightforward modification of the statements and proofs. To avoid notation overload, however, we do not pursue this level of generality here. In Section~\ref{sec:contX}, where we consider non-finitely valued covariates, restrictions on the set of decision rules will be imposed.
\end{remark}

\subsection{Goal of the DM}\label{sec:goal}

We have not yet formally defined the actual \emph{goal} of the DM. In order to decide which decision rules~$\bm{\delta}$ are preferable in a specific context, we 
\begin{enumerate}
\item need to investigate the structure of the population cdf (cumulative distribution function) implied by rolling out a decision rule~$\bm{\delta}$, and
\item single out a criterion on the basis of which to compare different population cdfs. 
\end{enumerate}
This is addressed in the following subsections: After introducing some notation in Section~\ref{ssec:notation}, we address the first item in the above enumeration in Section~\ref{ssec:distrroll} and the second in Section~\ref{ssec:obj}. An illustrative toy example is studied in Section~\ref{ssec:ex} to illustrate the concepts.

\subsubsection{Some notation}\label{ssec:notation}

We denote the joint probability mass function of~$(X_j, Z_j, D_j)$ by~$p_{X, Z, D}(x, z, i) = \mathbb{P}(X_j = x, Z_j = z, D_j = i)$, write~$p_{X, Z}$,~$p_X$, and~$p_Z$ for the corresponding marginal probability mass functions of~$(X_j, Z_j)$, $X_j$ and~$Z_j$, respectively, and set
\begin{equation}\label{eqn:cprob}
p_{X \mid Z = z}(x) := \frac{p_{X, Z}(x, z)}{p_Z(z)} \quad \text{ and } \quad p_{Z \mid X = x}(z) := \frac{p_{X, Z}(x, z)}{p_X(x)}.
\end{equation}
On top of Assumptions~\ref{as:iid} and~\ref{as:cia}, \emph{we shall assume} (until Section~\ref{sec:contX}) that the common support of~$(X_j, Z_j, D_j)$ is~$\mathcal{X} \times \mathcal{Z} \times \{1, \hdots, K\}$ so that, in particular, the quotients in~\eqref{eqn:cprob} are well defined:
\begin{assumption}\label{as:posprob}
We have~$p_{X, Z, D}(x, z, i) > 0$ for every $x \in \mathcal{X}$, $z \in \mathcal{Z}$, and $i \in \{1, \hdots, K\}$.
\end{assumption}
For~$x \in \mathcal{X}$ and~$z \in \mathcal{Z}$ we denote by~$F^{i}(\cdot \mid x, z)$ the cumulative distribution function (cdf) of~$$Y_{i, j} \mid X_j = x, Z_j = z.$$ Furthermore, we denote by~$F^i(\cdot \mid x)$ the cdf of~$Y_{i, j} \mid X_j = x,$ and by~$F^i$ the cdf of~$Y_{i, j}$. We write~$\bm{F} = (F^1, \hdots, F^K)'$, and analogously define for every~$x$ and~$z$ the vectors of conditional cdfs~$\bm{F}(\cdot \mid x)$, and~$\bm{F}(\cdot\mid x, z)$. For notational convenience, we shall collect the conditional cdfs of a random vector~$\bm{Y}_j'$ given~$X_j = x$ and~$Z_j = z$ and the corresponding probability mass functions in an array 
\begin{equation}\label{eqn:array}
\mathcal{F} := \left[(F^i(\cdot\mid x, z), p_{X, Z}(x, z)): i = 1, \hdots, K;~x \in \mathcal{X};~z \in \mathcal{Z}\right].
\end{equation}
Given~$\mathcal{F}$ as in~\eqref{eqn:array}, i.e., an array consisting of cdfs and a probability mass function over~$\mathcal{X} \times \mathcal{Z}$, we shall sometimes write~$$(\bm{Y}_j', X_j, Z_j)' \sim \mathcal{F}$$ to signify that~$\mathcal{F}$ is the array \emph{corresponding}\footnote{Note that~$\mathcal{F}$ corresponds to but does not specify completely the joint distribution of~$(\bm{Y}_j', X_j, Z_j)'$, because it does not specify the dependence between the potential outcomes, an aspect that is irrelevant for our results.} to the distribution of~$(\bm{Y}_j', X_j, Z_j)'$. Generically, we shall \emph{throughout denote} by~$\mathcal{F}$ the array corresponding to the distribution of~$(\bm{Y}_j', X_j, Z_j)'$, which is \emph{unknown} to the DM. 

\subsubsection{Distributions generated by rolling out a decision rule}\label{ssec:distrroll}

We now consider the effect of rolling out a (fixed) decision rule~$\bm{\delta} \in \mathbb{S}_K$ to the whole population via randomized assignment. More specifically, consider a new subject independently drawn from the same population as the training sample in the sense that
\begin{equation}\label{eqn:rollo}
(\bm{Y}', X, Z)' \sim \mathcal{F},
\end{equation} 
and assume that its assignment~$D$ (chosen by the DM), say, satisfies:
\begin{enumerate}
	\item conditionally on~$X$, it is distributed according to~$\bm{\delta}(X)$, \emph{and}
	\item  conditionally on $(X, Z)$, it is independent of every outcome~$Y_{i}$ for~$i = 1, \hdots, K$;
\end{enumerate}
which, for every~$x \in \mathcal{X}, y \in \R, z \in \mathcal{Z}$ and treatment~$i$, we summarize in the property\footnote{\label{foot:concrass}For concreteness, let~$U$ be a uniformly distributed random variable on~$(0, 1)$ that is independent of the training sample and of~$(\bm{Y}', X, Z)'$. Set~$\gamma_{0}(x) := 0$ and~$\gamma_{i}(x) = \sum_{j = 1}^i\delta_{j}(x)$ for~$i = 1, \hdots, K$ and~$x \in \mathcal{X}$, and assign the new subject to treatment~$i$ if $U \in [\gamma_{i-1}(X),  \gamma_i(X)).$ Then~\eqref{eqn:inddraw} is satisfied (note that it also remains satisfied conditionally on the training sample in case~$\bm{\delta}$ depends on the training sample).}
\begin{equation}\label{eqn:inddraw}
P(D = i, Y_i \leq y \mid X = x, Z = z) = \delta_i(x) P(Y_i \leq y \mid X = x, Z = z).
\end{equation} 
Denoting~$Y_D$ by~$\tilde{Y}$, it is elementary to verify that the cdf of~$\tilde{Y}$ then equals
\begin{equation}\label{eqn:mixroll}
\begin{aligned}
\mathbb{P}\left(\tilde{Y} \leq \cdot\right) 
=  &\sum_{x \in \mathcal{X}}  \sum_{i = 1}^K \delta_i(x) p_X(x) F^i(\cdot\mid x)  =: \langle \bm{\delta}, \mathcal{F} \rangle.
\end{aligned}
\end{equation}
For every~$z \in \mathcal{Z}$, we can analogously consider the cdf of~$\tilde{Y}$ conditional on~$Z = z$, i.e., the cdf in the \emph{subpopulation} of subjects with protected characteristic equal to~$z$, which is given by 
\begin{equation}\label{eqn:mixrollsub}
\mathbb{P}\left(\tilde{Y} \leq \cdot \mid Z = z\right) =  \sum_{x \in \mathcal{X}} \sum_{i = 1}^K \delta_i(x) p_{X \mid  Z = z}(x) F^i(\cdot\mid x, z) =: \langle \bm{\delta}, \mathcal{F} \rangle_z,
\end{equation}
and for which we thus have~$\langle \bm{\delta}, \mathcal{F} \rangle = \sum_{z \in \mathcal{Z}} p_Z(z) \langle \bm{\delta}, \mathcal{F} \rangle_z$.

\subsubsection{Objective: Target functional penalized for unfairness}\label{ssec:obj}

In the previous subsection we illustrated how the choice of~$\bm{\delta}$ implies a population cdf~\eqref{eqn:mixroll}, which is further decomposed as a weighted average of subpopulation cdfs~\eqref{eqn:mixrollsub}. Which types of population cdfs are desirable to the DM depends strongly on the context. Typically, ``desirability'' will be measured through a user-specified real-valued functional~$\mathsf{T}$, say, of the cdf generated; such as its mean, a quantile such as the median, or a trimmed mean, possibly augmented by a poverty or inequality measure. 

Given the functional~$\mathsf{T}$---the concrete form of which we leave unspecified for now to ease the presentation---we consider the problem where the DM wants to trade off two potentially conflicting objectives described next:
\begin{enumerate}
\item The DM wants to choose~$\bm{\delta} \in \mathbb{S}_K$ to \emph{maximize} $\mathsf{T}(\langle \bm{\delta}, \mathcal{F} \rangle)$, i.e., the functional~$\mathsf{T}$ evaluated at the outcome distribution~$\langle \bm{\delta}, \mathcal{F} \rangle$. 
\item The DM wants to roll out a~$\bm{\delta}$ that is fair in the sense that the subpopulation cdfs~$\langle \bm{\delta}, \mathcal{F} \rangle_z$ are all ``close" to the population cdf~$\langle\bm{\delta}, \mathcal{F} \rangle$ singled out by the first objective. If this condition is violated, there are subpopulations defined by the protected characteristic, the cdf of which substantially deviates from the ``global" outcome cdf~$\langle \bm{\delta}, \mathcal{F} \rangle$, which the DM wants to avoid.
\end{enumerate}
While the first objective is governed by a functional~$\mathsf{T}$, we need to make the second part of the DM's objective more precise. To this end, because the similarity of cdfs can be measured in different ways, we introduce another functional~$\mathsf{S}$ that: 
\begin{enumerate}[label=(\alph*)]
\item maps \emph{pairs} of cdfs to~$[0, \infty)$,
\item equals~$0$ if the two input cdfs coincide,
\item is intended to measure the ``similarity" between two cdfs, high values corresponding to a strong degree of dissimilarity.
\end{enumerate}  
For concreteness, one may think of~$\mathsf{S}$ as the Kolmogorov-Smirnov distance (or more generally, any pseudo-distance) on the set of cdfs, or the absolute difference in the functional~$\mathsf{T}$ evaluated at two input cdfs. One could also introduce some ``directionality effects" by working, e.g., with a one-sided Kolmogorov-Smirnov pseudo-distance (for cdfs~$F$ and~$G$ consider~$\mathsf{S}(F, G) = \sup_{x \in \R} \max(F(x) - G(x), 0)$ instead of~$\mathsf{S}(F, G) = \sup_{x \in \R} |F(x) - G(x)|$). These are important examples, but the theory developed below applies more generally.

The remaining question now is how the DM could balance the two optimization objectives. In the present article, we consider rules~$\bm{\delta}^{\lambda} \in \mathbb{S}_K$, say, that maximize, for a chosen penalty~$\lambda \in [0, 1]$, the \emph{penalized} objective
\begin{equation}\label{eqn:omegadef}
\Omega_{\lambda, \mathcal{F}}(\bm{\delta}) := (1-\lambda) \mathsf{T}\left( \langle \bm{\delta}, \mathcal{F} \rangle \right) - \lambda \max_{z \in \mathcal{Z}} \mathsf{S} \left(\langle \bm{\delta}, \mathcal{F} \rangle_z , \langle \bm{\delta}, \mathcal{F} \rangle \right).
\end{equation} 

The objective~$\Omega_{\lambda, \mathcal{F}}(\bm{\delta})$ in~\eqref{eqn:omegadef} is composed of two terms: The first summand depends on the functional~$\mathsf{T}$ evaluated at the cdf generated by the decision rule~$\bm{\delta} \in \mathbb{S}_K$. The DM's primary goal is to maximize this quantity. The second term adds the maximal dissimilarity between a subpopulation cdf~$ \langle \bm{\delta}, \mathcal{F} \rangle_z$ and the population cdf~$\langle \bm{\delta}, \mathcal{F} \rangle$. Note that if~$\mathsf{S}$ is chosen suitably (e.g., when~$\mathsf{S}$ is a pseudo-metric on the set of cdfs), and~$$\max_{z \in \mathcal{Z}} \mathsf{S} \left(\langle \bm{\delta}, \mathcal{F} \rangle_z , \langle \bm{\delta}, \mathcal{F} \rangle \right)$$ is large, this corresponds to a situation where the cdf generated by rolling out~$\bm{\delta}$ leads to very different cdfs between subpopulations defined by the protected characteristics and the population cdf. In this sense, rolling out the treatments according to~$\bm{\delta}$ may be considered unfair. 

\begin{remark}\label{rem:groupsize}
Note that we have chosen the penalty term~$\max_{z \in \mathcal{Z}} \mathsf{S} \left(\langle \bm{\delta}, \mathcal{F} \rangle_z , \langle \bm{\delta}, \mathcal{F} \rangle \right)$ not to incorporate the ``sizes''~$p_Z(z)$ of the protected groups. Alternatively, one could also have used a penalty term of the form~$\max_{z \in \mathcal{Z}} p_Z(z)\mathsf{S} \left(\langle \bm{\delta}, \mathcal{F} \rangle_z , \langle \bm{\delta}, \mathcal{F} \rangle \right),$ which, however, has the disadvantage of down-weighing small marginalized groups.  
\end{remark}

The degree at which unfairness is penalized depends on the \emph{preference parameter}~$\lambda$, which has to be chosen by the DM. Choosing~$\lambda \approx 0$ corresponds to a situation where fairness is essentially not taken into account, and where the DM is mostly interested in determining which treatment combination maximizes~$\mathsf{T}$. On the other hand, choosing~$\lambda \approx 1$ corresponds to a situation where unfairness is severely penalized. 

\begin{remark}\label{rem:tunefirst}
Typically, the DM will determine solutions to (suitable approximations of) the optimization problem~\eqref{eqn:omegadef2} for different values of~$\lambda$, and will then determine the decision rule to eventually roll out by comparing aspects of the corresponding path of solutions~$\bm{\delta}^{\lambda}$. That is, the DM will study aspects of (estimates of) the functions~$$\lambda \mapsto \mathsf{T}\left(\langle \bm{\delta}^{\lambda}, \mathcal{F} \rangle \right) \quad \text{ and } \quad \lambda \mapsto \mathsf{S} \left(\langle \bm{\delta}^{\lambda}, \mathcal{F} \rangle_z, \langle \bm{\delta}^{\lambda}, \mathcal{F} \rangle \right),$$ cf.~Section~\ref{sec:empirics} for some illustration in the context of two empirical examples. Based on these functions, the DM can decide which~$\lambda$ to choose and which corresponding~$\bm{\delta}^{\lambda}$ to roll out. The decision rule a DM applies proceeding in this way incorporates a \emph{data-driven} choice of~$\lambda$. Because this choice may be highly complicated or subjective, we shall establish regret guarantees \emph{irrespective} of what the data-driven preference parameter choice of the DM is.
\end{remark}

\begin{remark}\label{rem:alternreg}
	Alternatively to the objective in~\eqref{eqn:omegadef}, one could consider an objective function that maximizes~$\mathsf{T}\left( \langle \bm{\delta}, \mathcal{F} \rangle \right)$ subject to the constraint that~$\bm{\delta}$ satisfies~$$\max_{z \in \mathcal{Z}}\mathsf{S} (\langle \bm{\delta}, \mathcal{F} \rangle_z , \langle \bm{\delta}, \mathcal{F} \rangle ) \leq \varepsilon.$$ As mentioned in the introduction, an advantage of working with the penalized criterion~\eqref{eqn:omegadef} over this constrained maximization problem is that a solution to the latter may fail to exist for a given~$\varepsilon$, because no~$\bm{\delta}$ may satisfy the restriction.
\end{remark}

Proposition~\ref{prop:exmax} below shows that under weak assumptions the optimization problem 
\begin{equation}\label{eqn:omegadef2}
\max_{\bm{\delta} \in \mathbb{S}_K}	\Omega_{\lambda, \mathcal{F}}(\bm{\delta}),
\end{equation} 
permits a (possibly non-unique) solution for every~$\lambda \in [0, 1]$. Note that the optimization problem in~\eqref{eqn:omegadef2} cannot directly be solved by the DM, because~$\mathcal{F}$ is unknown. Therefore, the DM first has to estimate~$\mathcal{F}$ and will then solve an approximation to that optimization problem. We address this aspect further below.

A detailed example that illustrates the concepts introduced so far can be found in Section~\ref{sec:numex}.

\subsection{Main technical assumption concerning the functionals~$\mathsf{T}$ and~$\mathsf{S}$ and some preliminary observations}

To establish theoretical guarantees, \emph{we impose throughout} the assumption that all cdfs (and conditional cdfs) we are working with are supported on a common interval~$[a, b]$, where~$a < b$ are real numbers (a cdf~$F$ is supported on~$[a,b]$ if~$F(a-) = 0$ and~$F(b) = 1$). We denote the set of cdfs supported on~$[a,b]$ by~$D_{cdf}([a,b])$. Furthermore, to allow the researcher to incorporate known structural properties of the cdfs, we shall introduce a ``parameter space''~$\mathscr{D} \subseteq D_{cdf}([a,b])$, in which the conditional cdfs appearing in~$\mathcal{F}$ given in~\eqref{eqn:array} are assumed to lie in. The following assumption on the target functional~$\mathsf{T}$ and the functional measuring fairness~$\mathsf{S}$ in relation to~$\mathscr{D}$ is crucial and \emph{will be assumed to hold throughout the article}.

\begin{assumption}\label{as:MAIN}
The functionals
\begin{equation}\label{eqn:functs}
\begin{aligned}
\mathsf{T}: D_{cdf}([a,b]) \to \R \quad \text{ and } \quad
\mathsf{S}:  D_{cdf}([a,b]) \times  D_{cdf}([a,b]) \to [0, \infty)
\end{aligned} 
\end{equation}
and the non-empty convex set~$\mathscr{D} \subseteq D_{cdf}([a,b])$ satisfy 
\begin{equation}\label{eqn:lipcondAS}
\begin{aligned}
|\mathsf{T}(F) - \mathsf{T}(G)| \leq \|F- G\|_{\infty} \quad \text{ and } \quad |\mathsf{S}(F, \tilde{F}) - \mathsf{S}(G, \tilde{G})|\leq \|F- G\|_{\infty} + \|\tilde{F}- \tilde{G}\|_{\infty}
\end{aligned}
\end{equation}
for every~$(F, \tilde{F}) \in \mathscr{D} \times \mathscr{D}$ and every~$(G, \tilde{G}) \in D_{cdf}([a,b]) \times D_{cdf}([a,b])$. Furthermore, it holds that~$\mathsf{S}(F, F) = 0$ for every~$F \in D_{cdf}([a,b])$. 
\end{assumption}

Assumption~\ref{as:MAIN} imposes an asymmetric Lipschitz-type condition on the functionals~$\mathsf{T}$ and~$\mathsf{S}$ that will allow us, e.g., to establish concentration results for plug-in estimators based on empirical cdfs through the Dvoretzky-Kiefer-Wolfowitz-Massart (DKWM) inequality, cf.~\cite{massart1990}.

It is not difficult to verify that the functionals discussed in the example presented in Section~\ref{ssec:ex} (i.e.,~where~$\mathsf{T}$ is a normalized Gini-welfare measure and~$\mathsf{S}$ the Kolmogorov-Smirnov distance) satisfy Assumption~\ref{as:MAIN}.\footnote{Use the reverse triangle inequality for the Kolmogorov-Smirnov distance to verify the condition on~$\mathsf{S}_{KS}$ there, and the discussion after Lemma E.9 in \cite{kpv1} to verify the condition on~$\mathsf{T}_{Gini}$.} Note that functionals that satisfy a variant of~\eqref{eqn:lipcondAS}, but where constants appear in front of the norms in the upper bounds can be brought into the framework of Assumption~\ref{as:MAIN} by normalization. We could also work with a version of Assumption~\ref{as:MAIN} with constants appearing in the upper bounds, but this would be notationally more cumbersome. These constants would then enter the regret upper bounds established below multiplicatively, but would not enter the policies introduced (hence their knowledge is not required by the DM to carry out the policy). Appendices F and G in~\cite{kpv1} verify the latter version of the above assumption for various poverty, inequality and welfare measures, as well as for functionals such as quantiles, U-statistics, L-statistics, and more. 

\subsubsection{Some consequences}

An important first
question is whether the optimization problem in~\eqref{eqn:omegadef2} permits a solution and
if this is the case, how it
depends on~$\lambda$. For an array~$\mathcal{F}$ as in~\eqref{eqn:array}, the following result establishes, among other useful properties, that the objective~$\Omega_{\lambda, \mathcal{F}}$ defined in Equation~\eqref{eqn:omegadef} indeed has a maximizer (which need not be unique, cf.~the example discussed in Section~\ref{ssec:ex}). For an array~$\mathcal{F}$ as in~\eqref{eqn:array} and~$\mathscr{H}$ a set of cdfs, we write~$\mathcal{F} \sqsubset \mathscr{H}$ if \emph{all} cdfs contained in the array~$\mathcal{F}$ are elements of~$\mathscr{H}$, i.e., 
\begin{equation*}
\mathcal{F} \sqsubset \mathscr{H} \quad \Leftrightarrow \quad  F^i(\cdot \mid x, z) \in \mathscr{H} \text{ for every } i = 1, \hdots, K,~x \in \mathcal{X}, \text{ and } z \in \mathcal{Z}.
\end{equation*} For a definition of upper hemicontinuity of a set-valued function (correspondence) we refer to Section 17.2 in~\cite{hhg}.
\begin{proposition}\label{prop:exmax}
The following statements hold for~$\mathcal{F} \sqsubset \mathscr{D}$ (cf.~Assumption~\ref{as:MAIN}):
\begin{enumerate}
\item $\argmax_{\bm{\delta} \in \mathbb{S}_K}\Omega_{\lambda, \mathcal{F}}(\bm{\delta})$ is a non-empty and compact subset of~$\mathbb{S}_K$ for every~$\lambda \in [0, 1]$;
\item the function~$\lambda \mapsto \max_{\bm{\delta} \in \mathbb{S}_K}\Omega_{\lambda, \mathcal{F}}(\bm{\delta})$ is continuous and convex on~$[0, 1]$;
\item the correspondence~$\lambda \mapsto \argmax_{\bm{\delta} \in \mathbb{S}_K}\Omega_{\lambda, \mathcal{F}}(\bm{\delta})$ is upper hemicontinuous on~$[0, 1]$;
\item for every array~$\mathcal{G}$, say, as in~\eqref{eqn:array} that satisfies~$\mathcal{G} \sqsubset D_{cdf}([a,b])$ (but not necessarily~$\mathcal{G} \sqsubset \mathscr{D}$), it holds that $\sup_{\bm{\delta} \in \mathbb{S}_K}	\Omega_{\lambda, \mathcal{G}}(\bm{\delta})$ is finite.
\end{enumerate} 
\end{proposition}
As a consequence of Proposition~\ref{prop:exmax}, if~$\mathcal{F}$ entering the objective~$\Omega_{\lambda, \mathcal{F}}(\bm{\delta})$ were known, the DM could simply attempt to solve the optimization problem~$\max_{\bm{\delta} \in \mathbb{S}_K}\Omega_{\lambda, \mathcal{F}}(\bm{\delta})$ and would then roll out an (approximate) solution to the population (possibly after comparing solutions for different values of the preference parameter~$\lambda$). However, note that this is \emph{not} feasible, because the conditional cdfs entering the problem are not available to the DM. Nevertheless, the DM can use the training sample to estimate those quantities and use a plug-in approach to empirically determine an (approximate) best policy, which we shall detail in Section~\ref{sec:esr}.

\subsection{Regret}\label{sec:reg}

The goal of the DM is to use the training sample~$\tilde{\bm{W}}_j = (\tilde{Y}_{j}, X_j, Z_j, D_j)$ for~$j = 1, \hdots, n$, for which~$(\bm{Y}_j', X_j, Z_j)' \sim \mathcal{F}$, unknown, to learn an element of
\begin{equation*}
\arg\max_{\bm{\delta} \in \mathbb{S}_K} \Omega_{\lambda, \mathcal{F}}(\bm{\delta}), \quad \text{ for } \lambda \in [0, 1],
\end{equation*}
a non-empty set (given the assumptions imposed throughout) as just established in Proposition~\ref{prop:exmax}. Given a policy~$\bm{\pi}_n$, i.e., an algorithm to single out a decision rule in~$\mathbb{S}_K$ based on the training sample, the regret we shall work with is
\begin{equation}\label{eqn:regret}
r(\bm{\pi}_n; \lambda, \mathcal{F}) := \max_{\bm{\delta} \in \mathbb{S}_K} \Omega_{\lambda, \mathcal{F}}(\bm{\delta}) - \Omega_{\lambda, \mathcal{F}} \left(\bm{\pi}_n(\tilde{\bm{W}}_j, j = 1, \hdots, n)\right).
\end{equation}
The regret measures the quality of the recommendation the DM makes based on the sample of subjects~$j = 1, \hdots, n$ in terms of the value of the objective function. 

If the DM chooses from a set of policies~$\bm{\pi}_n^{\lambda}$ indexed by~$\lambda \in \Lambda \subseteq [0, 1]$ and the preference parameter~$  \hat{\lambda}_n \in \Lambda$ is chosen in a \emph{data-driven way} (i.e., as a function of~$\tilde{\bm{W}}_j$ for~$j = 1, \hdots, n$), the relevant regret notion becomes
\begin{equation}
r(\bm{\pi}^{  \hat{\lambda}_n}_n;   \hat{\lambda}_n, \mathcal{F}) = \max_{\bm{\delta} \in \mathbb{S}_K} \Omega_{\hat{\lambda}_n, \mathcal{F}}(\bm{\delta}) - \Omega_{\hat{\lambda}_n, \mathcal{F}} \left(\bm{\pi}^{\hat{\lambda}_n}_n(\tilde{\bm{W}}_j, j = 1, \hdots, n)\right),
\end{equation}
the behavior of which we study in such situations. That is, given the DM has decided to use the concrete value~$\hat{\lambda}_n$, we ask how good the performance of the corresponding policy~$\bm{\pi}^{  \hat{\lambda}_n}_n$ is relative to the best one for the selected value of the preference parameter. An alternative regret notion, which is interesting if~$\hat{\lambda}_n$ estimates a certain target preference parameter~$\lambda^*$ (cf.~the situation described in Section~\ref{sec:tune}) is briefly discussed in Appendix~\ref{sec:altreg}.

\section{Empirical success policies}\label{sec:esr}

We now consider \emph{empirical success policies}. Such policies depend on two quantities: First, the preference parameter~$\lambda \in [0, 1]$, which defines the target, and a tuning parameter~$\varepsilon > 0$ regulating the optimization accuracy. Given these two parameters, the DM proceeds in the following way (for some quantities, we abstain from signifying their dependence on~$n$ notationally to keep the expressions easy to read):
\begin{enumerate}
\item For every~$i \in \{1, \hdots, K\}$,~~$x \in \mathcal{X}$, and~$z \in \mathcal{Z}$, determine~$\hat{F}^i(\cdot\mid x, z)$, the empirical cdf based on all observations~$j$ in~
\begin{equation}\label{eqn:selobs}
\mathcal{M}^i_{x, z} := \{j = 1, \hdots, n: D_j = i, X_j = x, Z_j = z\},
\end{equation}
i.e.,~$$\hat{F}^i(\cdot\mid x, z) := |\mathcal{M}^i_{x, z}|^{-1} \sum_{j \in \mathcal{M}^i_{x, z}}  \mathds{1}( Y_{D_j, j} \leq \cdot) = |\mathcal{M}^i_{x, z}|^{-1} \sum_{j \in \mathcal{M}^i_{x, z}}  \mathds{1}( Y_{i,j} \leq \cdot),$$ which we set equal to, e.g., the cdf corresponding to point mass at~$b$ (the upper endpoint of the support of the cdfs one is working with) in case~$\mathcal{M}^i_{x, z}$ is empty.
\item For every~$x \in \mathcal{X}$ and~$z \in \mathcal{Z}$, obtain estimates~
\begin{align*}
&\hat{p}_{X}(x) :=\frac{ |\{j: X_j = x\}|}{n},~\hat{p}_{Z}(z) :=\frac{ |\{j: Z_j = z\}|}{n},~ \text{ and }   \hat{p}_{X}(x, z) := \frac{|\{j: X_j = x, Z_j = z\}|}{n}, 
\end{align*}
and set~$\hat{p}_{Z\mid X = x}(z) := \hat{p}_{X, Z}(x, z)/\hat{p}_{X}(x)$, which we set~$0$ if~$\hat{p}_{X}(x) = 0$.
\item Set~$\hat{\mathcal{F}}_n :=  \left[(\hat{F}^i(\cdot\mid x, z), \hat{p}_{X, Z}(x, z)): i = 1, \hdots, K;~x \in \mathcal{X};~z \in \mathcal{Z}\right]$.
\item Choose
\begin{equation}\label{eqn:approxmax}
\bm{\pi}^{\varepsilon, \lambda}_n(\tilde{\bm{W}}_j, j = 1, \hdots, n) \in \bigg\{\bm{\gamma} \in \mathbb{S}_K : \sup_{\bm{\delta} \in \mathbb{S}_K} \Omega_{\lambda, \hat{\mathcal{F}}_n}(\bm{\delta}) - \Omega_{\lambda, \hat{\mathcal{F}}_n}(\bm{\gamma}) \leq \varepsilon\bigg\}.
\end{equation}
\end{enumerate}

Some technical remarks are discussed in Appendix~\ref{sec:techpol}.

\begin{remark}[Data-driven preference parameters]\label{rem:ddtune}
A data-driven preference parameter is any function~$\hat{\lambda}_n$, say, that maps the training sample to~$[0, 1]$, i.e.,~$$\hat{\lambda}_n: \R^n \times \mathcal{X}^n \times \mathcal{Z}^n \times \{1, \hdots, K\}^n \to [0, 1].$$ To extract a decision rule, the function~$\hat{\lambda}_n$ is applied to the data to obtain a value in~$[0, 1]$, which is then used as the input in the above description of the policy to obtain the decision rule~$$\bm{\pi}_n^{\varepsilon, \hat{\lambda}_n(\tilde{\bm{W}}_j, j = 1, \hdots, n)}(\tilde{\bm{W}}_j, j = 1, \hdots, n);$$ we shall (with some abuse of notation) drop the dependence on the training sample and simply write~$\bm{\pi}_n^{\varepsilon, \hat{\lambda}_n}$ if no confusion can arise. For example, one could proceed in three steps:
\begin{enumerate}
\item Fix a finite set of preference parameters~$\Lambda \subseteq [0, 1]$, which one intends to consider, and obtain (approximate) solutions~$\bm{\pi}_n^{\varepsilon, \lambda}$ as in~\eqref{eqn:approxmax} for every~$\lambda \in \Lambda$.
\item Compare the solutions according to some criterion (cf.~Section~\ref{sec:tune} for a specific suggestion) and choose~$\hat{\lambda}_n$ accordingly.
\item Roll out~$\bm{\pi}_n^{\varepsilon, \hat{\lambda}_n}$.
\end{enumerate}
\end{remark}

\subsection{Regret bounds}

We now establish an upper bound on the expected regret of any empirical success policy with data-dependent preference parameter. In addition to the bound on expected regret, the result also contains high-probability bounds on the regret.

\begin{theorem}\label{thm:upreg2}
Let~$(\bm{Y}_j', X_j, Z_j)' \sim \mathcal{F} \sqsubset \mathscr{D}$ for~$j =1, \hdots, n$ and abbreviate
\begin{equation}
\eta(\mathcal{F}) := \max_{x, z} \sum_{i = 1}^K\frac{1}{ \sqrt{\mathbb{P}\left(D_j = i \mid X_j = x, Z_j = z\right)}} \times \sum_{\overline{z} \in \mathcal{Z}} \frac{1}{\sqrt{p_Z(\overline{z})}}.
\end{equation} 
Then, for \emph{any} data-driven preference parameter~$  \hat{\lambda}_n: \R^n \times \mathcal{X}^n \times \mathcal{Z}^n \to [0, 1]$, the expected regret 
\begin{equation}\label{eqn:regupmain}
\mathbb{E}^*\left(r(\bm{\pi}^{\varepsilon,   \hat{\lambda}_n}_n;   \hat{\lambda}_n, \mathcal{F}) \right) \leq 22 \eta(\mathcal{F}) \sqrt{ \frac{\log(K)|\mathcal{X}|}{n}}  +  4 \sqrt{\frac{|\mathcal{Z}|}{n}} + \varepsilon;
\end{equation}  
furthermore, for every~$\rho > 0$ it holds that
\begin{equation}\label{eqn:hprobreg}
\mathbb{P}^*\left(
r(\bm{\pi}^{\varepsilon,  \hat{\lambda}_n}_n;  \hat{\lambda}_n, \mathcal{F})
\geq \rho + \varepsilon
\right) \leq 4 | \mathcal{Z}||\mathcal{X}|K^2 \max_{x, z, i}  e^{- \frac{nq_{x,z,i} \rho^2}{512(|\mathcal{X}| \vee |\mathcal{Z}|)^2}},
\end{equation}
and where~$q_{x,z,i} := p_{X, Z, D}(x, z, i)$.
\end{theorem}

It is important to emphasize that the upper bound just given decays at the parametric rate~$1/\sqrt{n}$ in the size~$n$ of the training sample, given that one chooses the optimization procedure in~\eqref{eqn:approxmax} in such a way that the optimization error is of the order~$1/\sqrt{n}$. By similar methods to the ones in~\cite{kpv3}, this could formally be achieved, e.g., by optimizing the empirical objective function~$\bm{\delta} \mapsto \Omega_{\lambda, \hat{\mathcal{F}}_n}(\bm{\delta})$ over a fine enough grid in~$\mathbb{S}_K$ (exploiting that the true underlying objective function~$\bm{\delta} \mapsto \Omega_{\lambda, \mathcal{F}}(\bm{\delta})$ is Lipschitz continuous, cf.~Lemma~\ref{lem:exmax}). We have decided not to specify the optimization method explicitly in this article. In practice, running an optimization algorithm such as a Nelder-Mead search (possibly multiple times at randomly chosen initial values) with a choice of accuracy parameters that one can afford in a particular application (in terms of runtime) is more convenient than optimizing over a grid and is what we do in our numerical and empirical results. The question which optimization heuristic should be used in terms of efficiency is interesting and deserves further investigation, but goes beyond the scope of the present article.

Next, we comment on the quantity~$\eta(\mathcal{F})$ that enters the upper bound in Theorem~\ref{thm:upreg2}. This quantity is composed of two factors. The first, \begin{equation}\label{eqn:firstqbd}
\max_{x, z} \sum_{i = 1}^K\frac{1}{ \sqrt{\mathbb{P}\left(D_j = i \mid X_j = x, Z_j = z\right)}},
\end{equation}
is related to what is typically referred to as an overlap condition. It is particularly large if there exists a pair~$x$ and~$z$ and a treatment~$i$, say, that is only assigned with a very low probability whenever~$X_j = x$ and~$Z_j = z$; this then leads to less accurate estimation of the cdf~$\hat{F}^i(\cdot \mid x, z)$. On the other hand, if $$ \mathbb{P}\left(D_j = i \mid X_j = x, Z_j = z\right) \approx  1/K,$$ i.e., if the assignment is balanced, then~\eqref{eqn:firstqbd} is approximately bounded from above by~$K^{3/2}.$ The second factor~$$\sum_{\overline{z} \in \mathcal{Z}} \frac{1}{\sqrt{p_Z(\overline{z})}}$$ is smallest if the protected subgroups are of equal size, in which case it equals~$|\mathcal{Z}|^{3/2}$ and becomes large if a protected subgroup is rather small. This is not surprising, because the objective (deliberately) does not down-weigh unfairness against small groups (cf.~the discussion in Remark~\ref{rem:groupsize}), and therefore needs accurate estimates for the cdfs also within small groups. Hence, if there are small groups, the problem of determining the optimal policy becomes more difficult, which is reflected in the larger upper bound. 

We note that the theoretical guarantees given in Theorem~\ref{thm:upreg2} are informative if the cardinalities of~$\mathcal{X}$ and~$\mathcal{Z}$ are small relative to the size of the training sample. For a discussion of the setting where~$\mathcal{X}$ is not finite, we refer the reader to Section~\ref{sec:contX}. In Appendix \ref{sec:valinterpol} we discuss how the value function $\lambda \mapsto \max_{\bm{\delta} \in \mathbb{S}_K} \Omega_{\lambda, \mathcal{F}}(\bm{\delta})$ can be estimated by interpolation, which may be of some independent interest.

\subsection{Consistency}\label{ssec:consi}

As the next theoretical result in this article, we give conditions under which the~$d_1$-distance of the policy~$\bm{\pi}^{\varepsilon_n, \hat{\lambda}_n}$ to the (non-empty) \emph{random set} of maximizers\footnote{Here, as usual, the distance of a point~$x$ in a metric space~$(X, d)$ to a non-empty subset~$A \subseteq X$ is defined as~$\inf_{z \in A} d(x, z)$.}
\begin{equation}\label{eqn:constarget1}
\arg\max_{\bm{\delta} \in \mathbb{S}_K} \Omega_{  \hat{\lambda}_n, \mathcal{F}}(\bm{\delta})
\end{equation}
converges to zero in probability as~$n \to \infty$; i.e., we give conditions under which the policy~$\bm{\pi}^{\varepsilon_n,   \hat{\lambda}_n}_n$ is consistent. Furthermore, we shall also quantify the rate of convergence to zero of the probability that the decision rule deviates from its target in~\eqref{eqn:constarget1} by more than a given distance~$\rho > 0$. 

As in the previous section, we allow for data-driven preference parameters~$\hat{\lambda}_n$, so that the set of maximizers in the previous display is data-dependent (unless~$\hat{\lambda}_n$ is constant). In the context of establishing consistency, this introduces some technical difficulties that---in contrast to Theorem~\ref{thm:upreg2}---need to be traded off with additional assumptions on the data-driven preference parameter choice. In Appendix~\ref{sec:auxMest} we exhibit the subtleties that arise for consistency results in the presence of data-driven preference parameters~$\hat{\lambda}_n$, which explains why additional restrictions on~$\hat{\lambda}_n$ need to be imposed for a consistency result. 

\subsubsection{Consistency of the policy}

Recall that all random variables are defined on an underlying probability space~$(\Omega, \mathcal{A}, \mathbb{P})$. We denote convergence in~$\mathbb{P}$-probability by~``$\cp$" and convergence in outer~$\mathbb{P}$-probability by~``$\cps$". Furthermore, for~$\rho > 0$ and~$\lambda \in [0, 1]$  we denote
\begin{equation}\label{eqn:Mrhodef}
\mathbb{M}_{\mathcal{F}, \lambda}^{\rho} = \big\{\bm{\gamma} \in \mathbb{S}_K : d_1\big( 
\bm{\gamma},
\argmax_{\bm{\delta} \in \mathbb{S}_K} \Omega_{\lambda, \mathcal{F}}(\bm{\delta})
\big) \geq \rho
\big \}.
\end{equation}
For every~$\rho > 0$ we denote
\begin{equation}\label{eqn:cFdef}
c_{\mathcal{F}}(\rho, \lambda) :=  \max_{\bm{\delta} \in \mathbb{S}_K} \Omega_{\lambda, \mathcal{F}}(\bm{\delta}) - \max_{\bm{\delta} \in \mathbb{M}_{\mathcal{F},  \lambda}^{\rho}} \Omega_{\lambda, \mathcal{F}}(\bm{\delta}),
\end{equation}
if~$\mathbb{M}_{\mathcal{F},  \lambda}^{\rho} \neq \emptyset$ and we set~$c_{\mathcal{F}}(\rho, \lambda) = 0$ in case~$\mathbb{M}_{\mathcal{F},  \lambda}^{\rho}$ is empty. Proposition~\ref{prop:exmax} shows that the quantities appearing in~\eqref{eqn:Mrhodef} and \eqref{eqn:cFdef} are well defined (noting that~$\mathbb{M}_{\mathcal{F}, \lambda}^{\rho}$ is a compact set). Based on Lemma~\ref{lem:mtune}, the following result exhibits a general upper bound on the probability that the~$d_1$-distance between~$\bm{\pi}_{n}^{\varepsilon_n, \hat{\lambda}_{n}}$ and~$\argmax_{\bm{\delta} \in \mathbb{S}_K} \Omega_{\lambda, \mathcal{F}}(\bm{\delta})$ exceeds~$\rho$, which then leads to consistency statements under further conditions on the quantities appearing in that upper bound. 
\begin{theorem}\label{thm:consi}
Let~$(\bm{Y}_j', X_j, Z_j)' \sim \mathcal{F} \sqsubset \mathscr{D}$ for~$j =1, \hdots, n$. Let a sequence of preference parameters~$$\hat{\lambda}_n: \R^n \times \mathcal{X}^n \times \mathcal{Z}^n \to \Lambda \subseteq [0, 1]$$ be given, where~$\Lambda \neq \emptyset$ does not depend on~$n$. Then, for every~$\rho > 0$, the function~$\lambda \mapsto c_{\mathcal{F}}(\rho, \lambda)$ is Borel measurable, and for every positive real number~$\zeta$, it holds that
\begin{equation}
\begin{aligned}\label{eqn:firstconsup}
\mathbb{P}^* \left( \bm{\pi}_{n}^{\varepsilon_n, \hat{\lambda}_{n}}  \in \mathbb{M}_{\mathcal{F},  \hat{\lambda}_{n}}^{\rho}\right) 
\leq 2 |\mathcal{X}|
\left(K^2 + 1\right) & \sum_{z \in \mathcal{Z}} \max_{x, j}  e^{-\frac{nq_{x,z,j} \zeta^2}{8 |\mathcal{X}|^2 }} + 2 |\mathcal{Z}| e^{-\frac{n\zeta^2}{2 |\mathcal{Z}|^2}} \\
& + \mathbb{P}^*\left( 0 <   c_{\mathcal{F}}(\rho, \hat{\lambda}_n) <\varepsilon_n + 6 \zeta  \right),
\end{aligned}
\end{equation}
for~$q_{x,z,i} := p_{X, Z, D}(x, z, i)$; and where the event in the latter probability is measurable if~$\hat{\lambda}_n$ is measurable. Therefore, if there exists a sequence~$\zeta_n > 0$ such that for every~$\rho > 0$ we have~$$n\zeta_n^2 \to \infty \quad \text{ and }  \quad \mathbb{P}^*\left( 0 <   c_{\mathcal{F}}(\rho, \hat{\lambda}_n) <\varepsilon_n + 6 \zeta_n \right) \to 0,$$ then the policy~$\bm{\pi}_{n}^{\varepsilon_n, \hat{\lambda}_{n}}$ is consistent, i.e.,
\begin{equation}\label{eqn:conspro}
d_1\left(\bm{\pi}_{n}^{\varepsilon_n, \hat{\lambda}_{n}}, \arg \max_{\bm{\delta}\in\mathbb{S}_K} \Omega_{  \hat{\lambda}_{n}, \mathcal{F}}(\bm{\delta})\right) \cps 0, \quad \text{ as } n \to \infty.
\end{equation}
In particular, the policy~$\bm{\pi}_{n}^{\varepsilon_n, \hat{\lambda}_{n}}$ is consistent if~$\Lambda$ consists of finitely many elements and~$\varepsilon_n \to 0$.
\end{theorem}
We note that the upper bound in~\eqref{eqn:firstconsup} holds under minimal assumptions on the preference parameter. While the first two summands in that upper bound converge to~$0$ as~$n \to \infty$ (due to Assumption~\ref{as:posprob}), there is no guarantee that the third summand converges to~$0$, in general. This is only true under additional conditions, e.g., finiteness of~$\Lambda$. Note that the finiteness condition does not present a serious practical limitation, as one can typically only consider a finite grid of $\lambda$ values.

The final question that remains is how to choose the preference parameter~$\lambda$. Before we address this question, we show that a policy that is consistent for~$\arg\max_{\bm{\delta} \in \mathbb{S}_K} \Omega_{  \hat{\lambda}_n, \mathcal{F}}(\bm{\delta})$ and is based on a data-driven preference parameter~$\hat{\lambda}_n$ that converges to a non-random~$\lambda^* \in [0, 1]$, is also consistent for~$\arg\max_{\bm{\delta} \in \mathbb{S}_K} \Omega_{ \lambda^*, \mathcal{F}}(\bm{\delta})$; i.e., the policy is consistent for the target in which~$\hat{\lambda}_n$ is replaced by the limiting value~$\lambda^*$. A similar statement is established if the preference parameter approaches a random variable with finite support. Such a result is interesting, in case an \emph{infeasible} ``oracle'' preference parameter~$\lambda^* = \lambda^*(\mathcal{F})$ depending on the unknown~$\mathcal{F}$ is targeted by a consistent data-driven preference parameter selection mechanism, as will be done in Section~\ref{sec:tune}.
\begin{proposition}\label{prop:conslamb}
Let~$(\bm{Y}_j', X_j, Z_j)' \sim \mathcal{F} \sqsubset \mathscr{D}$ for~$j =1, \hdots, n$. Let a sequence of preference parameters~$\hat{\lambda}_n: \R^n \times \mathcal{X}^n \times \mathcal{Z}^n \to [0, 1]$ be given that converges in (outer) probability to a (non-random)~$\lambda^* \in [0, 1]$ and let~$\bm{\pi}_n$ be a sequence of policies. Then
\begin{equation}\label{eqn:conslamb}
d_1\left( \bm{\pi}_n, \arg\max_{\bm{\delta} \in \mathbb{S}_K} \Omega_{  \hat{\lambda}_n, \mathcal{F}}(\bm{\delta}) \right) \cps 0 \quad \Rightarrow \quad 
d_1\left( \bm{\pi}_n, \arg\max_{\bm{\delta} \in \mathbb{S}_K} \Omega_{ \lambda^*, \mathcal{F}}(\bm{\delta}) \right) \cps 0.
\end{equation}
If~$\hat{\lambda}_n \in \Lambda$, a finite subset of~$[0, 1]$ that does not depend on~$n$, and if~$\hat{\lambda}_n - \hat{\lambda}_n^* \cps 0$ for a sequence random variables~$\hat{\lambda}_n^* \in \Lambda$, then~\eqref{eqn:conslamb} holds with~$\lambda^*$ replaced by~$\hat{\lambda}_n^*$.
\end{proposition}  
Proposition~\ref{prop:conslamb} follows from upper-hemicontinuity of~$\lambda \mapsto 	\arg\max_{\bm{\delta} \in \mathbb{S}_K} \Omega_{  \lambda, \mathcal{F}}(\bm{\delta})$ that was established in Proposition~\ref{prop:exmax}.  Proposition~\ref{prop:conslamb} establishes that if one can single out a preferable theoretical preference parameter~$\lambda^*$ that itself is infeasible but consistently estimable, and if the policy~$\bm{\pi}_n$ is consistent for the \emph{random} set of maximizers~$\arg\max_{\bm{\delta} \in \mathbb{S}_K} \Omega_{  \hat{\lambda}_n, \mathcal{F}}(\bm{\delta})$, then the policy is also consistent for the \emph{non-random} set of maximizers~$\arg\max_{\bm{\delta} \in \mathbb{S}_K} \Omega_{  \lambda^*, \mathcal{F}}(\bm{\delta})$.

\section{Choosing the preference parameter}\label{sec:tune}

Theorem~\ref{thm:upreg2} shows that the expected regret of the policies considered decays at the parametric rate regardless of how~$\hat{\lambda}_n$ is chosen. Furthermore, we have seen in Theorem~\ref{thm:consi} that whenever the preference parameter is selected from a finite set of candidate values, the corresponding policy is consistent in the sense that $$d_1\left(\bm{\pi}_{n}^{\varepsilon_n, \hat{\lambda}_{n}}, \arg \max_{\bm{\delta}\in\mathbb{S}_K} \Omega_{  \hat{\lambda}_{n}, \mathcal{F}}(\bm{\delta})\right) \cps 0, \quad \text{ as } n \to \infty,$$ and Proposition~\ref{prop:conslamb} even shows that if~$\hat{\lambda}_n$ ``converges'' in (outer) probability to~$\hat{\lambda}^*_n$ (in the sense that their distance converges to~$0$ in (outer) probability), then $$d_1\left(\bm{\pi}_{n}^{\varepsilon_n, \hat{\lambda}_{n}}, \arg \max_{\bm{\delta}\in\mathbb{S}_K} \Omega_{  \hat{\lambda}_n^*, \mathcal{F}}(\bm{\delta})\right) \cps 0, \quad \text{ as } n \to \infty.$$ Those guarantees are not in place if~$\Lambda$ is not finite, but this is not a strong restriction: In practice, a DM will initially fix a \emph{finite} set of preference parameters~$\Lambda$, say, from which an element is then selected in a data-driven way, possibly by comparing properties of the inferred policies $\bm{\pi}_{n}^{\varepsilon_n, \lambda}$ for~$\lambda \in \Lambda$. In this section, we assume that~$\Lambda$ is finite and consider in more detail the question of which preference parameter~$\lambda^*$ in~$\Lambda$ should be used. A first answer to this question was already briefly touched upon in Remark~\ref{rem:tunefirst}, where it was recommended that the DM study aspects of (estimates of) the functions~$$\lambda \mapsto \mathsf{T}\left(\langle \bm{\pi}_{n}^{\varepsilon_n, \lambda}, \mathcal{F} \rangle \right) \quad \text{ and } \quad \lambda \mapsto \mathsf{S} \left(\langle \bm{\pi}_{n}^{\varepsilon_n, \lambda}, \mathcal{F} \rangle_z, \langle \bm{\pi}_{n}^{\varepsilon_n, \lambda}, \mathcal{F} \rangle \right).$$ The empirical examples in Section~\ref{sec:empirics} will illustrate this approach.

To provide another answer to this question, we adapt the ``price of fairness'' criterion in \cite{berk2017convex} to our context. Suppose that the DM has obtained the policies~$\bm{\pi}_{n}^{\varepsilon_n, \lambda}$ for every~$\lambda \in \Lambda$, which we shall assume to contain~$0$ in what follows, and now wants to decide which one among those to roll out. Recall that if the DM decides to roll out the policy~$\bm{\pi}_{n}^{\varepsilon_n, \lambda}$ for a given~$\lambda \in \Lambda$, this results in the \emph{population} cdf~$\langle \bm{\pi}_{n}^{\varepsilon_n, \lambda}, \mathcal{F} \rangle$, which is unknown to the DM, because~$\mathcal{F}$ is unknown. Ignore this ``estimation step'' for a moment, and define for every~$\lambda \in \Lambda$ the difference $$\Delta_n(\lambda, \mathcal{F}) := \mathsf{T}(\langle \bm{\pi}_{n}^{\varepsilon_n, 0}, \mathcal{F} \rangle) - \mathsf{T}(\langle \bm{\pi}_{n}^{\varepsilon_n, \lambda}, \mathcal{F} \rangle),$$ which depends on the data due to its dependence on the policies. The difference~$\Delta_n(\lambda, \mathcal{F})$ measures, for a given~$\lambda$, the difference in the target functional~$\mathsf{T}$ applied to the policy~$\bm{\pi}_{n}^{\varepsilon_n, 0}$ (that does not penalize unfairness) and the policy~$\bm{\pi}_{n}^{\varepsilon_n, \lambda}$ (that penalizes unfairness with penalty~$\lambda$). Note that if~$\Delta_n(\lambda, \mathcal{F}) > 0$, then penalizing unfairness leads to a decrease/loss in the target functional.\footnote{Note that~$\Delta_n(\lambda, \mathcal{F})$ can in principle also be negative, due to the policy being based on estimates of~$\mathcal{F}$.} Therefore, one may interpret~$\Delta(\lambda, \mathcal{F})$ as the ``price'' the DM is paying for penalizing unfairness by rolling out~$\bm{\pi}_n^{\varepsilon_n, \lambda}$ instead of~$\bm{\pi}_n^{\varepsilon_n, 0}$. We now assume that there is a maximal ``budget'' the DM can afford to spend in penalizing for unfairness, i.e., a maximal decrease in the
target functional~$\mathsf{T}$ that can be
tolerated. Call this budget~$\beta > 0$. Given that the DM intends to select the preference parameter~$\lambda$ from~$\Lambda$, we thus define the targeted preference parameter as 
\begin{equation}\label{eqn:lstardef}
\lambda^{*}_n(\beta, \mathcal{F}) := \max \{\lambda \in \Lambda: \Delta_n(\lambda, \mathcal{F}) \leq \beta\};
\end{equation} note that~$\lambda^{*}(\beta, \mathcal{F})$ exists because~$\Lambda$ is finite and contains~$0$. The so-defined ``oracle'' target~$\lambda^{*}_n(\beta, \mathcal{F})$ according to budget~$\beta$ is the maximal unfairness penalty that satisfies the budget constraint. Because~$\mathcal{F}$ is unknown, we have to estimate~$\lambda^{*}_n(\beta, \mathcal{F})$.

Taking care of the estimation step is more difficult than it may perhaps seem at first sight. To see this, note that the function~$h$, say, that maps~$z = (z_0, z_1, \hdots, z_M) \in \{0\} \times \R^M$ ($M \in \N$) to~$\max\{i : z_i \leq \beta\}$ is not everywhere continuous; the set of discontinuity points of~$h$ is given by
\begin{equation}\label{eqn:disconth}
D(h) := \{z \in \{0\} \times \R^M: z_{h(z)} = \beta\}.
\end{equation}
Therefore, replacing~$\mathcal{F}$ by~$\hat{\mathcal{F}}_n$ in the definition of the preference parameter in the penultimate display in a plug-in fashion, and even though $\Delta_n(\lambda, \hat{\mathcal{F}}_n) \approx \Delta_n(\lambda, \mathcal{F}_n)$ for every~$\lambda \in \Lambda$ actually holds under weak assumptions (cf.~the proof of Theorem~\ref{thm:consitune}), we can in general not easily conclude (e.g., by a suitable version of the continuous mapping theorem) that the plug-in preference parameter that replaces~$\mathcal{F}$ by~$\hat{\mathcal{F}}_n$ in~\eqref{eqn:lstardef} \emph{consistently} estimates its target without further assumptions.

What can be achieved---without additional conditions beyond Assumptions~\ref{as:posprob} and~\ref{as:MAIN} and a finiteness condition on~$\Lambda$---is to consistently \emph{under}estimate the oracle and to consistently satisfy the budget constraint. This is achieved by replacing~$\mathcal{F}$ by~$\hat{\mathcal{F}}_n$ in~\eqref{eqn:lstardef} \emph{and} by introducing some ``slack'', i.e., we use the estimator \begin{equation}\label{eqn:preftune}
\hat{\lambda}_n^*(\beta, \hat{\mathcal{F}}_n) = \max \left\{\lambda \in \Lambda: \Delta_n(\lambda, \hat{\mathcal{F}}_n) \leq \beta (1-c_n)\right \} \quad \text{ with } \quad c_n := \sqrt{\log(n)/n},
\end{equation} where~$$\Delta_n(\lambda, \hat{\mathcal{F}}_n) := \mathsf{T}(\langle \bm{\pi}_{n}^{\varepsilon_n, 0}, \hat{\mathcal{F}}_n \rangle) - \mathsf{T}(\langle \bm{\pi}_{n}^{\varepsilon_n, \lambda}, \hat{\mathcal{F}}_n \rangle).$$ The precise result is as follows.

\begin{theorem}\label{thm:consitune}
Let~$(\bm{Y}_j', X_j, Z_j)' \sim \mathcal{F} \sqsubset \mathscr{D}$ for~$j =1, \hdots, n$. Let~$\Lambda = \{0, \lambda_1, \hdots, \lambda_M\} \subseteq [0, 1]$ with $0 < \lambda_1 < \lambda_2 < \hdots < \lambda_M \leq 1$ for some fixed~$M\in \N$. Fix a budget~$\beta > 0$. If~$\varepsilon_n \to 0$, it follows that
\begin{equation}\label{eqn:conscons}
\mathbb{P}^* \left(\hat{\lambda}_n^*(\beta, \hat{\mathcal{F}}_n) > \lambda_n^*(\beta, \mathcal{F}) \right) \to 0 \quad \text{ and } \quad \mathbb{P}^* \left(\Delta_n\left(\hat{\lambda}_n^*(\beta, \hat{\mathcal{F}}_n), \mathcal{F}\right) > \beta \right) \to 0,
\end{equation}
i.e.,~$\hat{\lambda}_n^*(\beta, \hat{\mathcal{F}}_n)$ consistently underestimates~$\lambda_n^*(\beta, \mathcal{F})$ and consistently satisfies the budget constraint. Under the additional condition that for some~$\alpha > 0$ it holds that 
\begin{equation}\label{eqn:addcond}
\mathbb{P}^*\left( \beta - \Delta_n\left(\lambda_n^*(\beta, \mathcal{F}), \mathcal{F}\right) \leq [\alpha + \beta]c_n \right) \to 0,
\end{equation}
it even holds that $$\hat{\lambda}_n^*(\beta, \hat{\mathcal{F}}_n) - \lambda_n^*(\beta, \mathcal{F}) \cps 0,$$ from which it then also follows that
\begin{equation}\label{eqn:consitune}
d_1\left( \bm{\pi}_n^{\varepsilon_n, \hat{\lambda}_n^*(\beta, \hat{\mathcal{F}}_n)}, \arg\max_{\bm{\delta} \in \mathbb{S}_K} \Omega_{ \lambda_n^{*}(\beta, \mathcal{F}), \mathcal{F}}(\bm{\delta}) \right) \cps 0.
\end{equation}
\end{theorem}

Theorem~\ref{thm:consitune} provides conditions under which the data-driven preference parameter~$\hat{\lambda}_n^*(\beta, \hat{\mathcal{F}}_n)$ converges in outer probability to its target, and that selecting the policy from a set of empirical success policies indexed over~$\Lambda$ based on~$\hat{\lambda}_n^*(\beta, \hat{\mathcal{F}}_n)$ is consistent for the target that is defined by spending a maximal budget to penalize for unfairness. 

The main condition needed for consistency is the one in~\eqref{eqn:addcond}. It requires that the quantity~$\Delta_n\left(\lambda_n^*(\beta, \mathcal{F}), \mathcal{F}\right)$, which never exceeds~$\beta$ by definition, takes its values in an~$(\alpha + \beta)c_n$ neighborhood of~$\beta$ with outer probability converging to~$0$, thus avoiding the discontinuity problems implied by the function~$h$ as described around~\eqref{eqn:disconth}. That~$\Delta_n\left(\lambda_n^*(\beta, \mathcal{F}), \mathcal{F}\right)$ concentrates at the budget constraint~$\beta$ so that this condition is violated may arise in special cases but is not too likely given that~$\Lambda$ consists only of finitely many values chosen by the user. Hence, the extra condition needed for consistency does not seem to be a very strong restriction. 

If the extra condition~\eqref{eqn:addcond} for consistency is not satisfied, the data-driven preference parameter is ``robust'' in the sense that it consistently satisfies the budget constraint and is at least consistently not larger than the oracle, which is defined as the maximal penalty for unfairness that satisfies the budget constraint. 

That we can (only) consistently underestimate the oracle in general is conceptually in line with general findings in \cite{lsun}, where impossibility results in the related setting of empirical welfare maximization under (estimated) constraints are studied, and ways to take the corresponding estimation error into account are suggested.

\section{Generalization to non-discrete covariates}\label{sec:contX}

Throughout this article, we consider the case where~$\mathcal{Z}$ is finite. So far, we focused on a setting where~$\mathcal{X}$ is finite as well. In case~$\mathcal{X}$ contains infinitely many elements, one could simply ``discretize''~$\mathcal{X}$, and then apply the policy developed and studied for the case of finitely supported covariates in Section~\ref{sec:esr}. Granted the assumptions used to establish theoretical guarantees in earlier sections of this article are satisfied for the \emph{discretized} covariates, the theoretical guarantees (expected regret bounds, consistency, etc.) carry over immediately. One disadvantage of this approach is that the performance guarantees developed, e.g., in Theorem~\ref{thm:upreg2}, get weaker as the discretization gets finer. Yet, the finer the discretization, the more reasonable is, e.g., the conditional independence Assumption~\ref{as:cia}, conditioning on the discretized covariates, which restricts the scope of such a ``direct'' application of our results.

An alternative approach, which is not based on discretization, but can be viewed as a generalization of the approach of~\cite{kitagawa2019equality} to our context, is discussed in the present section. The non-finiteness of the number of elements of~$\mathcal{X}$ necessarily requires some notational and conceptual modifications, which are only maintained locally in this section and the corresponding proofs in Appendix~\ref{app:contX}. The objects studied (e.g., the training sample, policies, or the distributions generated in the roll-out phase) have the same conceptual meaning as before, and hence their interpretation does not require additional discussion on top of that already given earlier. In this section, we maintain throughout the following \emph{list of assumptions and notational conventions, without further mentioning them in the theoretical results}:

\begin{enumerate}
\item We assume that~$\mathcal{Z}$ is finite, but we do not assume that~$\mathcal{X}$ is finite (although this is not ruled out formally).
\item We impose Assumptions~\ref{as:iid},~\ref{as:cia}, and~\ref{as:MAIN} throughout, but no longer impose Assumption~\ref{as:posprob} (which is replaced, to some extent, by~\ref{ass:knownprop} below).
\item We assume that~$\bm{W}_j = (\bm{Y}_j', X_j, Z_j, D_j)'$ takes its values in~$\R^{K} \times  \mathcal{X} \times \mathcal{Z} \times \{1, \hdots, K\}$ equipped with the product sigma algebra, where finite sets (in particular~$\mathcal{Z}$) are equipped with the power set (set of all subsets),~$\R^K$ is equipped with the Borel sigma algebra, and the non-empty set~$\mathcal{X}$ is equipped with some sigma algebra. 
\item We denote a regular conditional distribution of $Y_{i,j}$ given $(X_j, Z_j)$ by~$\mathsf{K}^i_{x, z}(\cdot)$.\footnote{Because~$Y_{i,j}$ is real-valued, the existence of a regular conditional distribution is guaranteed, cf.~Definition A.7 and Theorem A.37 in \cite{liese}.} For~$x \in \mathcal{X}$ and~$z \in \mathcal{Z}$ the conditional cdf~$F^{i}(\cdot \mid x, z) := \mathsf{K}^i_{x, z}((-\infty, \cdot])$. Furthermore, we denote by~$F^i$ the cdf of~$Y_{i, j}$. We collect the conditional cdfs of a random vector~$\bm{Y}_j'$ given~$X_j = x$ and~$Z_j = z$ and $\mathbb{P}_{X, Z}$, the joint distribution of $X_j$ and $Z_j$, in an array
\begin{equation}\label{eqn:array2}
\mathcal{F} := \left[(F^i(\cdot\mid x, z), \mathbb{P}_{X, Z}): i = 1, \hdots, K;~x \in \mathcal{X};~z \in \mathcal{Z}\right].
\end{equation}
The symbol $\mathbb{P}_X$ denotes the distribution of $X_j$, $\mathbb{P}_Z$ denotes the distribution of $Z_j$ with corresponding probability mass function~$p_Z$, which we assume to be strictly positive at every element of~$\mathcal{Z}$, and $\mathbb{P}_{X \mid Z = z}(\cdot) = \mathbb{P}_{X, Z}( \cdot, \{z\})/p_Z(z)$ is a regular conditional distribution of $X_j$ given~$Z_j$ evaluated at $z \in \mathcal{Z}$ (recall that~$\mathcal{Z}$ is assumed finite).
\item A decision rule~$\bm{\delta}$ is a \emph{measurable} function from~$\mathcal{X}$ to~$\mathscr{S}_K$, the simplex in~$\R^K$. We consider situations, where the decision rules are chosen from a non-empty set $\Pi$ of measurable functions from~$\mathcal{X}$ to~$\mathscr{S}_K$.
\item For every decision rule~$\bm{\delta}$ and every~$z \in \mathcal{Z}$, we define (analogously to~\eqref{eqn:mixrollsub}) the cdf
\begin{equation}\label{eqn:langrang}
\langle \bm{\delta}, \mathcal{F} \rangle_z(y) := \int \sum_{i = 1}^K \delta_i(x) F^i( y \mid x, z) d\mathbb{P}_{X \mid Z = z}(x),
\end{equation}
and set (analogously to~\eqref{eqn:mixroll})~$$\langle \bm{\delta}, \mathcal{F} \rangle := \sum_{z \in \mathcal{Z}} \langle \bm{\delta}, \mathcal{F} \rangle_z p_{Z}(z).$$ Analogously to Section~\ref{ssec:distrroll}, the cdfs~$\langle \bm{\delta}, \mathcal{F} \rangle_z$ and~$\langle \bm{\delta}, \mathcal{F} \rangle$ are the conditional and unconditional cdfs obtained by rolling out the decision rule~$\bm{\delta}$.\footnote{More precisely, consider a new draw from the underlying population as in~\eqref{eqn:rollo}, and suppose that its assignment, $D$, say, satisfies Equation~\eqref{eqn:inddraw} for every~$i$,~$x \in \mathcal{X}$,~$y \in \R$ and~$z \in \mathcal{Z}$, e.g., because the mechanism as discussed in Footnote~\ref{foot:concrass} was implemented by the DM. Then, for every~$z \in \mathcal{Z}$,~$\langle \bm{\delta}, \mathcal{F} \rangle_z$ is the cdf of~$Y_D$ given~$Z = z$, and the cdf of~$Y_D$ is~$\langle \bm{\delta}, \mathcal{F} \rangle$.}
\item  We write~$\mathcal{F} \sqsubset_{\Pi} \mathscr{H}$ if for every~$\bm{\delta} \in \Pi$ and every~$z \in \mathcal{Z}$ the cdf $\langle \bm{\delta}, \mathcal{F}\rangle_z$ is an element of the closure of~$\mathscr{H}$ (w.r.t.~the supremum metric).
\end{enumerate}
In addition, we impose the following ``overlap'' condition throughout this section (in place of working with~Assumption~\ref{as:posprob}):
\begin{assumption}[Overlap condition]\label{ass:knownprop}
Let $$e_i(x, z) := \mathbb{P}(D_j = i \mid X_j = x, Z_j = z) \quad \text{ for } i = 1, \hdots, K,$$ define a regular conditional distribution of $D_j$ given $X_j$ and $Z_j$, and assume that (it can be chosen such that) $0 < e_i(x, z)$ for every $i = 1, \hdots, K$, every $x \in \mathcal{X}$, and every $z \in \mathcal{Z}$, and such that~$\mathbb{E}(e_i^{-1}(X_j, Z_j)) < \infty$ for every~$i = 1, \hdots, K$.
\end{assumption}

Granted all the assumptions above, we introduce the class of (non-negative) measurable functions~$$\mathcal{G}_{\Pi} := \{f_{\bm{\delta}, c, z}:\bm{\delta} \in \Pi, c \in \R, z\in\mathcal{Z} \},$$ where $f_{\bm{\delta}, c, z} : \R \times \mathcal{X} \times \mathcal{Z} \times \{1, \hdots, K\} \to \R$ is defined as (recall that~$p_Z(z) > 0$ is assumed for every~$z \in \mathcal{Z}$)
\begin{align*}f_{\bm{\delta}, c, z} (y, x, z^*, d) &:=   \delta_d(x) \times \mathds{1}(y \leq c) \times \frac{\mathds{1}( z^* = z)}{e_d(x, z)p_Z(z)}.
\end{align*}
That the expectation of~$f_{\bm{\delta}, c, z}(Y_{D_j, j}, X_j, Z_j, D_j)$ equals~$\langle \bm{\delta}, \mathcal{F} \rangle_z(c)$ is shown in Lemma~\ref{lem:expf} in Appendix~\ref{app:contX}, delivering (recall~$\tilde{\bm{W}}_j$ from~\eqref{eqn:obs}) the unbiased estimator~$n^{-1} \sum_{j = 1}^n f_{\bm{\delta}, c, z}(\tilde{\bm{W}}_j)$  of $\langle \bm{\delta}, \mathcal{F} \rangle_z(c)$, cf.~also the discussion in Section~3 of \cite{kitagawa2019equality} for a similar approach.
We abbreviate, $$\| P_n - P \|_{\mathcal{G}_{\Pi}} := \sup_{f \in \mathcal{G}_{\Pi}} |n^{-1} \sum_{i = 1}^n f(\tilde{\bm{W}}_j) - \mathbb{E}(f(\tilde{\bm{W}}_j))|.$$ 
Note that suitably ``small'', i.e., ``parametric'', sets of decision rules~$\Pi$ (and hence~$\mathcal{G}_{\Pi}$) allow one to establish, for some finite constant~$c(\mathcal{G}_{\Pi}, P) > 0$, the bound
\begin{equation}\label{eqn:maxineq}
\mathbb{E}^*(\| P_n - P \|_{\mathcal{G}_{\Pi}}) \leq c(\mathcal{G}_{\Pi}, P)/\sqrt{n} < \infty.
\end{equation}
For example, if one additionally assumes that $0 < \eps < e_i(x, z)$ for every $i = 1, \hdots, K$, every $x \in \mathcal{X}$, and every $z \in \mathcal{Z}$, the class of (non-negative) functions~$\mathcal{G}_{\Pi}$ is bounded from above by~$\max_{z \in \mathcal{Z}} 1/(\eps \times p_Z(z))$, so that Theorem 2.14.1 in~\cite{vdVW} delivers~\eqref{eqn:maxineq} under (i) finiteness of a uniform entropy integral (satisfied in particular for VC classes) and (ii) suitable measurability conditions and some structural conditions of the underlying probability space are assumed.

We abstain from formulating expected regret bounds in the present setting in terms of various other possible sufficient conditions for~\eqref{eqn:maxineq}, but develop an upper bound on the regret itself that depends on~$\| P_n - P \|_{\mathcal{G}_{\Pi}}$. This approach has the benefit of also being informative if~\eqref{eqn:maxineq} is not satisfied, e.g., because~$\mathcal{G}_{\Pi}$ is nonparametric, and a different rate of decay holds for the quantity to the left in~\eqref{eqn:maxineq}. We refer the reader to, e.g., \cite{vdVW} and~\cite{gn} for tools to establish~\eqref{eqn:maxineq} under low-level assumptions on~$\Pi$. 

The goal of the decision maker is to choose $\bm{\delta}$ such as to approximate~$$\sup_{\bm{\delta} \in \Pi} \Omega_{\mathcal{F}, \lambda}(\bm{\delta}),$$ with the penalized objective $\Omega_{\mathcal{F}, \lambda}(\bm{\delta})$ as defined in~\eqref{eqn:omegadef} (Lemma~\ref{lem:bdf} in Appendix~\ref{app:contX} shows that, under the maintained assumptions, the supremum in the previous display is finite). Similar to the definition in~\eqref{eqn:poly}, a \emph{policy} is a function~$\bm{\pi}_n: \R^n \times \mathcal{X}^n \times \mathcal{Z}^n \times \{1, \hdots, K\}^n \to \Pi$, and its regret is defined as 
\begin{equation}\label{eqn:regret2}
r(\bm{\pi}_n; \lambda, \mathcal{F}) := \sup_{\bm{\delta} \in \Pi} \Omega_{\lambda, \mathcal{F}}(\bm{\delta}) - \Omega_{\lambda, \mathcal{F}} \left(\bm{\pi}_n(\tilde{\bm{W}}_j, j = 1, \hdots, n)\right),
\end{equation}
where we replace~$\lambda$ by~$\hat{\lambda}_n$ if the preference parameter is selected in a data-driven way.

\subsection{Empirical success type policy}\label{sec:contXes}

For a given preference parameter $\lambda \in [0, 1]$ and tuning parameter $\varepsilon > 0$, the policy $\bm{\pi}^{\varepsilon, \lambda}_n$ we shall consider here is a suitably adapted version of the policy discussed in Section~\ref{sec:esr}. Conceptually, we do no longer work with a plug-in version of the objective (simply replacing $\mathcal{F}$ by an estimate), but we somewhat more carefully estimate the quantities through which the objective actually depends on directly. Here, we exploit that the expectation of~$f_{\bm{\delta}, c, z}(Y_{D_j, j}, X_j, Z_j, D_j)$ equals~$\langle \bm{\delta}, \mathcal{F} \rangle_z(c)$, which is in line with the approach in~\cite{kitagawa2019equality}. 

We work with the ``empirical target'' $\hat{\Omega}_{\lambda}(\bm{\delta})$, say, which, for $\bm{\delta} \in \Pi$ is defined as (cf.~Assumption~\ref{as:MAIN}) $$\hat{\Omega}_{\lambda}(\bm{\delta}) := (1-\lambda) \mathsf{T}(\widehat{\langle \bm{\delta}, \mathcal{F} \rangle} ) - \lambda \max_{z \in \mathcal{Z}} \mathsf{S}(\widehat{\langle \bm{\delta}, \mathcal{F} \rangle_z}, \widehat{\langle \bm{\delta}, \mathcal{F} \rangle}),$$ for estimates $\widehat{\langle \bm{\delta}, \mathcal{F} \rangle_z}$ and $\widehat{\langle \bm{\delta}, \mathcal{F} \rangle}$ defined in~\eqref{eqn:surro} below. 

To obtain~$\widehat{\langle \bm{\delta}, \mathcal{F} \rangle_z}$, we project the function~$c \mapsto n^{-1} \sum_{j = 1}^n f_{\bm{\delta}, c, z}(Y_{D_j, j}, X_j, Z_j, D_j)$, which (although being non-negative, non-decreasing, càdlàg, and pointwise unbiased for $\widehat{\langle \bm{\delta}, \mathcal{F} \rangle_z}$) is not guaranteed to be a cdf, cf.~also the discussion in Section~3 of \cite{kitagawa2019equality}, onto the set of cdfs; here, the projection of a non-decreasing, non-negative, càdlàg function $G: \R \to \R$ onto $D_{cdf}([a,b])$ (equipped with the Kolmogorov-Smirnov distance) is denoted by $M_{a,b}G$, i.e., $M_{a,b}G(x)$ equals $0$ for $x < a$, equals $\min(G(x), 1)$ for $x \in [a,b]$, and equals $1$ for $x \geq b$.
\begin{enumerate}
\item For every $z \in \mathcal{Z}$ and every $\bm{\delta} \in \mathbb{S}_K$ define the following cdfs in~$D_{cdf}([a,b])$ \begin{equation}\label{eqn:surro}
\widehat{\langle \bm{\delta}, \mathcal{F} \rangle_z} := M_{a,b} \left[n^{-1} \sum_{j = 1}^n f_{\bm{\delta}, \cdot, z}(Y_{D_j, j}, X_j, Z_j, D_j) \right], ~ \text{ and } ~ \widehat{\langle \bm{\delta}, \mathcal{F} \rangle} := \sum_{z \in \mathcal{Z}} p_Z(z) \widehat{\langle \bm{\delta}, \mathcal{F} \rangle_{z}}.
\end{equation} 
\item Choose
\begin{equation}\label{eqn:approxmax2}
\bm{\pi}^{\varepsilon, \lambda}_n(\tilde{\bm{W}}_j, j = 1, \hdots, n) \in \bigg\{\bm{\gamma} \in \Pi : \sup_{\bm{\delta} \in \Pi} \hat{\Omega}_{\lambda}(\bm{\delta}) - \hat{\Omega}_{\lambda}(\bm{\gamma}) \leq \varepsilon\bigg\}.
\end{equation}
\end{enumerate}

Lemma~\ref{lem:bdf} in Appendix~\ref{app:contX} shows that the supremum defining the set in~\eqref{eqn:approxmax2} is finite.

If the probabilities~$p_Z$ and the propensities~$e_i(x,z)$ are unknown, a DM needs to replace them by estimates~$\hat{p}_Z(z)$ and~$\hat{e}_{i,n}(x, z)$ (the former defined as in Section~\ref{sec:esr}, while we keep the latter abstract), respectively. In this situation, set $$\tilde{f}_{\bm{\delta}, c, z}(y, x, z^*, d) := \delta_d(x) \times \mathds{1}(y \leq c) \times \frac{\mathds{1}( z^* = z)}{\hat{e}_d(x, z) \hat{p}_Z(z)},$$ which we leave undefined if the denominator equals~$0$. Then, we consider the adapted policy~$\tilde{\bm{\pi}}^{\varepsilon,   \lambda}_n$, say, that differs from~$\bm{\pi}^{\varepsilon, \lambda}_n$ in that it is not based on the quantities defined in~\eqref{eqn:surro} when computing the empirical target~$\hat{\Omega}_{\lambda}(\bm{\delta})$, but replaces them by
\begin{equation}\label{eqn:surro2}
\widetilde{\langle \bm{\delta}, \mathcal{F} \rangle_z} := M_{a,b} \left[n^{-1} \sum_{j = 1}^n \tilde{f}_{\bm{\delta}, \cdot, z}(Y_{D_j, j}, X_j, Z_j, D_j) \right], ~ \text{ and } ~ \widetilde{\langle \bm{\delta}, \mathcal{F} \rangle} := \sum_{z \in \mathcal{Z}} \hat{p}_Z(z) \widetilde{\langle \bm{\delta}, \mathcal{F} \rangle_z}.
\end{equation}

\subsection{Performance bound}

We now provide an upper bound on the regret of  the policies just defined.
\begin{theorem}\label{thm:upregcont}
Let~$(\bm{Y}_j', X_j, Z_j)' \sim \mathcal{F}$ for~$j =1, \hdots, n$, and assume that~$\mathcal{F} \sqsubset_{\Pi} \mathscr{D}$. Then, for \emph{any} data-driven preference parameter~$  \hat{\lambda}_n: \R^n \times \mathcal{X}^n \times \mathcal{Z}^n \to [0, 1]$, we have
\begin{equation}\label{eqn:regupmaincont}
r(\bm{\pi}^{\varepsilon,   \hat{\lambda}_n}_n;   \hat{\lambda}_n, \mathcal{F}) \leq 6 \times \|P_n - P\|_{\mathcal{G}_{\Pi}} + \varepsilon,
\end{equation}  
and~$r(\tilde{\bm{\pi}}^{\varepsilon,   \hat{\lambda}_n}_n;   \hat{\lambda}_n, \mathcal{F})$ is bounded from above by the upper bound in~\eqref{eqn:regupmaincont} plus
\begin{equation}\label{eqn:paraprop}
\|\hat{\bm{p}}_Z - \bm{p}_Z \| + 6 \times \max_{z \in \mathcal{Z}} n^{-1}  \sum_{j = 1}^n  \left |\hat{p}^{-1}_Z(z)\hat{e}^{-1}_{D_j}(X_j, 
z) - p^{-1}_Z(z) e^{-1}_{D_j}(X_j, 
z) \right|,
\end{equation}
which is to be interpreted as~$\infty$ if one of the denominator vanishes.
\end{theorem}
Because~$\E \|\hat{\bm{p}}_Z - \bm{p}_Z \| = O(n^{-1/2})$, e.g., cf.~Lemma~\ref{lem:inez}, even if the propensity scores are unknown but (correctly) specified through a suitably regular parametric model, the expectation of the term in~\eqref{eqn:paraprop} will typically be of order~$O(n^{-1/2})$. Together with the behavior of the (outer) expectation of~$ \| P_n - P \|_{\mathcal{G}_{\Pi}}$ being of order~$O(n^{-1/2})$ if~$\Pi$ is sufficiently constrained (cf.~the discussion around~\eqref{eqn:maxineq}), this then leads to a~$O(n^{-1/2})$ upper bound on the expected regret of the policy in the context studied in this section provided~$\varepsilon = O(n^{-1/2})$. 

It is also apparent from Theorem~\ref{thm:upregcont} that ``large'' nonparametric classes of policies or propensities will lead to slower rates of convergence to~$0$ of the expectation of the upper bound on the regret developed in Theorem~\ref{thm:upregcont}.

\section{An example and numerical results}\label{sec:numex}

\subsection{A toy example}\label{ssec:ex}

In this section, we provide a toy example that illustrates the setting outlined in Section~\ref{sec:setting}. We shall re-investigate the example considered here in the numerical results below. Consider the case where~$\mathsf{T}$ is the Gini-welfare functional, which is a popular welfare measure in the economics literature; cf.~\cite{sen1976}, and for results in an optimal policy choice context \emph{without} fairness considerations see~\cite{kitagawa2019equality}. For a cdf~$F$, the Gini-welfare functional is defined (assuming that the Lebesgue-Stieltjes integrals involved in its definition exists) as 
\begin{equation}\label{eqn:giniw}
\mathsf{W}(F) := \int x dF(x) - \frac{1}{2} \int \int |x - y | dF(x)dF(y).
\end{equation}
This welfare measure trades-off the expected outcome~$\mu(F) = \int x dF(x)$, say, of~$F$ with~$\sigma(F) = \frac{1}{2} \int \int |x - y | dF(x)dF(y)$, say, the ``inequality'' inherent in the distribution~$F$, measured as its mean absolute difference (divided by 2). Hence, targeting the Gini-welfare measure, instead of simply the expected outcome, the decision maker not only aims at maximizing the expected outcome in the population considered, but also incorporates inequality concerns. We furthermore consider the similarity measure~$\mathsf{S}_{KS}$, say, which maps two cdfs~$F$ and~$G$ to the Kolmogorov-Smirnov distance between them, i.e.,~$$\mathsf{S}_{KS}(F, G) := \|F-G\|_{\infty} := \sup_{x \in \R} |F(x) - G(x)|.$$ We shall here consider the case where~$F \in \mathscr{D} =  D_{cdf}([0, 1])$. Denoting by~$\mathsf{T}_{Gini} = \mathsf{W}/2$ the Gini-welfare functional normalized by~$2$, the penalized objective we shall work with is defined as
\begin{equation*}
\Omega_{\lambda, \mathcal{F}}(\bm{\delta}) = (1-\lambda) \mathsf{T}_{Gini} \left(\langle \bm{\delta}, \mathcal{F} \rangle\right) - \lambda  \max_{z \in \mathcal{Z}}  \| \langle \bm{\delta}, \mathcal{F} \rangle_z - \langle \bm{\delta}, \mathcal{F} \rangle \|_{\infty};
\end{equation*}
the normalization guarantees that Assumption~\ref{as:MAIN} holds with~$\mathscr{D} = D_{cdf}([0, 1])$. 

\begin{figure}
\minipage{0.47\textwidth}
\includegraphics[width=\linewidth]{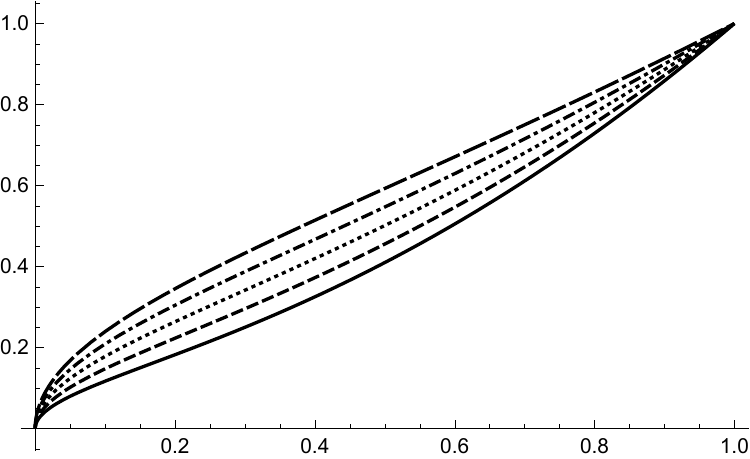}
\endminipage\hfill
\minipage{0.47\textwidth}
\includegraphics[width=\linewidth]{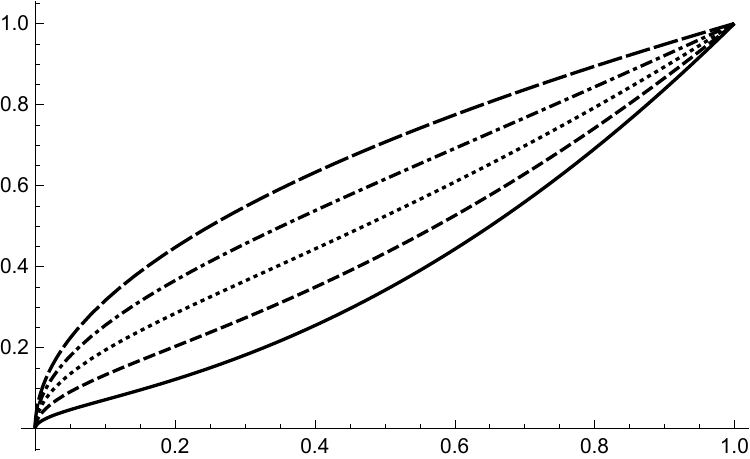}
\endminipage
\caption{The figure plots the cdf $\langle \bm{\delta}, \mathcal{F}\rangle$ (left panel) and the cdf $\langle \bm{\delta}, \mathcal{F}\rangle_0$ (right panel) for different values of~$\delta \in \{0, \frac{1}{4}, \frac{1}{2},\frac{3}{4}, 1\}$ and for~$p = 3/4$. Lower values of~$\delta$ lead to stochastically larger cdfs. Note that~$\langle \bm{\delta}, \mathcal{F}\rangle_1$ can also be read off from the right panel due to symmetry, but the dependence on~$\delta$ is now ``reverted'', cf.~Equation~\eqref{eqn:subpcdfs}.}\label{fig:popFdelta}
\end{figure}

\begin{figure}
\minipage{0.6\textwidth}
\includegraphics[width=\linewidth]{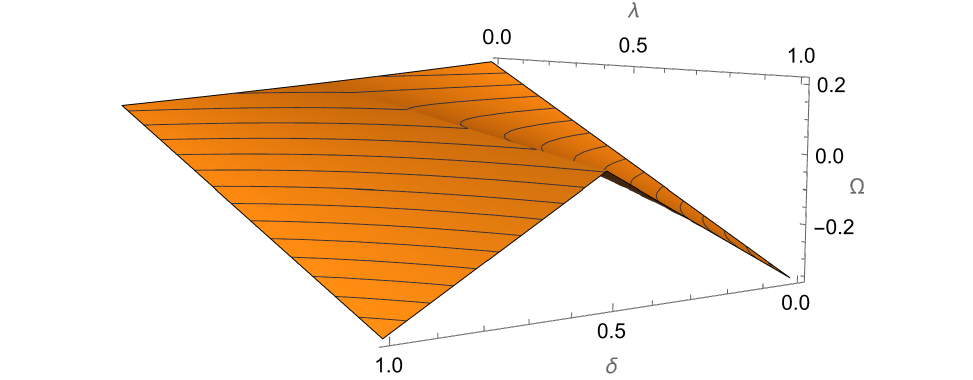}
\endminipage\hfill
\minipage{0.37\textwidth}
\includegraphics[width=\linewidth]{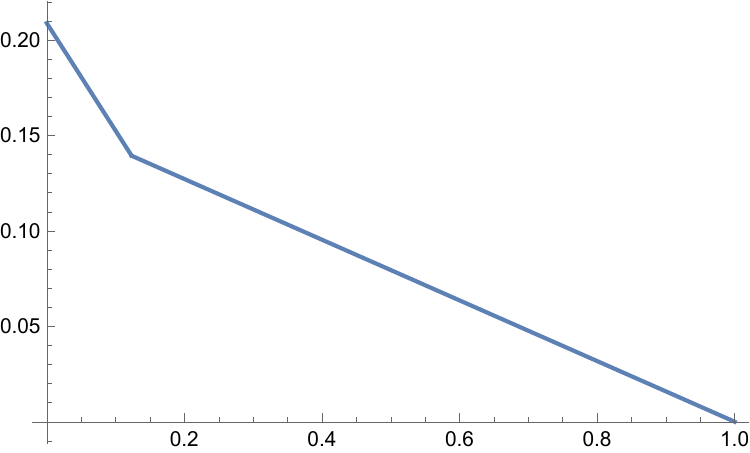}
\endminipage
\caption{Objective function~$\Omega_{\lambda, \mathcal{F}}(\bm{\delta})$ for~$p = 3/4$ (left panel), and the corresponding maximum value of the objective function for~$p=3/4$ in dependence on~$\lambda$ (right panel).}
\label{fig:objp34}
\end{figure}

Having fixed the functionals defining the target, we now consider the case where there are only~$K = 2$ treatments,~$\mathcal{X} = \{0\}$ and~$\mathcal{Z} = \{0, 1\}$. That is, for simplicity of discussion there is only one covariate value, but there are two possible values for the protected characteristic. To specify an array~$\mathcal{F}$ as in~\eqref{eqn:array}, we consider the cdfs
\begin{equation}\label{eqn:exFdef}
F^1(y\mid 0, 0) = F^2(y\mid 0, 1) = \sqrt{y} =: G(y)  ~~ \text{ and } ~~ F^2(y\mid 0, 0) =  F^1(y\mid 0, 1) = y^2 =: H(y),
\end{equation}
for~$y \in [0, 1]$ ($0$ for~$y < 0$ and $1$ for~$y > 1$), and the probability mass function~$$p_{X, Z}(0, 0) =: p > 1/2\quad \text{ and } \quad p_{X, Z}(0, 1) = (1-p),$$ i.e., the subpopulation corresponding to~$Z = 0$ constitutes the ``majority'' whereas the subpopulation corresponding to~$Z = 1$ is the ``minority.'' It holds that~$\mu(G) = 1/3 < 2/3 = \mu(H)$ and~$\sigma(G) = 1/6 > \sigma(H) = 2/15$, from which it follows that~$$\mathsf{T}_{Gini}(G) = (\mu(G) - \sigma(G))/2 = 1/12 \quad \text{ and } \quad \mathsf{T}_{Gini}(H) = (\mu(H) - \sigma(H))/2 = 4/15.$$ Hence, the cdf~$H$ has larger Gini-welfare than the cdf~$G$, and, from this point-of-view, is preferable. As a consequence, for the subpopulation defined by~$Z = 0$ the second treatment (the cdf of which is~$H$) is better than the first treatment (the cdf of which is~$G$) in terms of~$\mathsf{T}_{Gini}$; the contrary is the case for the subpopulation defined by~$Z = 1$. Note that in this example
\begin{equation*}
F^1(\cdot\mid 0) = p G(\cdot) + (1-p) H(\cdot) \quad \text{ and } \quad 
F^2(\cdot\mid 0) = p H(\cdot) + (1-p) G(\cdot).
\end{equation*}
The cdf generated by rolling out the decision rule~$\bm{\delta} = (\delta, 1-\delta)'$ for a~$\delta \in [0, 1]$ equals\footnote{Note that in case~$\mathcal{X}$ a singleton set and~$K = 2$, a decision rule~$\bm{\delta}$ is uniquely characterized by the probability at which the first treatment is assigned, which we denote by~$\delta$ throughout this example.}
\begin{equation*}
\begin{aligned}
\langle \bm{\delta}, \mathcal{F} \rangle &= 
\left[\delta p + (1-\delta)(1-p)\right]G + \left[\delta(1-p) + (1-\delta)p\right]H, 
\end{aligned}
\end{equation*}
with corresponding expectation~$\mu(\langle \bm{\delta}, \mathcal{F} \rangle) = (\delta +(1-2 \delta ) p+1)/3$ and~$$\sigma(\langle \bm{\delta}, \mathcal{F} \rangle) = \frac{1}{210} \left(\delta  (20-27 \delta )-27 (1-2 \delta )^2 p^2+2 (\delta  (54 \delta -47)+10) p+35\right),$$ so that $$\mathsf{T}_{Gini}(\langle \bm{\delta}, \mathcal{F} \rangle) = \frac{1}{420} \left(50 \delta +27 \delta ^2 (1-2 p)^2-2 \delta  p (54 p+23)+p (27 p+50)+35\right).$$ The subpopulation cdfs are given by
\begin{equation}\label{eqn:subpcdfs}
\langle \bm{\delta}, \mathcal{F} \rangle_0 = \delta G + (1-\delta) H \quad \text{ and } \quad \langle \bm{\delta}, \mathcal{F} \rangle_1 = \delta H + (1-\delta) G.
\end{equation}
Figure~\ref{fig:popFdelta} illustrates the cdfs~$\langle \bm{\delta}, \mathcal{F} \rangle$ and~$\langle \bm{\delta}, \mathcal{F} \rangle_0$ in dependence on~$\bm{\delta}$ in case~$p = 3/4.$ 
A simple computation now shows that the penalty term equals (recalling that~$p > 1/2$)
\begin{equation*}
\max_{z \in \mathcal{Z}}  \| \langle \bm{\delta}, \mathcal{F} \rangle_z - \langle \bm{\delta}, \mathcal{F} \rangle \|_{\infty} = p|2 \delta -1| \max_{y \in (0, 1)} y \left(1-y^{3}\right) = p\frac{3}{4\times 2^{2/3}} |2 \delta -1|.
\end{equation*}
To be explicit, the objective~$\Omega_{\lambda, \mathcal{F}}\left((\delta, 1-\delta)'\right)$ hence equals
\begin{equation}\label{eqn:objexs}
\frac{1}{420} (1-\lambda ) \left(50 \delta +27 \delta ^2 (1-2 p)^2-2 \delta  p (54 p+23)+p (27 p+50)+35\right) -\frac{3 \lambda  p | 2 \delta -1| }{4 \times 2^{2/3}},
\end{equation}
and is plotted in the left panel in Figure~\ref{fig:objp34} (for~$p = 3/4$). 

Fix a~$\lambda\in[0, 1)$. From~\eqref{eqn:objexs} we then see that the function~$\delta \mapsto \Omega_{\lambda, \mathcal{F}}((\delta, 1-\delta)')$ when restricted to~$[0, 1/2]$ and also when restricted to~$[1/2, 1]$, respectively, coincides with a quadratic polynomial with positive leading coefficient. It thus follows that~$\argmax_{\delta \in [0, 1]}\Omega_{\lambda, \mathcal{F}}((\delta, 1-\delta)) \subseteq \{0, 1/2, 1\}$. Using~$p > 1/2$ together with~\eqref{eqn:objexs}, we see that~$\Omega_{\lambda, \mathcal{F}}((0, 1)') > \Omega_{\lambda, \mathcal{F}}((1, 0)')$. It follows that either the argmax is~$\{0\}$, or~$\{1/2\}$, or their union~$\{0, 1/2\}$, the latter occurring whenever~$\Omega_{\lambda, \mathcal{F}}((0, 1)') = \Omega_{\lambda, \mathcal{F}}((1/2, 1/2)')$, which is the case precisely if~$\lambda = c(p) := 1-\frac{630 \sqrt[3]{2} p}{2 p \left(54 p+315 \sqrt[3]{2}+100\right)-127}$. For~$\lambda = 1$ it is easy to see that the argmax is~$\{1/2\}$. In general, the argmax is given by
\begin{equation}\label{eqn:argmaxex}
\argmax_{\delta \in [0,1]} \Omega_{\lambda, \mathcal{F}}\left((\delta, 1-\delta)'\right) = \begin{cases}
0 & 0 \leq \lambda < c(p) \\
\{0, 1/2\} & \lambda = c(p) \\
1/2 & c(p) < \lambda \leq 1,
\end{cases}
\end{equation}
cf.~Figure~\ref{fig:popFdelta} for the corresponding cdfs when~$p = 3/4$, with a maximum value of
\begin{equation*}
\max_{\delta \in [0,1]} \Omega_{\lambda, \mathcal{F}}\left((\delta, 1-\delta)'\right) = \begin{cases}
\frac{ p \left(-5 \left(20+63 \sqrt[3]{2}\right) \lambda -54 (\lambda -1) p+100\right)-70 (\lambda -1)}{840} & 0 \leq \lambda \leq c(p) \\
\frac{89}{560} (1-\lambda) & c(p) < \lambda \leq 1,
\end{cases}
\end{equation*}
which we depict for~$p = 3/4$ in Figure~\ref{fig:objp34} (note that~$c(3/4) \approx 0.123$). 

From the structure of the argmax in~\eqref{eqn:argmaxex} we see that there is a phase-transition phenomenon in the value of the preference parameter~$\lambda$. Up to the threshold~$c(p)$, the optimal policy always assigns the second treatment, which is the optimal treatment within the majority group, but is also the sub-optimal treatment within the minority group. However, if~$\lambda$ surpasses the threshold~$c(p)$, i.e., if the degree of fairness increases sufficiently, the optimal policy changes to a $50:50$ assignment rule. 

\begin{remark}\label{rem:ex}
In this example, the value function is convex (cf.~also Proposition~\ref{prop:exmax}), but is not differentiable in~$\lambda$, as there is a kink at~$c(p)$, which is related to the argmax not being a singleton set at~$c(p)$. Note also that the argmax does \emph{not} continuously depend on the preference parameter~$\lambda$, even in this simple example. Also note that for~$\lambda > c(p)$ the optimal policy is \emph{randomized}. Therefore, this example also illustrates that focusing on non-randomized decision rules would come with an efficiency loss.
\end{remark}

\subsection{Numerical results}\label{sec:num}

We now simulate data in the setting just considered in Section~\ref{ssec:ex}, and estimate the empirical success policy (as introduced in Section~\ref{sec:esr}) on a grid of preference parameters~$\lambda \in \{0, 1/49, 2/49, \hdots, 1\}$, where the optimization step in the policy is achieved via a Nelder-Mead algorithm (as it is implemented in R's ``constrOptim'' routine), and for which the starting points are selected as the best among 50 randomly drawn points (according to a uniform distribution) from~$\mathbb{S}_2$, respectively. Recall that in this example~$\mathcal{X}$ consists only of one element, whereas there are two possible values in~$\mathcal{Z}$. To be comparable to the theoretical results in Section~\ref{ssec:ex} we fix~$p  = 3/4$, recall that~$p$ corresponds to the size of the majority group.

\begin{figure}[h!]
\centering
\includegraphics[width=\linewidth]{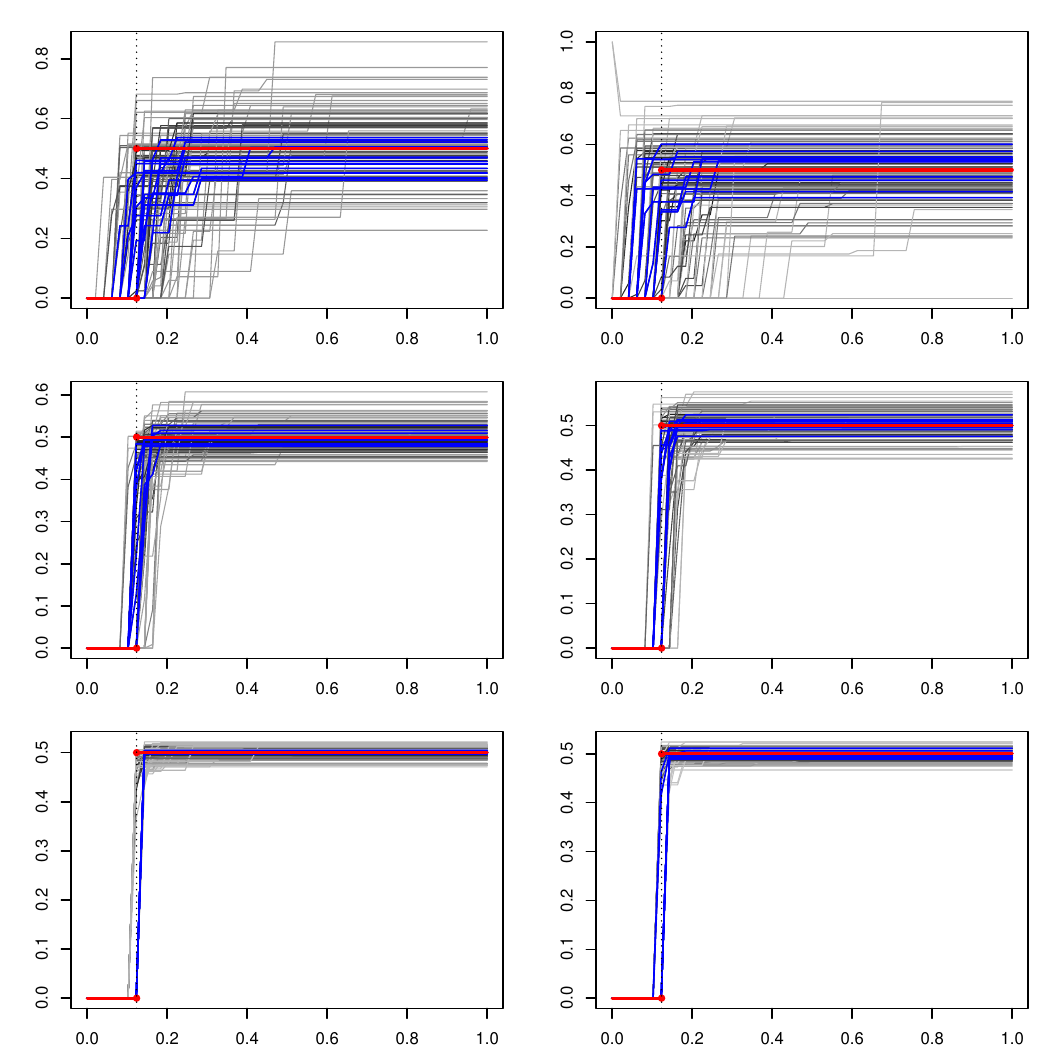}
\caption{Figures (rows 1, 2, and 3 are for differing sample sizes $n = 100, 1000, 10000$, respectively, columns for different assignment mechanisms A1 (left column) and A2 (right column), respectively) showing the inferred probability (y-axis) to assign Treatment 1 over all 100 replications, in dependence on the preference parameter~$\lambda$ (x-axis), with linear interpolation in between the points~$\{0, 1/49, 2/49, \hdots, 1\}$, at which the policy was actually estimated. The gray-scale coloring of the curves is chosen according to the depth of the curves, darker shades of gray corresponding to a stronger degree of typicality/centrality of the curve; see text for more explanation. The 10\% of most central curves are highlighted in blue color, whereas the true argmax in dependence on~$\lambda$ is highlighted in red color. The vertical dashed line intersects the abscissa at~$c(p) \approx 0.123$.}
\label{fig:sim1}
\end{figure}

\begin{figure}[h!]
\centering
\includegraphics[width=\linewidth]{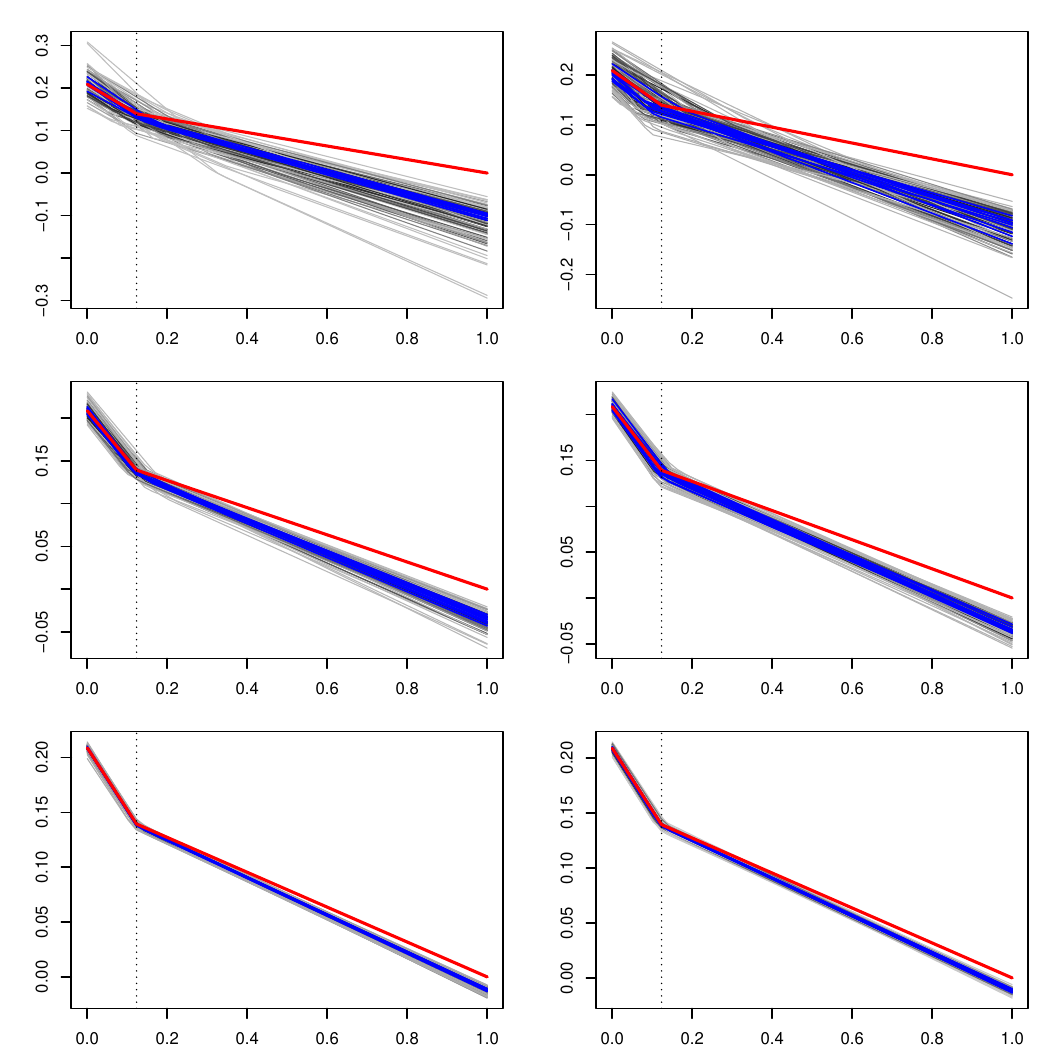}

\caption{Figures (rows 1, 2, and 3 are for differing sample sizes $n = 100, 1000, 10000$, respectively, columns for different assignment mechanisms A1 (left column) and A2 (right column), respectively) showing the empirical value function (y-axis) over all 100 replications, in dependence on the preference parameter~$\lambda$ (x-axis), with linear interpolation in between the points~$\{0, 1/49, 2/49, \hdots, 1\}$, at which the policy was actually estimated. The gray-scale coloring of the curves is chosen according to the depth of the curves, darker shades of gray corresponding to a stronger degree of typicality/centrality of the curve; see text for more explanation. The 10\% of most central curves are highlighted in blue color, whereas the true value function in dependence on~$\lambda$ is highlighted in red color. The vertical dashed line intersects the abscissa at~$c(p) \approx 0.123$.}
\label{fig:sim2}
\end{figure}

The treatment assignment mechanisms~$D_i$ studied in the numerical results (besides satisfying Assumption~\ref{as:cia}) were generated according to the 2 scenarios as follows (recall that~$Z_i = 0$ corresponds to being a member of the majority, whereas~$Z_i = 1$ corresponds to being a member of the minority):
\begin{description}
\item[A1] the conditional distribution of~$D_i$ given~$Z_i$ satisfies $P(D_i = 1 \mid Z_i = 0) = 1/4$ and $P(D_i = 1 \mid Z_i = 1) = 3/4$; thus, being assigned to Treatment 1 in the majority group is less likely than being assigned to Treatment 2, whereas being assigned Treatment 1 in the minority group is more likely than being assigned to Treatment 2.
\item[A2] the conditional distribution of~$D_i$ given~$Z_i$ satisfies $P(D_i = 1 \mid Z_i = 0) = 3/4$ and $P(D_i = 1 \mid Z_i = 1) = 1/4$; thus, being assigned to Treatment 1 in the majority group is more likely than being assigned to Treatment 2, whereas being assigned Treatment 1 in the minority group is less likely than being assigned to Treatment 2.
\end{description}
Recall from Section~\ref{ssec:ex} that in the majority group the second treatment is better, while the first treatment is better in the minority group. Therefore, the assignment mechanism~A1 corresponds to a setting where the respective optimal treatment was assigned more likely within each subgroup, whereas assignment mechanism A2 corresponds to the opposite of that situation. In addition to considering the mechanisms A1 and A2, we consider different sample sizes~$n \in \{100, 1 000, 10 000\}$, resulting in 6 different simulation scenarios. In each of these scenarios we generate~$100$ datasets and infer the optimal empirical-success policy together with the value function for every~$\lambda \in \{0, 1/49, 2/49, \hdots, 1\}$. The results are shown in Figures~\ref{fig:sim1} and~\ref{fig:sim2}, which we shall explain and discuss in the following paragraphs.

For each of the 6 scenarios considered, Figure~\ref{fig:sim1} shows 100 curves of policies (corresponding to the 100 datasets generated) extracted for the whole range of preference parameters~$\lambda \in [0, 1]$, where we interpolated linearly in between the grid-points~$\{0, 1/49, 2/49, \hdots, 1\}$. Note that because there are only two treatments, we can identify a policy with the corresponding probability to assign Treatment 1 (as we already did in Section~\ref{ssec:ex}). We decided to highlight the functions according to their ``typicality'', which we measured through Tukey's halfspace depth (cf.~\cite{tukdepth}) as implemented in the R package \textbf{DepthProc} by \cite{depthproc}: the darker a gray curve is in Figure~\ref{fig:sim1}, the higher its Tukey-depth (i.e., its ``typicality") among the 100 functions estimated; furthermore, the top 10\% of curves in terms of depth are colored in blue. Additionally, Figure~\ref{fig:sim1} also shows the true argmax correspondence in red and the cut-off~$c(3/4)$ is highlighted as a vertical dashed line (cf.~Section~\ref{ssec:ex} for additional explanation). Figure~\ref{fig:sim1} clearly illustrates the effect of a higher sample size: the larger the sample size, the closer are the estimated policy functions to the true argmax. Recalling from Section~\ref{ssec:ex} that there is a phase-transition phenomenon concerning the optimal policy present at~$c(p) = 0.123$, we note that this phenomenon is reflected quite accurately in the estimated policies for higher sample sizes, whereas for~$n = 100$ there is a lot of variation concerning the value of~$\lambda$ at which the estimated policy changes, which is even more pronounced in scenario A2 (i.e., the second column in Figure~\ref{fig:sim1}). In this context, we also note that the threshold~$c(p) =0.123$ is not an element of the grid~$\{0, 1/49, 2/49, \hdots, 1\}$ of~$\lambda$ values for which the policy actually has been estimated. Although using~$0.123$ as an additional grid point would have potentially improved the policy, this would give an overly optimistic impression, because the threshold depends on~$\mathcal{F}$, which is unknown in practice. Therefore, using~$0.123$ as an additional grid point would have distorted the results in favor of the method studied. To increase the accuracy, one could nevertheless work with a finer grid, at higher computational costs.

In Figure~\ref{fig:sim2}, we show the corresponding estimated value functions, with an analogous coloring scheme as in Figure~\ref{fig:sim1} and where the red curve depicts the true value function. Interestingly, the figure shows that (in the example considered) there appears to be a downward bias present in the estimation of the value function, which diminishes as sample size grows. In general, a larger sample size leads to more accurate results. 

\begin{figure}[h!]
\centering
\includegraphics[width=.8\linewidth]{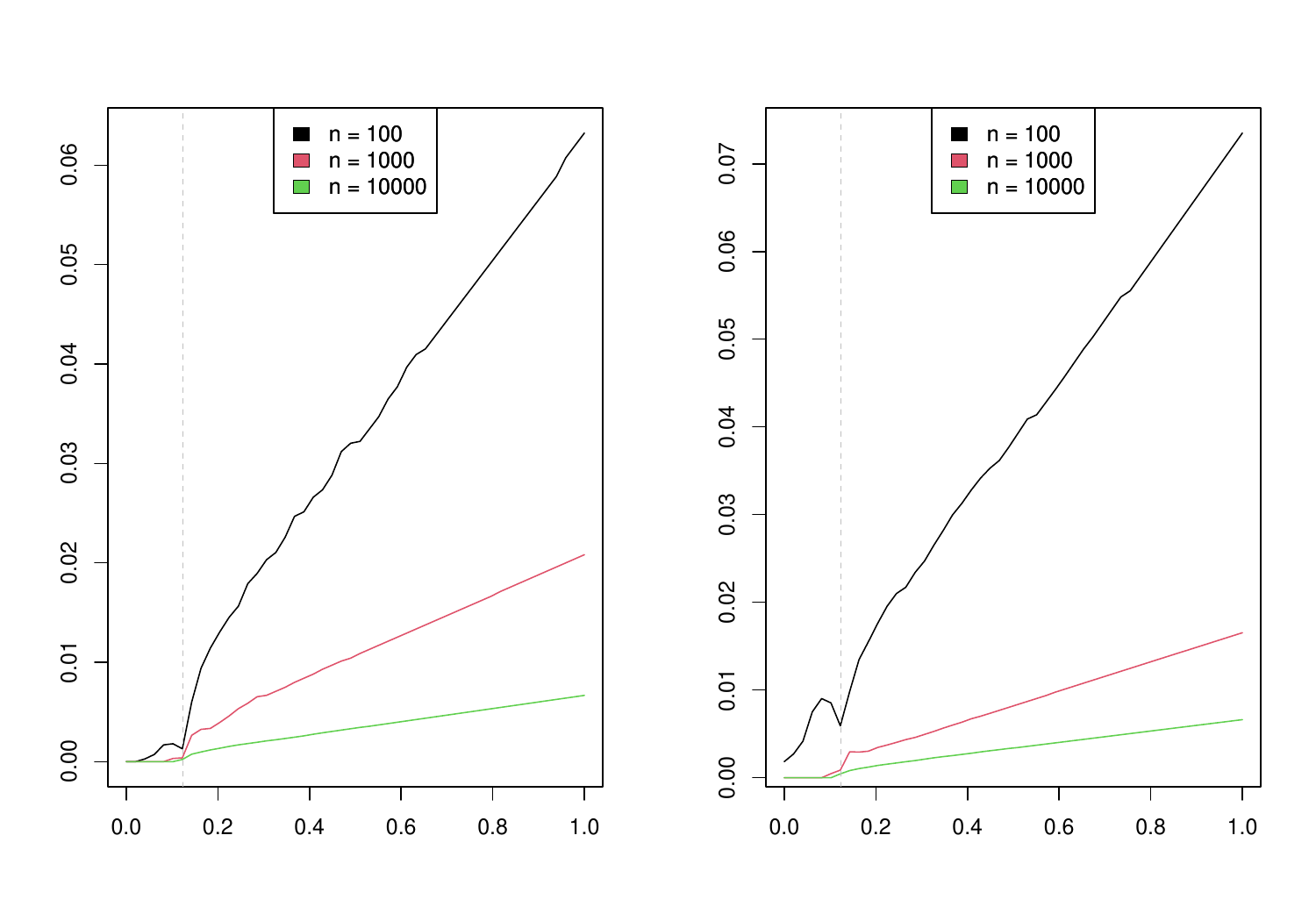}
\caption{Average regret for sample sizes $n = 100, 1000, 10000$, respectively, assignment mechanisms A1 (left column) and A2 (right column), in dependence on the preference parameter~$\lambda$ (x-axis), with linear interpolation in between the points~$\{0, 1/49, 2/49, \hdots, 1\}$, at which the policy was actually estimated. The vertical dashed line intersects the abscissa at~$c(p) \approx 0.123$.}
\label{fig:regre}
\end{figure}

For completeness, we also plot the average regret in all 6 scenarios and for all 50 values of the preference parameter in Figure~\ref{fig:regre}. As expected from the theoretical results, the regret gets smaller as sample size increases. Furthermore, larger values of the preference parameter correspond to relatively larger values of the average regret.

\section{Empirical illustrations} \label{sec:empirics}

To showcase the methods discussed in this article in an empirical context, we shall now apply them (more precisely the policies in Section~\ref{sec:esr}) to two datasets:
\begin{itemize}
\item data on the Pennsylvania reemployment bonus experiment underlying the analysis in~\cite{bilias} (the dataset can be obtained, e.g., via the replication material for the article \cite{1ddml} on the publisher's website\footnote{URL \url{https://doi.org/10.1111/ectj.12097} retrieved on 25.01.2024.});
\item the entrepreneurship program dataset from \cite{lyonszhang} (available in the replication material to that article via the publisher's website\footnote{URL \url{https://www.openicpsr.org/openicpsr/project/113492/version/V1/view} retrieved on 26.08.2023.}), which builds the basis of the empirical illustration in \cite{vivbra}.
\end{itemize}
We use the same target~$\mathsf{T}$ (Gini-welfare) and similarity measure~$\mathsf{S}$ (Kolmogorov-Smirnov distance) as in Section~\ref{sec:num} (after rescaling the outcome to~$[0, 1]$ in the first dataset). Furthermore, we used the same optimization heuristic (based on the same grid and starting value selection procedure). For both datasets we summarize the numerical computations in figures showing the following quantities in dependence on the preference parameter~$\lambda \in [0, 1]$ based on linear interpolation in between the grid points
\begin{equation}\label{eqn:empirics_functions}
\max_{\bm{\delta} \in \mathbb{S}_K} \Omega_{\lambda, \hat{\mathcal{F}}_n}(\bm{\delta}), \quad \mathsf{T}\left( \langle \bm{\pi}_n^{\lambda}, \hat{\mathcal{F}}_n \rangle \right), \quad \text{ and } \quad \max_{z \in \mathcal{Z}}\mathsf{S}\left( \langle \bm{\pi}_n^{\lambda}, \hat{\mathcal{F}}_n \rangle_z, \langle \bm{\pi}_n^{\lambda}, \hat{\mathcal{F}}_n \rangle \right);
\end{equation}
i.e., the (empirical) value function, the Gini-welfare corresponding to the predicted outcome distribution~$\langle \bm{\pi}_n^{\lambda}, \hat{\mathcal{F}}_n \rangle$ when implementing the policy~$\bm{\pi}_n^{\lambda}$, which (nearly) maximizes the empirical objective function~$\bm{\delta} \mapsto \Omega_{\lambda, \hat{\mathcal{F}}_n}(\bm{\delta})$, and the (empirical) similarity measure at the policy~$\bm{\pi}_n^{\lambda}$. 

\subsection{Pennsylvania reemployment bonus experiment}

In this dataset the treatment is a bonus for unemployment insurance claimants granted if they become employed within a given period of time. We here focus on treatment group 4 as the treatment vs.~a control group, cf.~\cite{bilias} for details. In particular, it follows that~$K = 2$. The total sample size then is~$n = 5099$. The outcome variable measures the duration (in weeks) of the first spell of unemployment. We incorporated the following covariates into our analysis: 
\begin{enumerate}
\item a categorical variable (with~$3$ categories) indicating the number of dependents of the claimant being $0$, $1$, or $\geq 2$.
\item a categorical variable (with~$3$ categories) indicating whether the claimant's age was $<35$, between $35$ and $54$, or higher;
\item a categorical variable (with~$3$ categories) indicating whether the claimant was working in the sector of durable manufacturing, the sector of nondurable manufacturing, or in another sector.
\end{enumerate}  
After combining them into a single factor (and dropping one category for which there was no observation) this resulted in a factor with 26 levels, i.e.,~$|\mathcal{X}| = 26$.

\begin{figure}[ht]
\centering
\includegraphics[width=\linewidth]{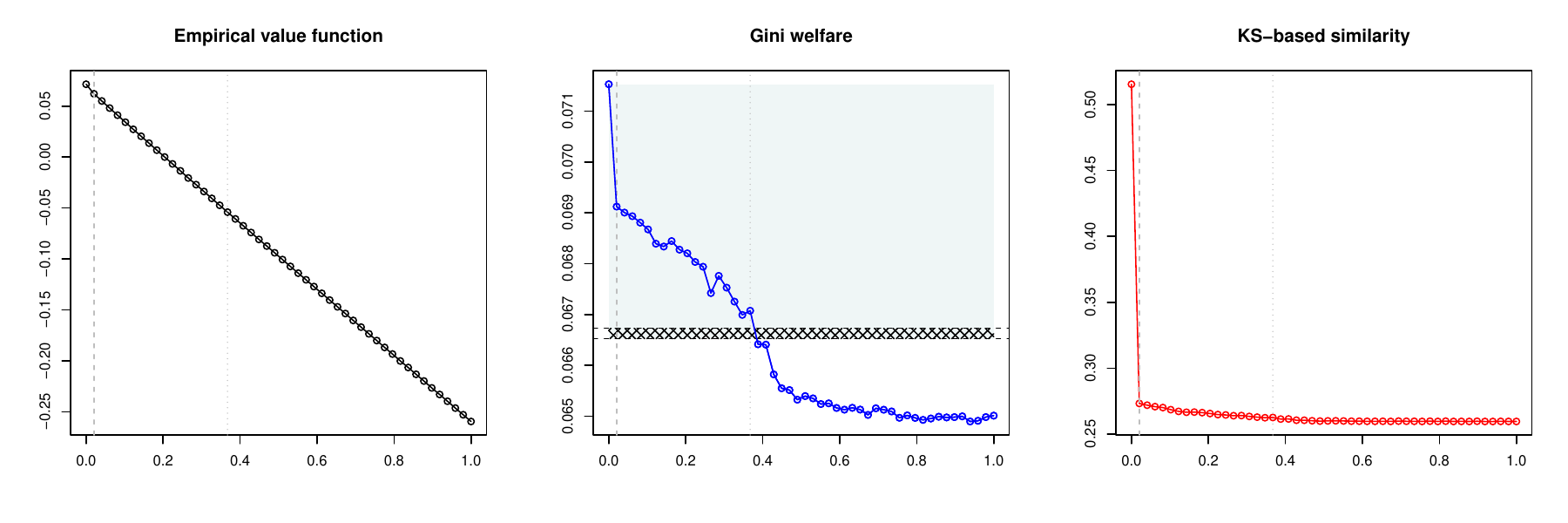}
\caption{Results for the Pennsylvania reemployment bonus experiment data. From left to right, the three plots show (in dependence on the preference parameter~$\lambda$, which is plotted on the $x$-axes): the empirical value function, the predicted Gini-welfare when the optimal policy is implemented, and the corresponding similarity measure. See Equation~\eqref{eqn:empirics_functions} for a precise description of the functions that are plotted. The dashed vertical line corresponds to~$\lambda = 1/49$. The dotted vertical line corresponds to~$\lambda = 18/49$. The dashed horizontal lines (figure in the middle) indicate the efficiency of the policy corresponding to~$\lambda = 0$ minus~$\beta = .05$ and minus $\beta(1-c_n) \approx 0.0048$, respectively. The crossed region between those lines corresponds to the slack. The colored region above the crossed region contains the predicted Gini-welfare of the ``affordable'' policies; cf.~the description in the text.}
\label{fig:penn}
\end{figure}

Furthermore, we worked with the following sensitive characteristics, after combining them into a single factor with 8 levels, i.e.,~$|\mathcal{Z}| = 8$:
\begin{enumerate}
\item a dummy variable indicating whether or not the claimant is female;
\item a categorical variable (4 categories) indicating whether the claimant is Black, Hispanic, White, or other; 
\end{enumerate}
Figure~\ref{fig:penn} summarizes the results. The figure to the right, which shows the KS-based similarity measure (low values corresponding to higher similarity, i.e., higher fairness), shows that the difference between the distributions in the subpoulations (defined by the sensitive characteristics in the above enumeration) decreases sharply when passing from~$\lambda = 0$ to~$\lambda = 1/49$, whereas it remains essentially constant for~$\lambda \geq 2/49$. Although there is also an initial drop (of about~$0.0024$) in Gini-welfare in the total population, which is shown in the center of Figure~\ref{fig:penn}, the Gini-welfare keeps decreasing at a still notable degree also for~$\lambda \geq 2/49$. If a DM can afford a decrease in total Gini-welfare of~$0.0024$, the empirical results therefore suggest to use the policy corresponding to~$\lambda = 1/49$, as this already leads to a major reduction of unfairness. Higher levels of the preference parameter~$\lambda$ seem to mainly decrease welfare while not increasing fairness much further.

In addition to the ``qualitative'' or inspection-based approach to choosing the preference parameter just outlined, Figure~\ref{fig:penn} (center) also illustrates the more quantitative and budget-based approach lined out in Section~\ref{sec:tune} for the example of~$\beta = 0.005$. That is, when the DM deems an efficiency loss of~$0.005$ spent on increasing the degree of fairness affordable. In such a situation, one would, according to~\eqref{eqn:preftune}, introduce the slack~$c_n = \sqrt{\log(n)/n} \approx 0.041$, and then choose the policy corresponding to the largest preference parameter that realizes an efficiency not smaller than the efficiency realized at the policy corresponding to~$\lambda = 0$ (which equals about~$0.072$ in this example) minus $\beta(1-c_n) \approx 0.0048$, resulting in~$0.0667$ (as opposed to just subtracting~$\beta$ itself from~$0.072$, which would result in~$0.0665$). This is depicted in Figure~\ref{fig:penn}, where the effect of introducing some slack is also illustrated. In this case, the figure shows that (according to this procedure) the DM would use the preference parameter~$18/49$.
\begin{figure}[ht]
\centering
\includegraphics[width=\linewidth]{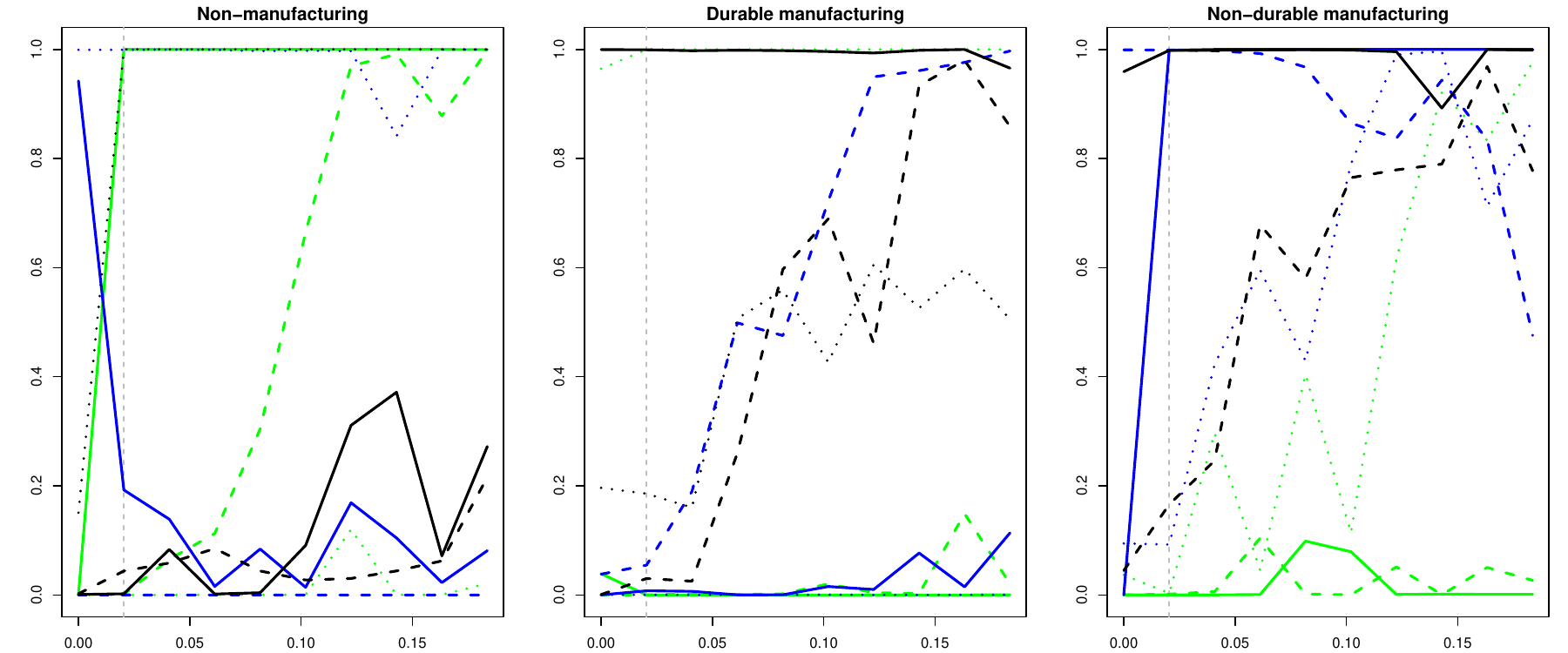}
\caption{Treatment assignment probabilities (y-axis) for all groups are plotted against the preference parameter~$\lambda$ (x-axis) for the Pennsylvania reemployment bonus experiment data. The three figures correspond to non manufacturing (left), durable manufacturing (middle), and nondurable manufacturing (right); the colors indicate age (green indicating $<35$, blue indicating an age between $35$ and $54$, and black indicating a higher age than~$54$); the line types solid, dashed and dotted indicate the number of dependents $0$, $1$ and $\geq 2$, respectively.}
\label{fig:pennpoly}
\end{figure}

The actual assignment probabilities to the treatment category in dependence on the covariates can be seen in Figure~\ref{fig:pennpoly}, where we focus on smaller levels of~$\lambda$, as this enhances readability. Given that introducing a small preference parameter of~$\lambda = 1/49$ already substantially increases the fairness in the output distributions, it is interesting to investigate that region of the plot in more detail:
\begin{itemize}
\item in the non-manufacturing sector (i.e., figure to the left in Figure~\ref{fig:pennpoly}), such a small change of preference parameter is related to a strong decrease in the treatment assignment probabilities in the ``0 dependents \& middle age" (solid blue) category, and to a strong increase in the treatment assignment probabilities in the ``0 dependents \& young age" (solid green) and the  ``2 dependents \& middle age" (dotted blue) categories.
\item in the durable manufacturing sector (i.e., middle figure in Figure~\ref{fig:pennpoly}) there is no comparable change in the assignment probabilities for very small~$\lambda$.
\item in the nondurable manufacturing sector (i.e., figure to the right in Figure~\ref{fig:pennpoly}) a small change of preference parameter is related to a strong increase in the assignment probabilities in the  ``0 dependents \& middle age" category.
\end{itemize}

\subsection{Entrepreneurship program}

\begin{figure}[h!]
\centering
\includegraphics[width=\linewidth]{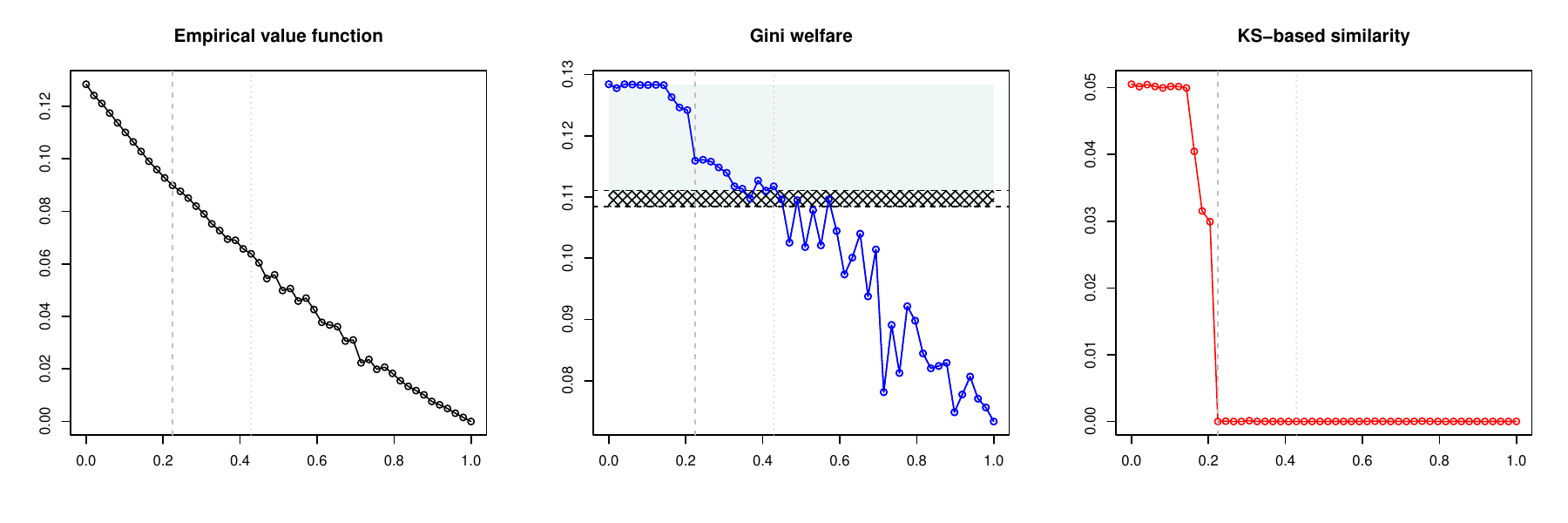}
\caption{Results for the entrepreneurship program data studied in~\cite{lyonszhang}. From left to right, the three plots show (in dependence on the preference parameter~$\lambda$, which is plotted on the $x$-axes): the empirical value function, the predicted Gini-welfare when the optimal policy is implemented, and the corresponding similarity measure. See Equation~\eqref{eqn:empirics_functions} for a precise description of the functions that are plotted. The dashed vertical line corresponds to~$\lambda = 11/49$. The dotted vertical line corresponds to~$\lambda = 21/49$. The dashed horizontal lines indicate the efficiency of the policy corresponding to~$\lambda = 0$ minus~$\beta = .02$ and minus $\beta(1-c_n) \approx 0.017$, respectively. The crossed region in between those lines corresponds to the slack. The colored region above the crossed region contains the predicted Gini-welfare of the ``affordable'' policies; cf.~the description in the text.}
\label{fig:lyon}
\end{figure}

The treatment variable here is an entrepreneurship training and incubation program for students and the outcome variable is subsequent entrepreneurial activity, which is a dummy variable, cf.~\cite{lyonszhang} and Section 6 of~\cite{vivbra} for more details. Also in this dataset~$K = 2$. The total sample size (after eliminating observations with missing values) is~$n = 335$. Following ``Case 2'' in \cite{vivbra} we incorporated the following covariates into our analysis:
\begin{enumerate}
\item a (rounded) score assigned to the candidate by
the interviewer taking values in $1, 2, \hdots, 10$ (cf.~Footnote 3 in~\cite{lyonszhang});
\item the school rank, taking values in $1, 2, 3, 4$.
\end{enumerate}  
In principle, this would result in 40 levels after combining the two variables into a single factor. However, there is no data available for 4 of the 40 levels, hence~$|\mathcal{X}| = 36$ in this application. The sensitive characteristic we worked with in this application is a dummy variable indicating whether the student is female, so that~$|\mathcal{Z}| = 2$. Figure~\ref{fig:lyon} contains the results of the computations. The similarity measure reveals three substantial drops. Each of them could correspond to a possible choice of preference parameter. Unless budget constraints do not allow for this possibility, the DM could implement the policy corresponding to~$\lambda = 11/49$: the KS-based similarity measure is essentially $0$ for the policy corresponding to this preference parameter and hence cannot reduce substantially for larger values of the preference parameter. This leads to a ``loss'' in the population Gini-welfare of about~$0.0124$ relative to the policy corresponding to~$\lambda = 0$. 

Figure~\ref{fig:lyon} (center) also illustrates the more quantitative and budget-based approach lined out in Section~\ref{sec:tune} for the example of~$\beta = 0.02$. That is, when the DM deems an efficiency loss of~$0.02$ spent on increasing the degree of fairness affordable. Here, the slack~$c_n = \sqrt{\log(n)/n} \approx 0.132$, and one would choose the policy corresponding to the largest preference parameter that realizes an efficiency not smaller than the efficiency realized at the policy corresponding to~$\lambda = 0$ (which equals about~$0.128$ in this example) minus $\beta(1-c_n) \approx 0.017$, resulting in~$0.111$ (as opposed to just subtracting~$\beta$ from~$0.128$, which would result in~$0.108$). This is depicted in Figure~\ref{fig:lyon}, where the effect of introducing some slack is also illustrated. In this example, the DM would conclude to use the preference parameter~$21/49$. Because using~$\lambda = 21/49$ instead of~$\lambda = 11/49$ (as we concluded in the previous paragraph) does not come with any notable decrease in unfairness (but decreases efficiency), a DM should perhaps stick to~$\lambda = 11/49$, even though a higher preference parameter could be afforded. This application hence also shows that even when using a budget-based approach, inspecting the fairness and efficiency implications of the policies graphically is advisable for practice.

\begin{figure}
\centering
\includegraphics[width=.8\linewidth]{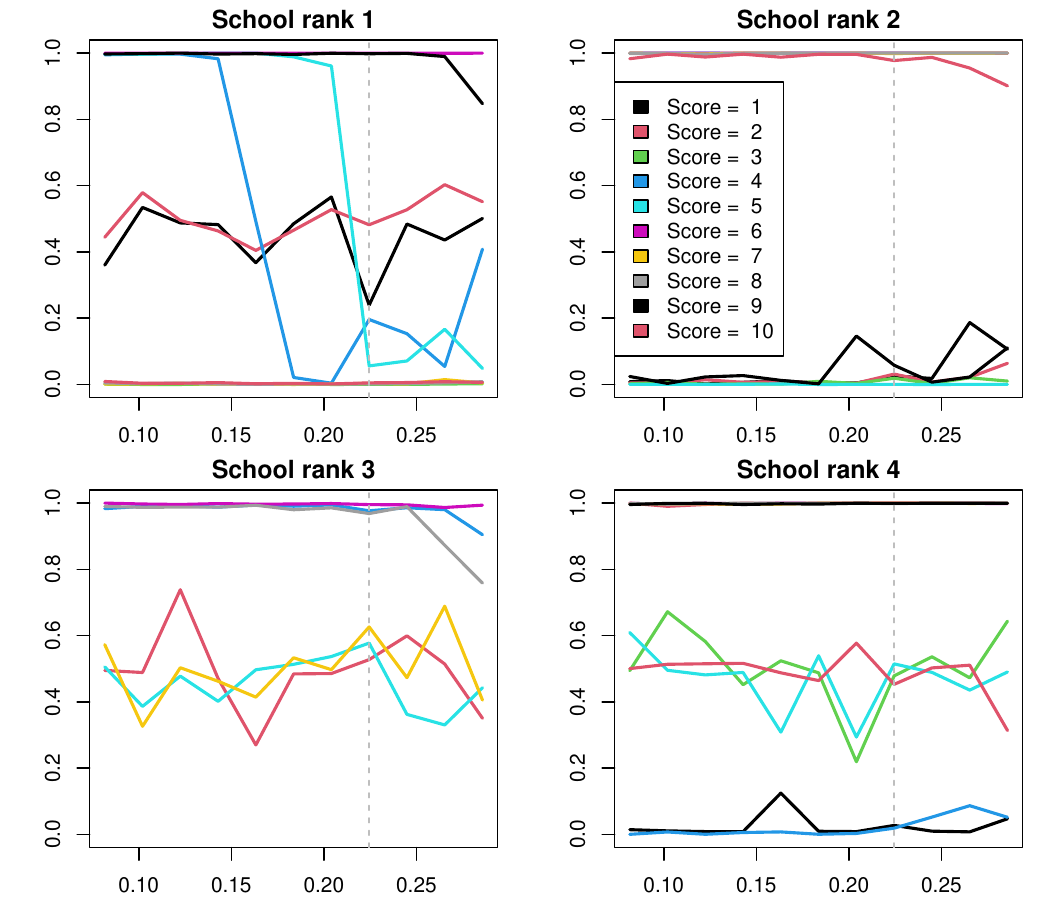}
\caption{Treatment assignment probabilities (y-axis) for all groups are plotted against the preference parameter~$\lambda$ (x-axis) for the entrepreneurship program data studied in~\cite{lyonszhang}. The four figures correspond to the school ranks and the colors indicate the score as shown in the legend.}
\label{fig:lyonspoly}
\end{figure}

The actual assignment probabilities to the treatment in dependence on the covariates can be seen in Figure~\ref{fig:lyonspoly}, where we focus on levels of~$\lambda$ close to~$11/49$ as this enhances readability and corresponds to the interesting region of preference parameters according to the above discussion. One can observe that apart from schools ranked first and fourth, there are no fundamental changes in the assignment probabilities when~$\lambda$ increases from $10/49$ to~$11/49$. In schools ranked first, one can observe that the assignment probabilities of students with a score of 5/10 and 9/10 drop, whereas the assignment probabilities of students with a score of 4/10 increase to some degree. In schools ranked fourth the assignment probabilities of students with scores 3/10 and 5/10 increase.

\clearpage

\section{Conclusion} \label{sec:conc}

We have developed tools that allow a DM to choose a policy targeting a general functional (e.g., the Gini-welfare) of the population cdf, while taking into account the objective of treating protected subgroups fairly. The magnitude by which unfairness is penalized can be regulated by the DM via a preference parameter. We study ways to select this preference parameter, and show that the policy's excess regret converges to~$0$ at the parametric rate. Furthermore, going beyond regret guarantees, we also establish the consistency of the policy under conditions that are satisfied in particular under the assumption that the preference parameter is selected (in an arbitrary way) from a finite set of preference parameters. We also introduced and discussed theoretical properties of a policy that applies in settings with non-discrete covariates, and illustrated our results numerically and in the context of two empirical applications. 

\bibliographystyle{ims} 
\bibliography{ref}		

\newpage

\appendix

\section{Relation to~\cite{kitagawa2019equality} and~\cite{kpv3}}\label{sec:reltokpv3}

\cite{kitagawa2019equality} study treatment choice problems where the target functional is an element of the class of ``rank-dependent'' social welfare functions, cf.~their Equation~(1). This class includes, in particular, the Gini-welfare measure. Their results extend optimal policy design in the spirit of~\cite{kitagawa2018should} to a more flexible class of targets, whereas the rest of the literature typically concerns expected outcome as the target measure. \cite{kitagawa2019equality} do not investigate how fairness regarding subgroups of the population can be achieved, or how fairness and efficiency properties can be traded-off against each other. Subpopulation distributions and their similarity are central to our approach, but play no role in their context. We also mention here that their approach to estimating outcome distribution functions provides an important guideline in our Section~\ref{sec:contX}, where we show how the assumption of discrete covariates, which is imposed in most parts of the present article, can be avoided. 

The observational setting we consider in the present article is related to that in \cite{kpv3}: we also consider a pure-exploration problem in which the decision maker can target a \emph{general} functional of the outcome distribution. In contrast to~\cite{kpv3} we
\begin{enumerate}
\item explicitly incorporate covariates into the decision making process to exploit observed heterogeneity in the population;
\item consider a more general objective that allows the DM to recommend a treatment (combination) that penalizes unfairness;
\item monitor how different degrees of fairness awareness affect the inferred decision, and investigate how to choose the relevant preference parameter in practice.
\end{enumerate}
The first item provides a practically relevant extension to~\cite{kpv3}. The second and third constitute the main contribution of the present article: we study the question how a DM can incorporate fairness considerations into the decision making process, and we give practical guidelines how the fairness-optimality trade-off can be monitored and chosen by the DM. In particular, our theory allows for data-dependent choices of the preference parameter that determines how strongly the fairness penalty enters the objective function. In contrast to \cite{kpv3} we do not consider sequential procedures, because we here want to leave the sampling mechanism imposed on the data available to the DM unspecified.

\section{Alternative regret notion: anchoring at a target preference parameter}\label{sec:altreg}

In situations where~$\hat{\lambda}_n$ estimates an underlying target~$\lambda_n^*$ (cf.~Section~\ref{sec:tune} for a concrete example), one could \emph{alternatively} consider anchoring the regret at the target~$\lambda_n^*$ itself instead of~$\hat{\lambda}_n$, i.e., 
\begin{equation*}
r(\bm{\pi}^{  \hat{\lambda}_n}_n; \lambda^*_n, \mathcal{F}) = \max_{\bm{\delta} \in \mathbb{S}_K} \Omega_{\lambda^*_n, \mathcal{F}}(\bm{\delta}) - \Omega_{\lambda^*_n, \mathcal{F}} \left(\bm{\pi}^{\hat{\lambda}_n}_n\right).
\end{equation*}
This regret notion evaluates the policy~$\bm{\pi}^{\hat{\lambda}_n}_n$ relative to the best one for the unknown value~$\lambda^*_n$. It is straightforward to verify (cf.~the proof of Lemma~\ref{lem:exmax}) that for a constant~$C>0$ depending on~$\mathsf{S}$,~$\mathsf{T}$, and~$\mathscr{D}$ it holds that
\begin{equation*}
\left| r(\bm{\pi}^{  \hat{\lambda}_n}_n; \hat{\lambda}_n, \mathcal{F}) - r(\bm{\pi}^{  \hat{\lambda}_n}_n; \lambda^*_n, \mathcal{F}) \right|
\leq 2 \max_{\bm{\delta} \in \mathbb{S}_K} |\Omega_{\hat{\lambda}_n, \mathcal{F}}(\bm{\delta})-\Omega_{\lambda^*, \mathcal{F}}(\bm{\delta})| \leq C |\hat{\lambda}_n - \lambda^*_n|,
\end{equation*}
so that
\begin{equation*}
\mathbb{E}^*(r(\bm{\pi}^{  \hat{\lambda}_n}_n; \lambda^*_n, \mathcal{F})) \leq \mathbb{E}^*(r(\bm{\pi}^{  \hat{\lambda}_n}_n; \hat{\lambda}_n, \mathcal{F})) + C \mathbb{E}^*|\hat{\lambda}_n - \lambda^*_n|.
\end{equation*}
This quantifies, to some extent, the (expected behavior) that for~$\hat{\lambda}_n \approx \lambda^*_n$ the two regret notions behave similarly, whereas they may behave quite differently in case~$\hat{\lambda}_n \not \approx \lambda^*_n$. 

Section~\ref{sec:tune} investigates consistency properties for a specific choice of~$\lambda^*_n$ and a corresponding estimate~$\hat{\lambda}_n$, under which $\hat{\lambda}_n \approx \lambda^*_n$ and hence~$r(\bm{\pi}^{  \hat{\lambda}_n}_n; \hat{\lambda}_n, \mathcal{F}) \approx r(\bm{\pi}^{  \hat{\lambda}_n}_n; \lambda^*_n, \mathcal{F})$ can be expected to hold. Nevertheless, theses results do not allow one to obtain regret guarantees for~$r(\bm{\pi}^{  \hat{\lambda}_n}_n; \lambda^*_n, \mathcal{F})$ from those of~$r(\bm{\pi}^{  \hat{\lambda}_n}_n; \hat{\lambda}_n, \mathcal{F})$ based on the above upper bound for two reasons: 

\begin{enumerate}
\item we do not study rates of convergence, so that the guarantees obtained for the regret we work with, i.e.,~$r(\bm{\pi}^{  \hat{\lambda}_n}_n;\hat{\lambda}_n, \mathcal{F})$, cannot be immediately shown to carry over to~$r(\bm{\pi}^{  \hat{\lambda}_n}_n; \lambda^*_n, \mathcal{F})$;
\item in the example studied in Section~\ref{sec:tune}, $\hat{\lambda}_n$ can only be expected to consistently \emph{underestimate} the target~$\lambda^*_n$, and the results in~\cite{lsun} suggest that this cannot be avoided, in general. Hence, $\mathbb{E}^*|\hat{\lambda}_n - \lambda^*_n|$ may not converge to~$0$ and may actually dominate the behavior of~$r(\bm{\pi}^{  \hat{\lambda}_n}_n; \lambda^*, \mathcal{F})$.
\end{enumerate} 

\section{Supplementary results and discussions for Section~\ref{sec:esr}}\label{sec:suppesr}

In Appendix~\ref{sec:suppesr} we maintain (without further mentioning) all assumptions and notational conventions that are maintained in Section~\ref{sec:esr}.

\subsection{Technical remarks concerning the empirical success policy}\label{sec:techpol}

We here summarize some technical remarks concerning the policy stated in the beginning of Section~\ref{sec:esr}.

\begin{remark}
The policy stated in the beginning of Section~\ref{sec:esr} allows for the possibility that the empirical version 
\begin{equation}\label{eqn:empv2}
\sup_{\bm{\delta} \in \mathbb{S}_K} \Omega_{\lambda, \hat{\mathcal{F}}_n}(\bm{\delta})
\end{equation}
of the maximization problem~$\max_{\bm{\delta} \in \mathbb{S}_K} \Omega_{\lambda, \mathcal{F}}(\bm{\delta})$ is only solved up to a given level of accuracy~$\varepsilon$, cf.~\eqref{eqn:approxmax}. While the problem~$\max_{\bm{\delta} \in \mathbb{S}_K} \Omega_{\lambda, \mathcal{F}}(\bm{\delta})$  is guaranteed to permit a solution by Proposition~\ref{prop:exmax}, there is no such guarantee for the problem in~\eqref{eqn:empv2}. The level of accuracy~$\varepsilon$ can be chosen by the user, and can be made arbitrarily small, balancing accuracy with computational (e.g., runtime) considerations. For the policy stated in the beginning of Section~\ref{sec:esr} to be feasible, in the sense that the set in~\eqref{eqn:approxmax} is non-empty, it is important that the supremum that is approximated is actually finite, which is guaranteed by the last item in Proposition~\ref{prop:exmax}.
\end{remark}

\begin{remark}
Note that the policy~$\bm{\pi}_n^{\varepsilon, \lambda}$ stated in the beginning of Section~\ref{sec:esr} is not required to be a measurable selection of the set in~\eqref{eqn:approxmax} and hence may not be measurable. Our theoretical results take care of this possibility.
\end{remark}

\begin{remark}
If decision rules are to be obtained for multiple values of~$\lambda$, Step 4 in the policy stated in the beginning of Section~\ref{sec:esr}  (but not the other steps) need to be repeated for all preference parameters one considers. Because the maximizer of the objective~$\max_{\bm{\delta} \in \mathbb{S}_K} \Omega_{\lambda, \mathcal{F}}(\bm{\delta})$ may not be unique (cf.~the example discussed in Section~\ref{ssec:ex} for an illustration), there is no guarantee that specific approximate solutions (as in~\eqref{eqn:approxmax}) to $\sup_{\bm{\delta} \in \mathbb{S}_K} \Omega_{\lambda_1, \hat{\mathcal{F}}_n}(\bm{\delta})$ and $\sup_{\bm{\delta} \in \mathbb{S}_K} \Omega_{\lambda_2, \hat{\mathcal{F}}_n}(\bm{\delta})$ are ``close'' (relative to~$\varepsilon$) even if~$\lambda_1 \approx \lambda_2$ and~$\hat{\mathcal{F}}_n \approx \mathcal{F}$. What is true is that the corresponding values of the optimization problem's value will be similar (due to a continuity property established in Lemma~\ref{lem:exmax} and results entailing~$\hat{\mathcal{F}}_n \approx \mathcal{F}$), i.e., $\sup_{\bm{\delta} \in \mathbb{S}_K} \Omega_{\lambda_1, \hat{\mathcal{F}}_n}(\bm{\delta}) \approx \sup_{\bm{\delta} \in \mathbb{S}_K} \Omega_{\lambda_2, \hat{\mathcal{F}}_n}(\bm{\delta})$.
\end{remark}

\subsection{Value function estimation by linear interpolation}\label{sec:valinterpol}

We here show how the value of the decision problem under consideration in Section~\ref{sec:esr} as a function of the preference parameter, i.e., the value function~
\begin{equation}\label{eqn:valfun1}\lambda \mapsto \max_{\bm{\delta} \in \mathbb{S}_K} \Omega_{\lambda, \mathcal{F}}(\bm{\delta}),
\end{equation}
which we know to be convex from~Proposition~\ref{prop:exmax}, can be uniformly consistently estimated by suitable interpolation methods (for pointwise estimation see Remark~\ref{rem:pointwval} below). An estimate of the value function allows the decision maker to investigate the effect of different preference parameters on the maximal value of the objective (cf.~the figures in 
Sections~\ref{sec:numex} and~\ref{sec:empirics} and their discussion).

Due to numerical restrictions, it will typically only be possible to determine the empirical success policy~$\bm{\pi}^{\varepsilon_n, \lambda}$ over a finite grid of preference parameters~$\lambda$, from which one then infers the value function~\eqref{eqn:valfun1}.  We now show that if the empirical success policy~$\bm{\pi}^{\varepsilon_n, \lambda}$ is determined for every~$\lambda \in G_{m} := \{0, \frac{1}{m}, \frac{2}{m}, \hdots, 1\}$ with~$m = m_n \to \infty$, and if~$\varepsilon_n \to 0$, then the value function is \emph{uniformly} consistently estimated in (outer) probability by linearly interpolating the values of the empirical success policies determined for all preference parameters in the grid. The result follows from a finite-sample probability bound on $$\sup_{\bm{\delta} \in \mathbb{S}_K, \lambda \in [0, 1]} \left|\Omega_{\lambda, \mathcal{F}}(\bm{\delta}) - \Omega_{\lambda, \hat{\mathcal{F}}_n}(\bm{\delta})  \right|$$ that is also used in the proof of Theorem~\ref{thm:upreg2} (and which is given in Lemma~\ref{lem:uzaux} in the Appendix). To define the interpolation method properly, denote the set of continuous functions from~$[0, 1] \to \R$ by~$\mathcal{C}([0, 1])$, let~$m \in \N$, and let the ``linear interpolation operator"~$$\mathrm{LIP}_{m}: \R^{m + 1} \to \mathcal{C}([0, 1])$$ associate to every~$v = (v_0, \hdots, v_m)' \in \R^{m+1}$ the piecewise linear function $$\lambda \mapsto \mathrm{LIP}_m(\lambda; v) :=  v_{i} + \frac{v_{i+1} - v_i}{(i+1)/m - i/m} \left(\lambda - i/m\right), \text{ with } 0 \leq i < m \text{ so that } \lambda \in [i/m, (i+1)/m],$$ i.e., it holds that~$\mathrm{LIP}_m(i/m; v) = v_i$ for every~$i = 0, \hdots, m$, with linear interpolation in between the grid points~$i/m$ for~$i = 0, \hdots, m$.
\begin{theorem}\label{thm:interpol}
Let~$(\bm{Y}_j', X_j, Z_j)' \sim \mathcal{F} \sqsubset \mathscr{D}$ for~$j =1, \hdots, n$,~$m_n \in \N$, and~$\varepsilon_n > 0$. Then, there exists a real number~$\alpha(\mathcal{F}) > 0$ such that
\begin{equation}\label{eqn:interpol}
\sup_{\lambda \in [0, 1]}\left|\mathrm{LIP}_{m_n}\left(\lambda; \left[\Omega_{i/m_n, \hat{\mathcal{F}}_n}(\bm{\pi}_n^{\varepsilon_n, i/m_n})\right]_{i = 0}^{m_n}\right) - \max_{\bm{\delta} \in \mathbb{S}_K} \Omega_{\lambda, \mathcal{F}}(\bm{\delta})\right|
\end{equation}
is bounded from above by $$2\alpha(\mathcal{F})/m_n + \sup_{\bm{\delta} \in \mathbb{S}_K, \lambda \in [0, 1]} |\Omega_{\lambda, \mathcal{F}}(\bm{\delta}) - \Omega_{\lambda, \hat{\mathcal{F}}_n}(\bm{\delta}) | + \varepsilon_n,$$ where for every~$\rho > 0$ it holds that 
\begin{equation}\label{eqn:valfun}
\mathbb{P}^*\left( \sup_{\bm{\delta} \in \mathbb{S}_K, \lambda \in [0, 1]} \left|\Omega_{\lambda, \mathcal{F}}(\bm{\delta}) - \Omega_{\lambda, \hat{\mathcal{F}}_n}(\bm{\delta})  \right| \geq \rho \right) \leq  4 | \mathcal{Z}||\mathcal{X}|K^2 \max_{x, z, j}  e^{- \frac{nq_{x,z,j} \rho^2}{128(|\mathcal{X}| \vee |\mathcal{Z}|)^2}},
\end{equation}
for~$q_{x,z,i} := p_{X, Z, D}(x, z, i)$. In particular, the expression in~\eqref{eqn:interpol} converges to~$0$ in outer probability if~$m_n \to \infty$ and~$\varepsilon_n \to 0$.
\end{theorem}
To prove Theorem~\ref{thm:interpol} we need an auxiliary lemma:
\begin{lemma}\label{lem:auxinterpol}
If~$f: [0, 1] \to \R$ is Lipschitz continuous with constant~$c > 0$ and~$m \in \N$, then $$\sup_{\lambda \in [0, 1]}|\mathrm{LIP}_m(\lambda; [f(i/m)]_{i = 0}^m) - f(\lambda)|\leq 2c/m.$$
\end{lemma}

\begin{proof}
We give the proof for the case~$c = 1$ (which one can then apply to~$\tilde{f} := f/c$ to obtain the general statement). Fix~$\lambda \in [0, 1]$ and let~$i \in \{0, \hdots, m-1\}$ be such that~$\lambda \in [i/m, (i+1)/m]$. By definition, $$ \mathrm{LIP}_m(\lambda; [f(i/m)]_{i = 0}^m) = f(i/m) + m(f((i+1)/m) - f(i/m)) \left(\lambda - i/m\right),$$ from which we obtain that~$|\mathrm{LIP}_m(\lambda; [f(i/m)]_{i = 0}^m) - f(\lambda)|$ is bounded from above by $$|f(i/m) - f(\lambda)| + |f((i+1)/m) - f(i/m)| \leq 2/m.$$
\end{proof}

\begin{proof}[Proof of Theorem~\ref{thm:interpol}:]
Proposition~\ref{prop:exmax} shows that the value function~$\lambda \mapsto \max_{\bm{\delta} \in \mathbb{S}_K} \Omega_{\lambda, \mathcal{F}}(\bm{\delta})$ is continuous and convex on~$[0, 1]$. With the same argument as in the proof of that proposition one can show that the value function is continuous and convex on, e.g.,~$[-1, 2]$. It follows that the value function is Lipschitz continuous on~$[0, 1]$ with finite constant~$\alpha(\mathcal{F}) > 0$, say. From Lemma~\ref{lem:auxinterpol} it hence follows that 
\begin{equation}\label{eqn:supINT}
\sup_{\lambda \in [0, 1]}\left|\mathrm{LIP}_{m_n}\left(\lambda; \left[\max_{\bm{\delta} \in \mathbb{S}_K} \Omega_{i/m_n, \mathcal{F}}(\bm{\delta})\right]_{i = 0}^{m_n}\right) - \max_{\bm{\delta} \in \mathbb{S}_K} \Omega_{\lambda, \mathcal{F}}(\bm{\delta})\right| \leq 2\alpha(\mathcal{F})/m_n.
\end{equation}
Next, we note that (by the definition of~$\mathrm{LIP}_{m}$) the expression
\begin{equation}\label{eqn:tsinterpol}
\sup_{\lambda \in [0, 1]}\left|\mathrm{LIP}_{m_n}\left(\lambda; \left[\max_{\bm{\delta} \in \mathbb{S}_K} \Omega_{i/m_n, \mathcal{F}}(\bm{\delta})\right]_{i = 0}^{m_n}\right) - \mathrm{LIP}_{m_n}\left(\lambda; \left[\Omega_{i/m_n, \hat{\mathcal{F}}_n}(\bm{\pi}^{\varepsilon_n, i/m_n})\right]_{i = 0}^{m_n}\right)\right|.
\end{equation}
coincides with~$$\max_{i = 0, \hdots, m} \left| \max_{\bm{\delta} \in \mathbb{S}_K} \Omega_{i/m_n, \mathcal{F}}(\bm{\delta})  - \Omega_{i/m_n, \hat{\mathcal{F}}_n}(\bm{\pi}^{\varepsilon_n, i/m_n}) \right|.$$ Fix an index~$i \in \{0, \hdots, m\}$ and upper bound~$| \max_{\bm{\delta} \in \mathbb{S}_K} \Omega_{i/m_n, \mathcal{F}}(\bm{\delta})  - \Omega_{i/m_n, \hat{\mathcal{F}}_n}(\bm{\pi}^{\varepsilon_n, i/m_n}) |$ by\footnote{Recall from the last item in Proposition~\ref{prop:exmax} that the supremum inserted is finite.} $$\left| \sup_{\bm{\delta} \in \mathbb{S}_K} \Omega_{i/m_n, \mathcal{F}}(\bm{\delta})  - \sup_{\bm{\delta} \in \mathbb{S}_K} \Omega_{i/m_n, \hat{\mathcal{F}}_n}(\bm{\delta})\right| + \left| \sup_{\bm{\delta} \in \mathbb{S}_K} \Omega_{i/m_n, \hat{\mathcal{F}}_n}(\bm{\delta}) -  \Omega_{i/m_n, \hat{\mathcal{F}}_n}(\bm{\pi}^{\varepsilon_n, i/m_n}) \right|,$$ which, by definition of the policy, is upper bounded by $$ \sup_{\bm{\delta} \in \mathbb{S}_K, \lambda \in [0, 1]} |\Omega_{\lambda, \mathcal{F}}(\bm{\delta}) - \Omega_{\lambda, \hat{\mathcal{F}}_n}(\bm{\delta}) | + \varepsilon_n.$$ Combining the just-obtained bound with~\eqref{eqn:supINT}, we can upper bound $$\sup_{\lambda \in [0, 1]}\left|\mathrm{LIP}_{m_n}\left(\lambda; \left[\Omega_{i/m_n, \hat{\mathcal{F}}_n}(\bm{\pi}_n^{\varepsilon_n, i/m_n})\right]_{i = 0}^{m_n}\right) - \max_{\bm{\delta} \in \mathbb{S}_K} \Omega_{\lambda, \mathcal{F}}(\bm{\delta})\right|$$ by
\begin{equation*}
2\alpha(\mathcal{F})/m_n +  \sup_{\bm{\delta} \in \mathbb{S}_K, \lambda \in [0, 1]} |\Omega_{\lambda, \mathcal{F}}(\bm{\delta}) - \Omega_{\lambda, \hat{\mathcal{F}}_n}(\bm{\delta}) | + \varepsilon_n.
\end{equation*}
Equation~\eqref{eqn:valfun} is established as~\eqref{eqn:filemEc62} in Lemma~\ref{lem:uzaux}. The final statement in the theorem is now obvious.
\end{proof}

\begin{remark}\label{rem:pointwval}
In a situation where one is not interested in uniform estimation of the value function, but is only interested in estimating the value function at a given (possibly data-dependent) preference parameter~$\hat{\lambda}_n \in [0, 1]$ (e.g., the one that is intended to be used), one trivially obtains a ``pointwise''  approximation result by directly applying the upper bound given in Equation~\eqref{eqn:valfun}, noting that
$$\sup_{\bm{\delta} \in \mathbb{S}_K} \left|\Omega_{\hat{\lambda}_n, \mathcal{F}}(\bm{\delta}) - \Omega_{\hat{\lambda}_n, \hat{\mathcal{F}}_n}(\bm{\delta})  \right| \geq \rho$$
implies
$$\sup_{\bm{\delta} \in \mathbb{S}_K, \lambda \in [0, 1]} \left|\Omega_{\lambda, \mathcal{F}}(\bm{\delta}) - \Omega_{\lambda, \hat{\mathcal{F}}_n}(\bm{\delta})  \right| \geq \rho.$$ 
\end{remark}

\subsection{An auxiliary result on M-estimators}\label{sec:auxMest}

The strategy we employ to establish consistency of the policy in Section~\ref{ssec:consi} is based on the standard consistency proof for M-estimators. We need to take care of the possibility, however, that our (population) objective function evaluated at the data-driven tuning parameter, i.e., ~$\bm{\delta} \mapsto \Omega_{  \hat{\lambda}_n, \mathcal{F}}(\bm{\delta})$,
\begin{enumerate}[label=(\roman*)]
\item may not permit a unique optimizer (a problem that is fairly standard), and that
\item the corresponding~$\arg\max_{\bm{\delta} \in \mathbb{S}_K} \Omega_{  \hat{\lambda}_n, \mathcal{F}}(\bm{\delta})$ depends on the data-driven tuning parameter~$\hat{\lambda}_n$ (a problem which substantially complicates the situation).
\end{enumerate} 
To exhibit the problem and to connect it to classical situations that do not suffer from (i) and (ii), cf., e.g., Chapter 5 of \cite{vdV}, we start with the following auxiliary lemma. The statement is not formulated within the framework laid out in Section~\ref{sec:setting}, but is put in a more general context. In that result~$\theta$ plays the role of a target parameter that one wishes to estimate consistently by making use of the sequence of estimators~$\hat{\theta}_n$, whereas~$\gamma$ plays the role of a tuning parameter that defines the objective function and which is selected in a data-driven way through~$\hat{\gamma}_n$. Recall that all random variables are assumed to be defined on an underlying probability space~$(\Omega, \mathcal{A}, \mathbb{P})$. As usual, we shall notationally suppress the dependence of a function on~$\omega \in \Omega$; e.g., for a function~$f: A \times \Omega \to \R$, say, we write~$f(a)$ instead of~$f(a, \omega)$ if there is no reason for confusion. 

\begin{lemma}\label{lem:mtune}
Let~$(\Theta,d)$ be a compact metric space ($\Theta$ non-empty),~$\Gamma \subseteq \R$ be non-empty, and let the functions~$$Q_n: \Theta \times \Gamma \times \Omega \to \R,~n \in \N, \quad \text{ and } \quad Q: \Theta \times \Gamma \to \R$$ be such that~$Q(\cdot, \gamma): \Theta \to \R$ is continuous for every~$\gamma \in \Gamma$. For~$\rho > 0$ and~$\gamma \in \Gamma$ define $$\mathbb{M}_{\gamma}^{\rho} := \{\theta' \in \Theta: d(\theta', \argmax_{\theta \in \Theta} Q(\theta, \gamma)) \geq \rho\},$$ and let~$\hat{\gamma}_n: \Omega \to \Gamma$ and~$\hat{\theta}_n: \Omega \to \Theta$. Then, the following holds:
\begin{enumerate}
\item For every~$\rho > 0$, it holds that $\hat{\theta}_n \in \mathbb{M}_{\hat{\gamma}_n}^{\rho}$ implies~$\max_{\theta \in \Theta} Q(\theta, \hat{\gamma}_n) - \max_{\theta \in \mathbb{M}_{\hat{\gamma}_n}^{\rho}} Q(\theta, \hat{\gamma}_n) > 0$ and
\begin{equation}\label{eqn:pcon}
\frac{2 \sup_{\theta \in \Theta} | Q_n(\theta, \hat{\gamma}_n) - Q(\theta, \hat{\gamma}_n) | + \left(\sup_{\theta \in \Theta} Q_n(\theta, \hat{\gamma}_n) - Q_n(\hat{\theta}_n, \hat{\gamma}_n)\right)}{\max_{\theta \in \Theta} Q(\theta, \hat{\gamma}_n) - \max_{\theta \in \mathbb{M}_{\hat{\gamma}_n}^{\rho}} Q(\theta, \hat{\gamma}_n)}  \geq 1.
\end{equation}
\item For every~$\rho > 0$, it holds that if~$\mathbb{M}_{\gamma}^{\rho} \neq \emptyset$ for all~$\gamma \in \Gamma$, if~$\Gamma$ is compact, and if $\gamma \mapsto \max_{\theta \in \Theta} Q(\theta, \gamma) - \max_{\theta \in \mathbb{M}_{\gamma}^{\rho}} Q(\theta, \gamma)$ is continuous (e.g., because~$Q: \Theta \times \Gamma \to \R$ and~$\gamma \mapsto \max_{\theta \in \mathbb{M}_{\gamma}^{\rho}} Q(\theta, \gamma)$ are continuous), then 
\begin{equation}\label{eqn:infdenom}
\inf_{\gamma \in \Gamma} \left[ \max_{\theta \in \Theta} Q(\theta, \gamma) - \max_{\theta \in \mathbb{M}_{\gamma}^{\rho}} Q(\theta, \gamma) \right] > 0.
\end{equation}
\item The relation \eqref{eqn:infdenom} for every~$\rho > 0$, together with~$$\sup_{\theta \in \Theta} | Q_n(\theta, \hat{\gamma}_n) - Q(\theta, \hat{\gamma}_n) | \quad \text{ and } \quad \sup_{\theta \in \Theta} Q_n(\theta, \hat{\gamma}_n) - Q_n(\hat{\theta}_n, \hat{\gamma}_n)$$ both converging (as~$n \to \infty$) to zero in outer probability, implies consistency of~$\hat{\theta}_n$, in the sense that for every~$\rho > 0$ it holds that $$\mathbb{P}^* \left( d(\hat{\theta}_n, \argmax_{\theta \in \Theta} Q(\theta, \gamma)) \geq \rho \right) = \mathbb{P}^* \left( \hat{\theta}_n \in \mathbb{M}_{\hat{\gamma}_n}^{\rho} \right) \to 0.$$
\end{enumerate}   
\end{lemma}

\begin{proof}
Let~$\rho > 0$ be given. Suppose that~$\omega \in \Omega$ satisfies~$\hat{\theta}_n(\omega) \in \mathbb{M}_{\hat{\gamma}(\omega)}^{\rho}$ (we carefully signify the dependence on~$\omega$ of all respective quantities in this proof). Then~$\mathbb{M}_{\hat{\gamma}(\omega)}^{\rho}$ cannot be empty. The set~$\argmax_{\theta \in \Theta} Q(\theta, \hat{\gamma}_n(\omega))$ is non-empty and closed due to compactness of~$\Theta$ and continuity of~$\theta \mapsto Q(\theta, \hat{\gamma}_n(\omega))$. Therefore,~$\theta' \mapsto d(\theta', \argmax_{\theta \in \Theta} Q(\theta, \hat{\gamma}_n(\omega)))$ is Lipschitz-continuous on~$\Theta$. It follows that~$\mathbb{M}_{\hat{\gamma}(\omega)}^{\rho}$ is a closed, and hence compact, non-empty subset of~$\Theta$. Therefore,~$\max_{\theta \in \mathbb{M}_{\hat{\gamma}_n(\omega)}^{\rho}} Q(\theta, \hat{\gamma}_n(\omega))$ exists and, using the definition of~$\mathbb{M}_{\hat{\gamma}(\omega)}^{\rho}$ together with~$\rho > 0$, satisfies
\begin{equation*}
\infty > \max_{\theta \in \Theta} Q(\theta, \hat{\gamma}_n(\omega)) - \max_{\theta \in \mathbb{M}_{\hat{\gamma}_n(\omega)}^{\rho}} Q(\theta, \hat{\gamma}_n(\omega)) =: \Delta(\rho, \omega) > 0,
\end{equation*}
which establishes the first statement in Part 1. From~$\hat{\theta}_n(\omega) \in \mathbb{M}_{\hat{\gamma}(\omega)}^{\rho}$ we therefore obtain
\begin{equation}\label{eqn:tiq}
\max_{\theta \in \Theta} Q(\theta, \hat{\gamma}_n(\omega)) - Q(\hat{\theta}_n(\omega), \hat{\gamma}_n(\omega)) \geq \Delta(\rho, \omega).
\end{equation}
We now note that the inequality in~\eqref{eqn:pcon} is trivially satisfied if the supremum~$\sup_{\theta \in \Theta} Q_n(\theta, \hat{\gamma}_n(\omega)) = \infty$, and we can therefore assume henceforth that this supremum is a real number (it obviously cannot equal~$-\infty$). Adding and subtracting~$\sup_{\theta \in \Theta} Q_n(\theta, \hat{\gamma}_n(\omega))$ and~$Q_n(\hat{\theta}_n(\omega), \hat{\gamma}_n(\omega))$, respectively, and using~$$\big|\max_{\theta \in \Theta} Q(\theta, \hat{\gamma}_n(\omega)) - \sup_{\theta \in \Theta} Q_n(\theta, \hat{\gamma}_n(\omega))\big| \leq  \sup_{\theta \in \Theta} \big|Q(\theta, \hat{\gamma}_n(\omega)) - Q_n(\theta, \hat{\gamma}(\omega))\big|,$$ we can upper bound the left-hand side in~\eqref{eqn:tiq} by 
\begin{equation*}
2
\sup_{\theta \in \Theta} \big|Q(\theta, \hat{\gamma}_n(\omega)) - Q_n(\theta, \hat{\gamma}_n(\omega))\big| + \left(\sup_{\theta \in \Theta} Q_n(\theta, \hat{\gamma}_n(\omega)) - Q_n(\hat{\theta}_n(\omega), \hat{\gamma}_n(\omega)) \right).
\end{equation*}
Using~$\Delta(\rho, \omega) > 0$ we now obtain the inequality in~\eqref{eqn:pcon}.

To verify the statement in Part 2, just note that under the continuity condition imposed the infimum in~\eqref{eqn:infdenom} is attained at~$\gamma^* \in \Gamma$, say, because~$\Gamma$ is assumed to be compact (and non-empty). This proves the statement, cf.~also the definition of~$\mathbb{M}_{\gamma^*}^{\rho}$. To see that the statement in parenthesis is sufficient for the continuity assumption in Part 2, we can use Berge's maximum theorem, e.g., Theorem 17.31 in \cite{hhg}. 

Part 3 is an immediate consequence of (the imposed assumptions and) Part 1.

\end{proof}

\begin{remark}
Note that the quotient on the left-hand side of~\eqref{eqn:pcon} relates the worst-case \emph{approximation error} and the \emph{optimization error}, i.e., $$\sup_{\theta \in \Theta} | Q_n(\theta, \hat{\gamma}_n) - Q(\theta, \hat{\gamma}_n) | \quad \text{ and } \quad \sup_{\theta \in \Theta} Q_n(\theta, \hat{\gamma}_n) - Q_n(\hat{\theta}_n, \hat{\gamma}_n)$$ to the minimal change in the value of the objective function~$Q(\cdot, \hat{\gamma}_n)$ when being~$\rho$-separated from its set of maximizers, i.e., to $$\max_{\theta \in \Theta} Q(\theta, \hat{\gamma}_n) - \max_{\theta \in \mathbb{M}_{\hat{\gamma}_n}^{\rho}} Q(\theta, \hat{\gamma}_n).$$  In principle, the optimization error can be controlled by the practitioner and is therefore often, but not always, ignored, i.e., assumed to be~$0$ when studying consistency of an estimator defined as the solution of an optimization problem. Under suitable assumptions (e.g., a uniform law of large numbers in case~$Q_n$ is an average), the approximation error typically converges to~$0$ in (outer) probability. 
\end{remark}

\begin{remark}
Part 3 of Lemma~\ref{lem:mtune} shows that~\eqref{eqn:infdenom} together with convergence to zero of the approximation and optimization errors in (outer) probability is enough to ensure consistency of the estimator. The second part of that lemma shows that~\eqref{eqn:infdenom} follows from a set of conditions  including a continuity assumption. The condition in~\eqref{eqn:infdenom}, however, is not generally met in the problem we consider in the present article \emph{without} imposing conditions on~$\Gamma$, e.g., finiteness. This is illustrated in the context of the example discussed in Section~\ref{ssec:ex} in the following remark, and implies that additional assumptions on~$\hat{\lambda}_n$ are necessary to deduce a consistency statement through the approach outlined above. 
\end{remark}

\begin{example}\label{ex:ctd}
Reconsider the example discussed in Section~\ref{ssec:ex}. Given~$\rho \in (0, 1/2)$, we also have to consider the maximum of the objective in~\eqref{eqn:objexs} over the set~$\mathbb{M}_{\mathcal{F}, \lambda}^{\rho}$, say, of all decision rules~$\bm{\eta} = (\eta, 1-\eta)'$,~$\eta \in [0, 1]$, that are separated away from the argmax in~$d_1$-distance, i.e., \begin{equation*}
\mathbb{M}_{\mathcal{F}, \lambda}^{\rho} := \big\{\bm{\eta} \in \mathbb{S}_K : d_1\big( 
\bm{\eta},
\argmax_{\bm{\delta} \in \mathbb{S}_K} \Omega_{\lambda, \mathcal{F}}(\bm{\delta})
\big) \geq \rho
\big \}.
\end{equation*} Due to the specific structure of the argmax in that example, cf.~\eqref{eqn:argmaxex}, and continuity properties of the objective, is not difficult to see that in that example
\begin{equation*}
0 \leq \left[\max_{\bm{\delta} \in \mathbb{S}_K} \Omega_{\lambda_m, \mathcal{F}}\left(\bm{\delta}\right) - \max_{\bm{\delta} \in \mathbb{M}_{\mathcal{F}, \lambda_m}^{\rho}} \Omega_{\lambda_m, \mathcal{F}}\left(\bm{\delta}\right)\right] \leq \Omega_{\lambda_m, \mathcal{F}}((0, 1)') - \Omega_{\lambda_m}((1/2, 1/2)')\to 0
\end{equation*}
along any sequence $\lambda_m \in [0, c(p))$ that converges to~$c(p)$, showing that~\eqref{eqn:infdenom} is violated in this example. \end{example}
Hence, the condition in~\eqref{eqn:infdenom} cannot be expected to hold in our framework when selecting tuning parameters from~$[0, 1]$, and therefore obtaining consistency by controlling the approximation and estimation error as outlined above requires further conditions on~$\hat{\lambda}_n$ that were not necessary in the context of Theorem~\ref{thm:upreg2}. 

\section{Notation and some preparatory results}

Recall that for an array~$\mathcal{F}$ as in Equation~\eqref{eqn:array}, i.e.,
\begin{equation*}
\mathcal{F}= \left[(F^i(\cdot\mid x, z), p_{X, Z}(x, z)): i = 1, \hdots, K;~x \in \mathcal{X};~z \in \mathcal{Z}\right],
\end{equation*} 
and a given (non-empty) set of cdfs~$\mathscr{H} \subseteq D_{cdf}([a,b])$ we write~
\begin{equation*}
\mathcal{F} \sqsubset \mathscr{H} \quad \Leftrightarrow \quad  F^i(\cdot \mid x, z) \in \mathscr{H} \text{ for every } i = 1, \hdots, K,~x \in \mathcal{X}, \text{ and } z \in \mathcal{Z}.
\end{equation*}
For a vector~$\bm{G} = (G_1, \hdots, G_l)'$ of functions~$G: \R \to \R$ we write~$$\|\bm{G}\|_{\infty} := \max_{i = 1}^l \|G_i\|_{\infty} = \max_{i = 1}^l \sup_{v \in \R} |G_i(v)|,$$ which may be infinite. Recall that the symbol~$\|\cdot\|_p$ denotes the~$p$-norm on a Euclidean space. 

For two arrays~$\mathcal{F}$ and~$\tilde{\mathcal{F}}$ as in Equation~\eqref{eqn:array} that satisfy~$\mathcal{F} \sqsubset D_{cdf}([a,b])$ and~$\tilde{\mathcal{F}}  \sqsubset D_{cdf}([a,b])$, we obtain (for~$x \in \mathcal{X}$ and~$z \in \mathcal{Z}$ fixed, and cf.~Section~\ref{ssec:notation} for the definition of~$\bm{F}(\cdot \mid  x)$ and similar quantities)
\begin{equation*}
\| \bm{F}(\cdot \mid  x) - \tilde{\bm{F}}(\cdot \mid  x)\|_{\infty} = \max_{i = 1}^K \| F^i(\cdot \mid  x) - \tilde{F}^i(\cdot \mid  x)\|_{\infty},
\end{equation*}
and 
\begin{equation*}
\|\bm{F}(\cdot \mid  x, z) - \tilde{\bm{F}}(\cdot \mid  x, z)\|_{\infty} = \max_{i = 1}^K \|F^i(\cdot \mid  x, z) - \tilde{F}^i(\cdot \mid  x, z)\|_{\infty};
\end{equation*}
furthermore, we associate to an array~$\mathcal{F}$ as in Equation~\eqref{eqn:array} the quantities
\begin{equation*}
\bm{p}_X = \left(p_X(x)\right)_{x\in \mathcal{X}} \quad \text{ and } \quad \bm{p}_{X, Z} = \left(p_{X, Z}(x, z)\right)_{x\in \mathcal{X}, z \in \mathcal{Z}}
\end{equation*}
and write, e.g.,~$\tilde{\bm{p}}_X$ or $\hat{\bm{p}}_X$, respectively for the same quantities but corresponding to~$\tilde{\mathcal{F}}$ or~$\hat{\mathcal{F}}_n$.  Similarly, we associate to~$\mathcal{F}$ (and analogously to~$\tilde{\mathcal{F}}$ or~$\hat{\mathcal{F}}_n$), for every fixed~$z \in \mathcal{Z}$, the vectors~$\bm{p}_{X\mid Z=z} = (p_{X\mid Z=z}(x))_{x \in \mathcal{X}}$ and $\bm{p}_Z = \left(p_Z(z)\right)_{z\in \mathcal{Z}}$.

Recall that for~$\bm{\delta} \in \mathbb{S}_K$ and~$z \in \mathcal{Z}$ we write
\begin{equation*}
\langle \bm{\delta}, \mathcal{F} \rangle = \sum_{x \in \mathcal{X}}  \sum_{i = 1}^K  \delta_i(x) p_X(x) F^i(\cdot\mid x) \quad \text{ and } \quad \langle \bm{\delta}, \mathcal{F} \rangle_z = \sum_{x \in \mathcal{X}} \sum_{i = 1}^K \delta_i(x) p_{X \mid  Z = z}(x) F^i(\cdot\mid x, z),
\end{equation*}
and denote 
\begin{equation*}
u({\mathcal{F}}, \tilde{\mathcal{F}}) := \sum_{x \in \mathcal{X}} \|p_X(x) \bm{F}(\cdot \mid  x) - \tilde{p}_X(x)\tilde{\bm{F}}(\cdot \mid  x)\|_{\infty}.
\end{equation*} 
and
\begin{equation}\label{eqn:uzdef}
u_z(\mathcal{F}, \tilde{\mathcal{F}}) := \sum_{x \in \mathcal{X}} \|p_{X\mid Z=z}(x) \bm{F}(\cdot \mid  x, z) - \tilde{p}_{X\mid Z=z}(x) \tilde{\bm{F}}(\cdot \mid  x, z)\|_{\infty}.
\end{equation}
\begin{remark}
Throughout, we shall use the symbols~$\langle \bm \gamma, \mathcal{F} \rangle$ and~$\langle  \bm \gamma, \mathcal{F} \rangle_z$ for any~$ \bm \gamma: \mathcal{X} \to \R^K$ (not necessarily a map to~$\mathbb{S}_K$). Note, however, that for such~$ \bm \gamma$, these quantities are no longer cdfs in general.
\end{remark}

With this notation (recalling~$d_1$ from~\eqref{eqn:dstS}), we obtain the following result.
\begin{lemma}\label{lem:auxn1}
For two arrays~$\mathcal{F}$ and~$\tilde{\mathcal{F}}$ as in Equation~\eqref{eqn:array} and~$\bm{\delta}$ and~$\tilde{\bm{\delta}}$ in~$\mathbb{S}_K$, we have 
\begin{equation}\label{eqn:up1}
\| \langle \bm{\delta}, \mathcal{F} \rangle - \langle \tilde{\bm{\delta}}, \tilde{\mathcal{F}} \rangle \|_{\infty} \leq d_1(\bm{\delta}, \tilde{\bm{\delta}}) + u({\mathcal{F}}, \tilde{\mathcal{F}}),
\end{equation}	
and for every~$z \in \mathcal{Z}$ it holds that 
\begin{equation}\label{eqn:up2}
\| \langle \bm{\delta}, \mathcal{F} \rangle_z - \langle \tilde{\bm{\delta}}, \tilde{\mathcal{F}} \rangle_z \|_{\infty} \leq d_1(\bm{\delta}, \tilde{\bm{\delta}}) + u_z(\mathcal{F}, \tilde{\mathcal{F}}).
\end{equation}
Furthermore, it holds that
\begin{equation}\label{eqn:utouz}
u(\mathcal{F}, \tilde{\mathcal{F}}) \leq \sum_{z \in \mathcal{Z}} p_Z(z) u_z(\mathcal{F}, \tilde{\mathcal{F}}) + \|\bm{p}_Z - \tilde{\bm{p}}_Z\|_1 \leq \max_{z \in \mathcal{Z}} u_z(\mathcal{F}, \tilde{\mathcal{F}}) + \|\bm{p}_Z - \tilde{\bm{p}}_Z\|_1.
\end{equation}
\end{lemma}

\begin{proof}
We first establish~\eqref{eqn:up1}. Writing $$\langle \tilde{\bm{\delta}} - \bm{\delta}, \tilde{\mathcal{F}} \rangle := \sum_{x \in \mathcal{X}} \sum_{i = 1}^K (\tilde{\delta}_i(x) - \delta_i(x)) \tilde{p}_X(x) \tilde{F}^i(\cdot \mid x),$$ the triangle inequality delivers $$\| \langle \bm{\delta}, \mathcal{F} \rangle - \langle \tilde{\bm{\delta}}, \tilde{\mathcal{F}} \rangle \|_{\infty} \leq \| \langle \bm{\delta}, \mathcal{F} \rangle - \langle \bm{\delta}, \tilde{\mathcal{F}} \rangle  \|_{\infty} + \|  \langle \tilde{\bm{\delta}} - \bm{\delta}, \tilde{\mathcal{F}} \rangle \|_{\infty} \leq \| \langle \bm{\delta}, \mathcal{F} \rangle - \langle \bm{\delta}, \tilde{\mathcal{F}} \rangle  \|_{\infty} +  d_1(\bm{\delta}, \tilde{\bm{\delta}}).$$ From~$\bm{\delta}(x) \in \mathscr{S}_K$ for every~$x \in \mathcal{X}$, we obtain $$\| \langle \bm{\delta}, \mathcal{F} \rangle - \langle \bm{\delta}, \tilde{\mathcal{F}} \rangle  \|_{\infty} \leq \sum_{x \in \mathcal{X}} \sum_{i = 1}^K \delta_i(x) \| p_X(x) F^i(\cdot \mid x) - \tilde{p}_X(x) \tilde{F}^i(\cdot \mid x) \|_{\infty} \leq u(\mathcal{F}, \tilde{\mathcal{F}}),$$ which delivers~\eqref{eqn:up1}. The inequality in~\eqref{eqn:up2} is obtained analogously. Concerning the inequalities in~\eqref{eqn:utouz}, we note that~$$p_X(x) \bm{F}(\cdot \mid x) = \sum_{z \in \mathcal{Z}} p_{X, Z}(x, z) \bm{F}(\cdot \mid x, z)$$ allows us to write \begin{align*}
u(\mathcal{F}, \tilde{\mathcal{F}}) &= \sum_{x \in \mathcal{X}} \| \sum_{z \in \mathcal{Z}} p_Z(z) \left \{
p_{X \mid Z = z}(x) \bm{F}(\cdot \mid x, z) - \tilde{p}_{X \mid Z = z}(x) \tilde{\bm{F}}(\cdot \mid x, z) 
\right\} \\ & \hspace{4cm} - \sum_{z \in \mathcal{Z}} \tilde{p}_{X|Z = z}(x) \left(\tilde{p}_Z(z) - p_Z(z)\right) \tilde{\bm{F}}(\cdot | x, z) \|_{\infty},
\end{align*}
which (using the triangle inequality) we can upper bound by $$\sum_{z \in \mathcal{Z}} p_Z(z) u_z(\mathcal{F}, \tilde{\mathcal{F}}) + \sum_{z \in \mathcal{Z}} \left[ \sum_{x \in \mathcal{X}} \tilde{p}_{X \mid Z = z}(x) \right] |p_Z(z) - \tilde{p}_Z(z)|,$$ from which two inequalities in~\eqref{eqn:utouz} follow.
\end{proof}

\section{Auxiliary results and proofs for Section~\ref{sec:setting}}\label{app:proofsmain}

For the following result, we recall from Equation~\eqref{eqn:omegadef} the definition of the objective function
\begin{equation*}
\Omega_{\lambda, \mathcal{F}}(\bm{\delta}) = (1-\lambda) \mathsf{T}\left( \langle \bm{\delta}, \mathcal{F} \rangle \right) - \lambda \max_{z \in \mathcal{Z}} \mathsf{S} \left(\langle \bm{\delta}, \mathcal{F} \rangle_z , \langle \bm{\delta}, \mathcal{F} \rangle \right).
\end{equation*} 

\begin{lemma}\label{lem:exmax}
Fix~$\lambda \in [0, 1]$, let~$\mathcal{F} \sqsubset \mathscr{D}$ and~$\tilde{\mathcal{F}}  \sqsubset D_{cdf}([a,b])$ be arrays as in Equation~\eqref{eqn:array}, and let~$\bm{\delta}$ and~$\tilde{\bm{\delta}}$ be elements of~$\mathbb{S}_K$. Then, 
\begin{equation}\label{eqn:uplem1}
\left|\Omega_{\lambda, \mathcal{F}}(\bm{\delta}) - \Omega_{\lambda, \tilde{\mathcal{F}}}(\tilde{\bm{\delta}})\right| \leq  (1+\lambda) \left[ d_1(\bm{\delta}, \tilde{\bm{\delta}}) + \max_{z \in \mathcal{Z}} u_z(\mathcal{F}, \tilde{\mathcal{F}})\right] + \|\bm{p}_Z - \tilde{\bm{p}}_Z\|_1.
\end{equation}
In particular, the function~$\bm{\delta} \mapsto \Omega_{\lambda, \mathcal{F}}(\bm{\delta})$ is Lipschitz continuous on~$\mathbb{S}_K$ with constant~$1+\lambda$. Furthermore, the function $(\lambda, \bm{\delta}) \mapsto \Omega_{\lambda, \mathcal{F}}(\bm{\delta})$ is Lipschitz continuous on~$[0, 1] \times \mathbb{S}_K$.
\end{lemma}

\begin{proof}[Proof of Lemma~\ref{lem:exmax}]
We upper bound~$$|\Omega_{\lambda, \mathcal{F}}(\bm{\delta}) - \Omega_{\lambda, \tilde{\mathcal{F}}}(\tilde{\bm{\delta}})| \leq (1-\lambda) A + \lambda B,$$ where~$$A :=  \left |\mathsf{T}(\langle \bm{\delta}, \mathcal{F} \rangle) - \mathsf{T}(\langle \tilde{\bm{\delta}}, \tilde{\mathcal{F}} \rangle) \right| \leq d_1(\bm{\delta}, \tilde{\bm{\delta}}) + \max_{z \in \mathcal{Z}} u_z({\mathcal{F}}, \tilde{\mathcal{F}}) + \|\bm{p}_Z - \tilde{\bm{p}}_Z\|_1,$$ the inequality follows from Assumption~\ref{as:MAIN} (note that~$\mathscr{D}$ is assumed to be convex and~$\mathcal{F} \sqsubset \mathscr{D}$ holds by assumption, implying that~$\langle \bm{\delta}, \mathcal{F} \rangle \in \mathscr{D}$) and~\eqref{eqn:up1} together with~\eqref{eqn:utouz}; and 
\begin{align*}
B :&=  \max_{z \in \mathcal{Z}} \left| \mathsf{S}\left(\langle \bm{\delta}, \mathcal{F} \rangle_z, \langle \bm{\delta}, \mathcal{F} \rangle\right) - \mathsf{S}\left(\langle \bm{\tilde{\delta}}, \tilde{\mathcal{F}} \rangle_z, \langle \bm{\tilde{\delta}}, \tilde{\mathcal{F}} \rangle \right)\right| \\ &\leq
\max_{z \in \mathcal{Z}} \|\langle \bm{\delta}, \mathcal{F} \rangle_z - \langle \bm{\tilde{\delta}}, \tilde{\mathcal{F}} \rangle_z \|_{\infty} + \|\langle \bm{\delta}, \mathcal{F} \rangle - \langle \bm{\tilde{\delta}}, \tilde{\mathcal{F}} \rangle\|_{\infty},
\end{align*}
where we used Assumption~\ref{as:MAIN}. Using Equations~\eqref{eqn:up1},~\eqref{eqn:up2}, and~\eqref{eqn:utouz}, we obtain
\begin{equation*}
B \leq 2 \left[ d_1(\bm{\delta}, \tilde{\bm{\delta}}) + \max_{z \in \mathcal{Z}} u_z({\mathcal{F}}, \tilde{\mathcal{F}}) \right] + \|\bm{p}_Z - \tilde{\bm{p}}_Z\|_1.
\end{equation*}
Combining the upper bounds on A and B just obtained, the inequality in~\eqref{eqn:uplem1} and the Lipschitz-continuity statement in the lemma follows. To see the last statement in the lemma, note that the upper bounds on A and B show that the functions~$$\bm{\delta} \mapsto \mathsf{T}(\langle \bm{\delta}, \mathcal{F} \rangle) \quad \text{ and } \quad \bm{\delta} \mapsto \max_{z \in \mathcal{Z}} \mathsf{S} \left(\langle \bm{\delta}, \mathcal{F} \rangle_z , \langle \bm{\delta}, \mathcal{F} \rangle \right)$$ are Lipschitz continuous on~$\mathbb{S}_K$, from which the Lipschitz continuity of~$(\lambda, \bm{\delta}) \mapsto \Omega_{\lambda, \mathcal{F}}(\bm{\delta})$ follows.
\end{proof}

\begin{proof}[Proof of Proposition~\ref{prop:exmax}]
Lemma~\ref{lem:exmax} shows that the function~$(\lambda, \bm{\delta}) \mapsto \Omega_{\lambda, \mathcal{F}}(\bm{\delta})$ is continuous on~$[0, 1] \times \mathbb{S}_K$. Therefore, all assumptions for Berge's maximum theorem (as in, e.g., \cite{hhg}, Theorem 17.31) are satisfied, which establishes the first three statements in the proposition apart from the convexity statement, which is a consequence of Danskin's theorem (cf., e.g., Proposition B.25 in~\cite{bertsekas}).

To establish the last statement, let~$\mathcal{F} \sqsubset \mathscr{D}$, let~$\bm{\delta} \in \mathbb{S}_K$ and use Lemma~\ref{lem:exmax} to obtain 
\begin{equation}\label{eqn:supfinb}
\Omega_{\lambda, \mathcal{G}}(\bm{\delta}) \leq  |\Omega_{\lambda, \mathcal{G}}(\bm{\delta}) - \Omega_{\lambda, \mathcal{F}}(\bm{\delta})| + \Omega_{\lambda, \mathcal{F}}(\bm{\delta}) \leq 2 \max_{z \in \mathcal{Z}} u_z(\mathcal{F}, \tilde{\mathcal{F}}) + 2 + \Omega_{\lambda, \mathcal{F}}(\bm{\delta}).
\end{equation}
Taking the supremum over~$\bm{\delta} \in \mathbb{S}_K$ in both sides of~\eqref{eqn:supfinb} and using the fact already established in the first part of this proof then delivers the last statement of the proposition.
\end{proof}

\section{Auxiliary results and proofs for Section~\ref{sec:esr}}

Throughout this section, we shall sometimes abbreviate the training sample from~\eqref{eqn:obs} as follows
\begin{equation}\label{eqn:notconv}\left(\tilde{\bm{W}}_j, j = 1, \hdots, n\right) =: \tilde{\bm{W}}.
\end{equation} 
Correspondingly, we can write
$$\bm{\pi}^{\varepsilon, \lambda}_n\left(\tilde{\bm{W}}_j, j = 1, \hdots, n\right) = \bm{\pi}^{\varepsilon, \lambda}_n(\tilde{\bm{W}}).$$
\begin{lemma}\label{lem:upreg}
Let~$\lambda \in [0, 1]$ and~$\mathcal{F}$ be an array as in Equation~\eqref{eqn:array} satisfying $\mathcal{F} \sqsubset \mathscr{D}$. Let~$\bm{\pi}^{\varepsilon, \lambda}_n$ be an empirical success policy as defined in Section~\ref{sec:esr}. Then,
\begin{align*}
r(\bm{\pi}^{\varepsilon, \lambda}_n; \lambda, \mathcal{F})
&\leq 2 \sup_{\bm{\delta} \in \mathbb{S}_K} |\Omega_{\lambda, \mathcal{F}}(\bm{\delta}) - \Omega_{\lambda, \hat{\mathcal{F}}_n}(\bm{\delta})| + \varepsilon \\ &\leq \varepsilon + 2(1+\lambda) \max_{z \in \mathcal{Z}} u_z(\mathcal{F}, \hat{\mathcal{F}}_n) + 2 \|\bm{p}_Z - \hat{\bm{p}}_Z\|_1.
\end{align*}

\end{lemma}

\begin{proof}
Using the last item in Proposition~\ref{prop:exmax} (guaranteeing that all suprema below are finite) and~\eqref{eqn:approxmax}, we write
\begin{align*}r(\bm{\pi}^{\varepsilon, \lambda}_n; \lambda, \mathcal{F}) =
&~~\max_{\bm{\delta} \in \mathbb{S}_K} \Omega_{\lambda, \mathcal{F}}(\bm{\delta}) - \sup_{\bm{\delta} \in \mathbb{S}_K} \Omega_{\lambda, \hat{\mathcal{F}}_n}(\bm{\delta}) + \sup_{\bm{\delta} \in \mathbb{S}_K} \Omega_{\lambda, \hat{\mathcal{F}}_n}(\bm{\delta}) - \Omega_{ \lambda, \mathcal{F}} \left(\bm{\pi}^{\varepsilon, \lambda}_n(\tilde{\bm{W}})\right) \\
\leq & ~~
\max_{\bm{\delta} \in \mathbb{S}_K} \Omega_{\lambda, \mathcal{F}}(\bm{\delta}) - \sup_{\bm{\delta} \in \mathbb{S}_K} \Omega_{\lambda, \hat{\mathcal{F}}_n}(\bm{\delta}) + \Omega_{\lambda, \hat{\mathcal{F}}_n}(\bm{\pi}^{\varepsilon,\lambda}_n(\tilde{\bm{W}})) - \Omega_{\lambda, \mathcal{F}} \left(\bm{\pi}^{\varepsilon, \lambda}_n(\tilde{\bm{W}})\right) + \varepsilon \\
\leq & ~~2 \sup_{\bm{\delta} \in \mathbb{S}_K} |\Omega_{\lambda, \mathcal{F}}(\bm{\delta}) - \Omega_{\lambda, \hat{\mathcal{F}}_n}(\bm{\delta})| + \varepsilon.
\end{align*}
This proves the first inequality; the second now follows from~\eqref{eqn:uplem1} in Lemma~\ref{lem:exmax}.
\end{proof}

From~\eqref{eqn:uzdef} and Lemma~\ref{lem:upreg} it follows that the quantities determining the behavior of the upper bound for the regret just developed in Lemma~\ref{lem:upreg} are the random variables~
\begin{equation}\label{eqn:pFdef}
\|p_{X\mid Z=z}(x) \bm{F}(\cdot \mid  x, z) - \hat{p}_{X\mid Z=z}(x) \hat{\bm{F}}(\cdot \mid  x, z)\|_{\infty}, \quad \text{ for } x \in \mathcal{X}, z \in \mathcal{Z},
\end{equation}
where~$\hat{p}_{X\mid Z=z}(x) = \hat{p}_{X, Z}(x, z)/\hat{p}_Z(z)$ whenever~$\hat{p}_Z(z) > 0$ and~$\hat{p}_{X\mid Z=z}(x) = 0$ else, and the random variable~$$\|\bm{p}_Z - \hat{\bm{p}}_Z\|_1.$$ We shall next develop suitable bounds for these quantities. 

Let us recall here that for every~$i \in \{1, \hdots, K\}$,~$x \in \mathcal{X}$ and~$z \in \mathcal{Z}$ we denote by~$\hat{F}^i(\cdot\mid x, z)$ the empirical cdf based on observations in~
\begin{equation}
\mathcal{M}^i_{x, z} := \{j = 1, \hdots, n: D_j = i, X_j = x, Z_j = z\},
\end{equation}
i.e.,~$$\hat{F}^i(\cdot\mid x, z) = |\mathcal{M}^i_{x, z}|^{-1} \sum_{j \in \mathcal{M}^i_{x, z}}  \mathds{1}( Y_{D_i, j} \leq \cdot) = |\mathcal{M}^i_{x, z}|^{-1} \sum_{j \in \mathcal{M}^i_{x, z}}  \mathds{1}( Y_{i,j} \leq \cdot),$$ which we set equal to the cdf corresponding to point mass at~$b$ (the upper endpoint of the support of the cdfs one is working with) in case~$\mathcal{M}^i_{x, z}$ is empty. 

Note that~$\mathcal{M}^i_{x, z}$ is a \emph{random} set, i.e., actually~$\mathcal{M}^i_{x, z} = \mathcal{M}^i_{x, z}(D_j, X_j, Z_j; j = 1, \hdots, n)$, which we suppress notationally. In the conditioning arguments in the following proof, given fixed numbers $(d_j, x_j, z_j) \in \{1, \hdots, K\} \times \mathcal{X} \times \mathcal{Z}$ for $j = 1, \hdots, n$ we shall analogously write 
\begin{equation}\label{eqn:nrs}
\mathcal{M}^i_{x, z}(d_j, x_j, z_j; j = 1, \hdots, n) = \{j = 1, \hdots, n: d_j = i, x_j = x, Z_j = z\},
\end{equation} which is a \emph{non-random} subset of~$\{1, \hdots, n\}$ (i.e.,~$\mathcal{M}^i_{x, z}(d_j, x_j, z_j; j = 1, \hdots, n)$ is a possible realization of~$\mathcal{M}^i_{x, z}$).

\begin{lemma}\label{lem:aux2}
Let~$(\bm{Y}_j', X_j, Z_j)' \sim \mathcal{F}$ for~$j =1, \hdots, n$, and let~$x \in \mathcal{X}$ and~$z \in \mathcal{Z}$. Then, %
\begin{equation}\label{eqn:upcdf}
\mathbb{E}\left(\| \hat{\bm{F}}(\cdot \mid x, z) - \bm{F}(\cdot\mid x, z)\|_{\infty} \right) \leq  2.75 \times \sqrt{\frac{\log(K)}{n}} \sum_{j = 1}^K (q_{x,z,j})^{-1/2},
\end{equation}
where~$q_{x,z,i} := \mathbb{P}(X_1 = x, Z_1 = z, D_1 = i)$ for~$i = 1, \hdots, K$; furthermore, for every~$\rho > 0$ it holds that
\begin{equation}\label{eqn:upcdfprob}
\mathbb{P}\left(\| \hat{\bm{F}}(\cdot \mid x, z) - \bm{F}(\cdot\mid x, z)\|_{\infty} \geq \rho \right) \leq  2K \sum_{i = 1}^K e^{-\frac{n \rho^2 q_{x, z, i}}{2}}.
\end{equation}
\end{lemma}

\begin{proof}
We first establish~\eqref{eqn:upcdf}. Fix~$x \in \mathcal{X}$ and~$z \in \mathcal{Z}$. Recall that~$q_{x,z,i} > 0$ for every~$i = 1, \hdots, K$ due to Assumption~\ref{as:posprob}. In this proof we abbreviate~$$M_i := \| \hat{F}^i(\cdot \mid  x, z) - F^i(\cdot\mid x, z)\|_{\infty},$$ and we define the random variable~$$\beta := \min_{i = 1}^K |\mathcal{M}_{x, z}^i|.$$ Define the set~$S \subseteq \left(\{1, \hdots, K\} \times \mathcal{X} \times \mathcal{Z}\right)^n$ via 
\begin{equation*}
S := \left\{(\overline{\bm{d}}, \overline{\bm{x}}, \overline{\bm{z}})\in \left(\{1, \hdots, K\} \times \mathcal{X} \times \mathcal{Z}\right)^n: \beta(\overline{\bm{d}}, \overline{\bm{x}}, \overline{\bm{z}})  > 0\right\},
\end{equation*}
where we denoted, cf.~the discussion surrounding~\eqref{eqn:nrs}, $$\beta(\overline{\bm{d}}, \overline{\bm{x}}, \overline{\bm{z}}) := \min_{i = 1}^K \left| \mathcal{M}_{x, z}^i(\overline{d_j}, \overline{x_j}, \overline{z_j}; j = 1, \hdots, n) \right|.$$ Use the law of total probability,~$\mathbb{P}(\beta = 0) = \mathbb{P}((\bm{D}, \bm{X}, \bm{Z}) \notin S)$, and~$M_i \leq 1$, to upper bound $$\mathbb{E} (\| \hat{\bm{F}}(\cdot \mid x, z) - \bm{F}(\cdot\mid x, z)\|_{\infty}) = \mathbb{E}(\max_{i = 1}^K  M_i)$$ by the sum
\begin{equation}\label{eqn:expa}
\mathbb{P}\left(\beta = 0\right) +  \sum_{(\overline{\bm{d}}, \overline{\bm{x}}, \overline{\bm{z}}) \in S} \mathbb{E}_{\overline {\bm d}, \overline {\bm x},  \overline {\bm z}} \left ( \max_{i = 1}^K M_i \right) \mathbb{P}(\bm D = \overline{\bm{d}}, \bm{X} = \overline{\bm{x}}, \bm{Z} = \overline{\bm{z}}),
\end{equation}
where~$\mathbb{E}_{\overline{\bm{d}}, \overline{\bm{x}}, \overline{\bm{z}}}$ denotes the expectation operator conditional on the event~$\{\bm{D} = \overline{\bm{d}},~\bm{X} =  \overline{\bm{x}},~\bm{Z} = \overline{\bm{z}}\}$, where~$\bm{X} := (X_1, \hdots, X_n)'$,~$\bm{Z} := (Z_1, \hdots, Z_n)'$ and~$\bm{D} := (D_1, \hdots, D_n)'$.

Fix a~$(\overline{\bm{d}}, \overline{\bm{x}}, \overline{\bm{z}}) \in S$, and note that~$\mathbb{P}(\bm D = \overline{\bm{d}}, \bm{X} = \overline{\bm{x}}, \bm{Z} = \overline{\bm{z}}) > 0$ by Assumptions~\ref{as:iid} and~\ref{as:posprob}. On the event~$\{\bm D = \overline{\bm{d}}, \bm{X} = \overline{\bm{x}}, \bm{Z} = \overline{\bm{z}}\} \subseteq \Omega$, the random set~$\mathcal{M}_{x, z}^i$ is constant and non-empty, i.e., there exist natural numbers~$j_1 < \hdots < j_{|\mathcal{M}_{x, z}^i|}$ for which~$\mathcal{M}_{x, z}^i = \{j_1, \hdots, j_{|\mathcal{M}_{x, z}^i|}\}$, and such that we can write~$$\hat{F}^i(\cdot\mid x, z) = |\mathcal{M}_{x, z}^i|^{-1} \sum_{j \in \mathcal{M}_{x, z}^i} \mathds{1}( Y_{D_j,j} \leq \cdot ) = |\mathcal{M}_{x, z}^i|^{-1} \sum_{l = 1}^{|\mathcal{M}_{x, z}^i|} \mathds{1}(Y_{i,j_l} \leq \cdot),$$ so that in particular $$M_i = \| |\mathcal{M}_{x, z}^i|^{-1} \sum_{l = 1}^{|\mathcal{M}_{x, z}^i|} \mathds{1}(Y_{i,j_l} \leq \cdot) - F^i(\cdot \mid x, z)\|_{\infty}.$$ Using the i.i.d.~assumption, for any real numbers~$c_1, \hdots, c_{|\mathcal{M}_{x, z}^i|}$ it holds that the probability
\begin{equation*}
\mathbb{P}\left( Y_{i,j_1} \leq c_1, \hdots, Y_{i,|\mathcal{M}_{x, z}^i|} \leq c_{|\mathcal{M}_{x, z}^i|} \mid  \bm D = \overline{\bm{d}}, \bm{X} = \overline{\bm{x}}, \bm{Z} = \overline{\bm{z}} \right),
\end{equation*}
can be written as the product 
\begin{equation}\label{eqn:indrel}
\prod_{l = 1}^{|\mathcal{M}_{x, z}^i|}\mathbb{P}\left( Y_{i,j_l} \leq c_l \mid  D_{j_l} = i, X_{j_l} = x, Z_{j_l} = z \right) = \prod_{l = 1}^{|\mathcal{M}_{x, z}^i|} F^i( c_l \mid x, z),
\end{equation}
the equality following from the conditional independence Assumption~\ref{as:cia}. That is, conditional on the event~$\{\bm D = \overline{\bm{d}}, \bm{X} = \overline{\bm{x}}, \bm{Z} = \overline{\bm{z}}\}$, the empirical cdf~$\hat{F}^i(\cdot\mid x, z)$ is based on~$|\mathcal{M}_{x, z}^i|$ i.i.d. random variables with cdf~$F^i(\cdot\mid x, z)$. Arguing analogously to the proof of Theorem~4.3 in \cite{kpv3} (from their Equation~C.2 to~C.3), but replacing the expectation operator there by the conditional expectation~$\mathbb{E}_{\overline{\bm{d}}, \overline{\bm{x}}, \overline{\bm{z}}}$, we hence obtain 
\begin{equation*}
\mathbb{E}_{\overline{\bm{d}}, \overline{\bm{x}}, \overline{\bm{z}}} \left(\max_{i = 1}^K M_i\right) \leq 1.505 \times \sqrt{\frac{\log(K)}{\beta(\overline{\bm{d}}, \overline{\bm{x}}, \overline{\bm{z}})}},
\end{equation*}
which together with~\eqref{eqn:expa} and upon denoting~$w_{x, z}(i) := \mathbb{P}\left(\min_{j = 1}^K |\mathcal{M}_{x, z}^j| = i\right),$ establishes
\begin{equation}\label{eqn:aux10add}
\mathbb{E}\left(\| \hat{\bm{F}}(\cdot \mid x, z) - \bm{F}(\cdot\mid x, z)\|_{\infty} \right) \leq  w_{x, z}(0) + 1.505 \sqrt{\log(K)} \sum_{i = 1}^n  i^{-1/2} w_{x, z}(i).
\end{equation}
To establish the inequality in~\eqref{eqn:upcdf}, we use the bound
\begin{equation*}
w_{x,z}(i) \leq \sum_{j = 1}^K \mathbb{P} \left( |\mathcal{M}_{x,z}^j| = i  \right),
\end{equation*}
and note that for every~$j = 1, \hdots, K$ the random variable~$|\mathcal{M}_{x,z}^j|$ is Bernoulli distributed with parameter~$n$ and success probability~$q_{x, z, j} := \mathbb{P}(X_1 = x, Z_1 = z, D_1 = j)$. From the upper bound in~\eqref{eqn:aux10add}, together with the inequality~
\begin{equation}\label{eqn:auxin}
(1-p)^m \leq e^{-pm} \leq (2epm)^{-1/2}
\end{equation}
(for every~$p \in (0, 1]$ and every~$m \in \N$), it follows that~$\mathbb{E} ( \| \hat{\bm{F}}(\cdot \mid x, z) - \bm{F}(\cdot \mid x, z) \|_{\infty} )$ is bounded from above by (recalling Assumption~\ref{as:posprob})
\begin{equation*}
\sum_{j = 1}^K \left\{ (2eq_{x, z, j}n)^{-1/2} +  1.505 \sqrt{\log(K)}\sum_{i = 1}^n i^{-1/2} \mathbb{P} ( |\mathcal{M}_{x,z}^j| = i  )\right\}.
\end{equation*}
To obtain~\eqref{eqn:upcdf} it remains to apply Lemma~\ref{lem:bin2} (with~$m = 1$ and~$\alpha = 1/2$) to bound
\begin{equation*}
\sum_{i = 1}^n i^{-1/2} \mathbb{P} \left( |\mathcal{M}_{x,z}^j| = i  \right) \leq  \sqrt{\frac{2}{q_{x, z, j} n}},
\end{equation*}
and to collect terms (recalling that~$K \geq 2$).

Next we establish~\eqref{eqn:upcdfprob}. Note that for~$\rho \geq 1$ the claimed inequality is trivially true and fix~$\rho \in (0, 1)$. Then, arguing as above and using analogous notation, with~$\mathbb{P}_{\overline {\bm d}, \overline {\bm x},  \overline {\bm z}}$ the conditional probability corresponding to~$\mathbb{E}_{\overline {\bm d}, \overline {\bm x},  \overline {\bm z}}$, the law of iterated expectations shows that~$\mathbb{P}(\| \hat{\bm{F}}(\cdot \mid x, z) - \bm{F}(\cdot\mid x, z)\|_{\infty} \geq \rho )$ is bounded from above by
\begin{equation}\label{eqn:pbd}
\mathbb{P}\left(\beta = 0\right) +  \sum_{(\overline{\bm{d}}, \overline{\bm{x}}, \overline{\bm{z}}) \in S} \mathbb{P}_{\overline {\bm d}, \overline {\bm x},  \overline {\bm z}} \left ( \max_{i = 1}^K M_i \geq \rho \right) \mathbb{P}(\bm D = \overline{\bm{d}}, \bm{X} = \overline{\bm{x}}, \bm{Z} = \overline{\bm{z}}),
\end{equation}
which by a union bound and the DKWM-inequality, cf.~\cite{massart1990}, exploiting the conditional independence relation established around~\eqref{eqn:indrel}, can be upper bounded by
\begin{align*}
& ~~\mathbb{P}\left(\beta = 0\right) + 2  K \sum_{(\overline{\bm{d}}, \overline{\bm{x}}, \overline{\bm{z}}) \in S} \mathbb{P}(\bm D = \overline{\bm{d}}, \bm{X} = \overline{\bm{x}}, \bm{Z} = \overline{\bm{z}})  e^{-2\beta(\overline{\bm{d}}, \overline{\bm{x}}, \overline{\bm{z}}) \rho^2}  \\
=&  ~~ w_{x, z}(0) + 2 K\sum_{i = 1}^n e^{-2 i \rho^2} w_{x, z}(i) \\
\leq&  ~~ 2K\sum_{j = 1}^K \sum_{i = 0}^n e^{-2 i \rho^2} \mathbb{P}(|\mathcal{M}^j_{x, z}| = i) = 2K \sum_{j = 1}^K (1-q_{x, z, j}(1-e^{-2\rho^2}))^n,
\end{align*}
the second equality following from $|\mathcal{M}_{x,z}^j|$ being binomially distributed with parameter~$n$ and success probability~$q_{x, z, j}$ and its moment-generating function (mgf). The inequality in~\eqref{eqn:upcdfprob} follows from Lemma~\ref{lem:auxinL} below.
\end{proof}
\begin{lemma}\label{lem:auxinL}
For every~$p \geq 0$ and~$\rho \in [0, 1]$, it holds that~$$1-p(1-e^{-2\rho^2}) \leq e^{-p\rho^2/2}.$$
\end{lemma}
\begin{proof}
Combine~$1 - e^{-2\rho^2} \geq \rho^2/2$ with~$1-z \leq e^{-z}$ for every~$z \in \R$.
\end{proof}
\begin{lemma}\label{lem:manyin}
For every~$z \in \mathcal{Z}$, it holds that
\begin{equation}\label{eqn:morein}
\mathbb{E}\left( \|\hat{\bm{p}}_{X\mid Z = z} - \bm{p}_{X\mid Z = z}\|_1\right) \leq 2 \sqrt{ \frac{|\mathcal{X}| }{p_Z(z) n}};
\end{equation}
furthermore, for every~$\rho > 0$ it holds that
\begin{equation}\label{eqn:pconc}
\mathbb{P}\left( \|\hat{\bm{p}}_{X\mid Z = z} - \bm{p}_{X\mid Z = z}\|_1 \geq \rho \right)  \leq  2 |\mathcal{X}| e^{-\frac{np_Z(z)\rho^2}{2|\mathcal{X}|^2}};
\end{equation}
cf.~the discussion after~\eqref{eqn:pFdef} for the definition of~$\hat{\bm{p}}_{X\mid Z = z}$. 
\end{lemma}

\begin{proof}
Fix~$z \in \mathcal{Z}$. We first claim that for every~$x \in \mathcal{X}$ we have
\begin{equation*}
\mathbb{E}\left(| \hat{p}_{X \mid  Z = z}(x) - p_{X \mid  Z = z}(x)|\right) \leq p_{X \mid  Z = z}(x) (1-p_Z(z))^n + \frac{\sqrt{2 p_{X\mid Z = z}(x)(1-p_{X\mid Z = z}(x))}}{\sqrt{p_Z(z) n}};
\end{equation*} 
given this inequality,~\eqref{eqn:morein} follows by summing over~$x \in \mathcal{X}$ using that for non-negative real numbers~$p_i$, $i = 1, \hdots, m$, adding up to~$1$ it holds that~$\sum_{i = 1}^m \sqrt{p_i (1-p_i)} \leq \sqrt{m-1}$ (by the Cauchy-Schwarz inequality),~\eqref{eqn:auxin}, and combining the resulting terms using~$(2e)^{-1/2} + \sqrt{2} \sqrt{|\mathcal{X}| - 1} \leq 2 \sqrt{ |\mathcal{X}|}$. To prove the claim, fix~$x \in \mathcal{X}$ and note that by the law of total probability and Jensen's inequality $$\mathbb{E}(| \hat{p}_{X \mid  Z = z}(x) - p_{X \mid  Z = z}(x)|) \leq \sum_{\bm{z}^* \in \mathcal{Z}^n} \sqrt{\mathbb{E}(( \hat{p}_{X \mid  Z = z}(x) - p_{X \mid  Z = z}(x))^2 | \bm{Z} = \bm{z}^*)} \mathbb{P}(\bm{Z} = \bm{z}^*).$$ Conditional on~$\bm{Z} = \bm{z}^* \in \mathcal{Z}^n$ for a~$\bm{z}^*$ such that~$\{i: z^*_i = z\} \neq \emptyset$, the estimator~$\hat{p}_{X\mid Z = z}(x)$ is the relative frequency of successes out of~$|\{i: z^*_i = z\}|$ independent Bernoulli trials with success probability~$p_{X\mid Z = z}(x)$ (use an argument similar to the one leading up to~\eqref{eqn:indrel}), and therefore satisfies $$\mathbb{E}\left(( \hat{p}_{X \mid  Z = z}(x) - p_{X \mid  Z = z}(x))^2 \mid \bm{Z} = \bm{z}^*\right) = \frac{p_{X\mid Z = z}(x)(1-p_{X\mid Z = z}(x))}{|\{i: z^*_i = z\}|}.$$  Because~$\hat{p}_{X \mid  Z = z}(x) := 0$ if none of the coordinates of~$\bm{Z}$ equals~$z$, we have that~$\mathbb{E}(| \hat{p}_{X \mid  Z = z}(x) - p_{X \mid  Z = z}(x)|)$ is bounded from above by~$$p_{X\mid Z = z}(x) (1-p_Z(z))^n + \sqrt{p_{X\mid Z = z}(x)(1-p_{X\mid Z = z}(x))} \sum_{l = 1}^n \frac{\mathbb{P}(|\{i: Z_i = z\}| = l)}{\sqrt{l}} .$$ By Lemma~\ref{lem:bin2} (with~$m = 1$ and~$\alpha = 1/2$) the latter sum is not larger than~$\frac{\sqrt{2}}{\sqrt{p_Z(z) n}}.$ 

To prove the second statement in the lemma, we start with the following observation: fix~$x \in \mathcal{X}$ and~$\gamma \in (0, 1]$ and use the law of total probability with a similar conditioning argument as above  and Hoeffding's inequality to bound~$\mathbb{P}( | \hat{p}_{X \mid  Z = z}(x) - p_{X \mid  Z = z}(x)| \geq \gamma  )$ by 
\begin{equation*}
2\sum_{m = 0}^n e^{-2\gamma^2m} \mathbb{P}(|\{i: \bm{Z}_i = z\}| = m) = 2\left(1-p_{Z}(z)(1-e^{-2\gamma^2})\right)^n \leq 2  e^{-np_Z(z)\gamma^2/2},
\end{equation*}
the equality following from the mgf of the Binomial distribution and the inequality from Lemma~\ref{lem:auxinL}. Now, we consider two cases: if~$|\mathcal{X}| = 1$, then~\eqref{eqn:pconc} trivially holds for~$\rho \geq 1$ and it holds by the bound just developed for all~$\rho \in (0, 1)$. If~$|\mathcal{X}| > 1$,~\eqref{eqn:pconc} trivially holds for~$\rho > 2$. Fix~$\rho \in (0, 2]$. Then, the above observation (noting $\rho / |\mathcal{X}| \leq 1$) gives $$\mathbb{P}\left( \|\hat{\bm{p}}_{X\mid Z = z} - \bm{p}_{X\mid Z = z}\|_1 \geq \rho \right) \leq \sum_{x \in \mathcal{X}} \mathbb{P}( | \hat{p}_{X \mid  Z = z}(x) - p_{X \mid  Z = z}(x)| \geq \rho / |\mathcal{X}|) \leq 2 |\mathcal{X}| e^{-\frac{np_Z(z)\rho^2}{2|\mathcal{X}|^2}}.$$ 
\end{proof}
A completely analogous argument establishes the following result (the result actually follows immediately from Lemma~\ref{lem:manyin} for the special case~$\mathcal{Z} = 1$ and ``inverting'' the roles of~$X$ and~$Z$). We abstain from spelling out the details.
\begin{lemma}\label{lem:inez}
It holds that
\begin{equation}\label{lem:inez1}
\mathbb{E}\left( \|\hat{\bm{p}}_{Z} - \bm{p}_{Z}\|_1\right) \leq 2 \sqrt{\frac{|\mathcal{Z}|}{n}}.
\end{equation}
Furthermore, for every~$\rho > 0$ it holds that
\begin{equation}\label{lem:inez2}
\mathbb{P}\left( \|\hat{\bm{p}}_{Z} - \bm{p}_{Z}\|_1 \geq \rho \right)  \leq  2 |\mathcal{Z}| e^{-\frac{n\rho^2}{2 |\mathcal{Z}|^2}}.
\end{equation}
\end{lemma}

\begin{lemma}\label{lem:uzaux}
Let~$(\bm{Y}_j', X_j, Z_j)' \sim \mathcal{F}$ for~$j =1, \hdots, n$, and let~$z \in \mathcal{Z}$. Then
\begin{equation}\label{eqn:filemc6}
\mathbb{E}\left(u_z(\mathcal{F}, \hat{\mathcal{F}}_n)\right) \leq 5.5 \kappa_z \sqrt{\frac{\log(K)|\mathcal{X}|}{p_{Z}(z)n}}, \quad \text{ where} \quad \kappa_z := \max_{x} \sum_{j = 1}^K \sqrt{\frac{p_{X, Z}(x, z)}{q_{x, z, j}}},
\end{equation}
with~$q_{x,z,j} := \mathbb{P}(X_1 = x, Z_1 = z, D_1 = j)$ for~$x \in \mathcal{X}$ and $j = 1, \hdots, K$, and for every~$\rho > 0$ it holds that
\begin{equation}\label{eqn:filemc62}
\mathbb{P}\left(u_z(\mathcal{F}, \hat{\mathcal{F}}_n) \geq \rho\right) \leq 2 |\mathcal{X}|
\left(K^2 + 1\right)  \max_{x, j}  e^{-\frac{nq_{x,z,j} \rho^2}{8 |\mathcal{X}|^2 }}.
\end{equation}
If~$\mathcal{F} \sqsubset \mathscr{D}$, it follows with~$\kappa := \max_z \kappa_z$ that
\begin{equation}\label{eqn:filemEc6}
\mathbb{E}^*\left( \sup_{\bm{\delta} \in \mathbb{S}_K, \lambda \in [0, 1]} \left|\Omega_{\lambda, \mathcal{F}}(\bm{\delta}) - \Omega_{\lambda, \hat{\mathcal{F}}_n}(\bm{\delta}) \right| \right) \leq 11 \kappa \sqrt{ \frac{\log(K)|\mathcal{X}|}{n}} \sum_{z \in \mathcal{Z}} \frac{1}{\sqrt{p_{Z}(z)}} +  2 \sqrt{\frac{|\mathcal{Z}|}{n}},
\end{equation}
and that for every~$\rho > 0$ we have
\begin{equation}\label{eqn:filemEc62}
\mathbb{P}^*\left( \sup_{\bm{\delta} \in \mathbb{S}_K, \lambda \in [0, 1]} \left|\Omega_{\lambda, \mathcal{F}}(\bm{\delta}) - \Omega_{\lambda, \hat{\mathcal{F}}_n}(\bm{\delta})  \right| \geq \rho \right) \leq 4 | \mathcal{Z}||\mathcal{X}|K^2 \max_{x, z, j}  e^{- \frac{nq_{x,z,j} \rho^2}{128(|\mathcal{X}| \vee |\mathcal{Z}|)^2}}.
\end{equation}
\end{lemma}
\begin{proof}
We start with the inequality in~\eqref{eqn:filemc6}. Recall that~$p_Z(z)$ and the~$q_{x,z,j}$ appearing in the upper bound are non-zero, because of Assumption~\ref{as:posprob}. From the definition of~$u_z$ in Equation~\eqref{eqn:uzdef} we obtain
\begin{equation}
\begin{aligned}\label{eqn:inrecuz}
u_z(\mathcal{F}, \hat{\mathcal{F}}_n) &\leq \left[\sum_{x \in \mathcal{X}} p_{X\mid Z= z}(x)\|\bm{F}(\cdot \mid  x, z) - \hat{\bm{F}}(\cdot \mid  x, z)\|_{\infty} \right] + \|\bm{p}_{X\mid Z=z} - \hat{\bm{p}}_{X\mid Z=z}\|_1.
\end{aligned}
\end{equation}
Lemmas~\ref{lem:aux2} and~\ref{lem:manyin} together hence imply, for every~$z \in \mathcal{Z}$, the bound
\begin{equation}\label{eqn:upl}
\mathbb{E}\left(u_z(\mathcal{F}, \hat{\mathcal{F}}_n)\right) \leq T_1 + T_2,
\end{equation}
where we set 
\begin{equation*}
T_1 := 2.75 \times \sqrt{\frac{\log(K)}{n}} \sum_{x \in \mathcal{X}} \sum_{j = 1}^K p_{X\mid Z=z}(x)  (q_{x,z,j})^{-1/2} \quad \text{ and } \quad 
T_2 := 2 \sqrt{ \frac{|\mathcal{X}|}{p_Z(z) n}}.
\end{equation*}
Recalling~$q_{x,z,j} = \mathbb{P}(X_1 = x, Z_1 = z, D_1 = j)$ and using the Cauchy-Schwarz inequality we get
\begin{equation*}
\sum_{x \in \mathcal{X}} \sum_{j = 1}^K p_{X\mid Z=z}(x) (q_{x,z,j})^{-1/2} \leq p^{-1}_{Z}(z) \kappa_z \sum_{x \in \mathcal{X}} \sqrt{p_{X,Z}(x, z)} \leq  \kappa_z \sqrt{|\mathcal{X}|/p_{Z}(z)},
\end{equation*}
so that
\begin{equation*}
T_1 \leq 2.75 \kappa_z  \sqrt{ \frac{\log(K)|\mathcal{X}|}{p_{Z}(z)n}}.
\end{equation*}
From~$\kappa_z \geq 1$ it follows that $$T_1 + T_2 \leq 5.5 \kappa_z \sqrt{ \frac{\log(K)|\mathcal{X}|}{p_{Z}(z)n}},$$ which establishes~\eqref{eqn:filemc6}. 

To verify~\eqref{eqn:filemc62}, we first use~\eqref{eqn:inrecuz} to obtain
\begin{equation*}
\mathbb{P}\left(u_z(\mathcal{F}, \hat{\mathcal{F}}_n) \geq \rho\right) \leq \sum_{x \in \mathcal{X}} \mathbb{P}\left( \|\bm{F}(\cdot \mid  x, z) - \hat{\bm{F}}(\cdot \mid  x, z)\|_{\infty}  \geq \rho/2 \right) + \mathbb{P}\left( \|\bm{p}_{X\mid Z=z} - \hat{\bm{p}}_{X\mid Z=z}\| \geq \rho/2 \right),
\end{equation*}
which we then estimate using the bounds in Equations~\eqref{eqn:upcdfprob} (Lemma~\ref{lem:aux2}) and~\eqref{eqn:pconc} (Lemma~\ref{lem:manyin}) to arrive at  (note that $q_{x, z, j} \leq p_Z(z)$) 
\begin{equation*}
2K \sum_{x \in \mathcal{X}} \sum_{j = 1}^K e^{-\frac{n \rho^2 q_{x, z, j}}{8}} + 2 |\mathcal{X}| e^{-\frac{np_Z(z)\rho^2}{8|\mathcal{X}|^2}} \leq 2 |\mathcal{X}|
\left(K^2 + 1\right)  \max_{x, j}  e^{-\frac{nq_{x,z,j} \rho^2}{8 |\mathcal{X}|^2 }}.
\end{equation*}
To verify~\eqref{eqn:filemEc6}, we recall from Lemma~\ref{lem:exmax} that  $$\sup_{\bm{\delta} \in \mathbb{S}_K, \lambda \in [0, 1]}\left|\Omega_{\lambda, \mathcal{F}}(\bm{\delta}) - \Omega_{\lambda, \hat{\mathcal{F}}_n}(\bm{\delta})\right| \leq  2 \max_{z \in \mathcal{Z}} u_z(\mathcal{F}, \hat{\mathcal{F}}_n) + \|\bm{p}_Z - \hat{\bm{p}}_Z\|_1.$$ Therefore, the already established inequality in~\eqref{eqn:filemc6} together with~\eqref{lem:inez1} in Lemma~\ref{lem:inez} establishes~\eqref{eqn:filemEc6}. 

To prove~\eqref{eqn:filemEc62} note that the previous display together with the already established relation~\eqref{eqn:filemc62} and~\eqref{lem:inez2} in Lemma~\ref{lem:inez} establishes the bound $$\mathbb{P}^*\left( \sup_{\bm{\delta} \in \mathbb{S}_K, \lambda \in [0, 1]} \left|\Omega_{\lambda, \mathcal{F}}(\bm{\delta}) - \Omega_{\lambda, \hat{\mathcal{F}}_n}(\bm{\delta})  \right| \geq \rho \right) \leq 2 |\mathcal{X}| \left(K^2 + 1\right) \sum_{z \in \mathcal{Z}} \max_{x, j}  e^{- \frac{nq_{x,z,j} \rho^2}{128|\mathcal{X}|^2}} + 2 |\mathcal{Z}| e^{-\frac{n\rho^2}{8 |\mathcal{Z}|^2}},$$ which can further be upper bounded by $$2 |\mathcal{X}| \left(K^2 + 1\right) |\mathcal{Z}| \max_{x, z, j}  e^{- \frac{nq_{x,z,j} \rho^2}{128|\mathcal{X}|^2}} + 2 |\mathcal{Z}| e^{-\frac{n\rho^2}{8 |\mathcal{Z}|^2}} \leq 
4 | \mathcal{Z}||\mathcal{X}|K^2 \max_{x, z, j}  e^{- \frac{nq_{x,z,j} \rho^2}{128(|\mathcal{X}| \vee |\mathcal{Z}|)^2}}.$$ 
\end{proof}
\begin{proof}[Proof of Theorem~\ref{thm:upreg2}:]
Lemma~\ref{lem:upreg} shows for every~$\lambda \in [0, 1]$ that $$r(\bm{\pi}^{\varepsilon, \lambda}_n; \lambda, \mathcal{F})
\leq  2 \sup_{\bm{\delta} \in \mathbb{S}_K} |\Omega_{\lambda, \mathcal{F}}(\bm{\delta}) - \Omega_{\lambda, \hat{\mathcal{F}}_n}(\bm{\delta})| + \varepsilon.$$ We therefore obtain
\begin{equation*}
\mathbb{E}^*(r(\bm{\pi}^{\varepsilon,  \hat{\lambda}_n}_n;  \hat{\lambda}_n, \mathcal{F}) ) \leq \mathbb{E}^*\left(\sup_{\lambda \in [0, 1] } r\left(\bm{\pi}^{\varepsilon, \lambda}_n; \lambda, \mathcal{F}\right) \right) \leq 2 \mathbb{E}^*\left(  \sup_{\bm{\delta} \in \mathbb{S}_K, \lambda \in [0, 1]} |\Omega_{\lambda, \mathcal{F}}(\bm{\delta}) - \Omega_{\lambda, \hat{\mathcal{F}}_n}(\bm{\delta})| \right)   + \varepsilon,
\end{equation*}
which by~\eqref{eqn:filemEc6} is bounded from above by
\begin{equation}
22  \sqrt{ \frac{\log(K)|\mathcal{X}|}{n}} \kappa \sum_{z \in \mathcal{Z}} \frac{1}{\sqrt{p_{Z}(z)}} +  4 \sqrt{\frac{|\mathcal{Z}|}{n}} + \varepsilon.
\end{equation}
Inserting the definition of~$\kappa$ delivers the bound in~\eqref{eqn:regupmain}.

The inequality in~\eqref{eqn:hprobreg}, follows immediately from the first display in this proof together with~\eqref{eqn:filemEc62}.
\end{proof}

\begin{proof}[Proof of Theorem~\ref{thm:consi}:]
Fix~$\rho > 0$. Recall from~\eqref{eqn:cFdef} that for every~$\lambda \in [0, 1]$ we defined
\begin{equation}
c_{\mathcal{F}}(\rho, \lambda)= 
\begin{cases}
\max_{\bm{\delta} \in \mathbb{S}_K} \Omega_{\lambda, \mathcal{F}}(\bm{\delta}) - \max_{\bm{\delta} \in \mathbb{M}_{\mathcal{F},  \lambda}^{\rho}} \Omega_{\lambda, \mathcal{F}}(\bm{\delta})  & \text{ if } \mathbb{M}_{\mathcal{F},  \lambda}^{\rho} \neq \emptyset, \\
0 & \text{ else.}
\end{cases}
\end{equation}
In the following steps, we show that~$\lambda \mapsto 	c_{\mathcal{F}}(\rho, \lambda)$ is Borel-measurable on~$[0, 1]$; if not mentioned otherwise \emph{all references to theorems, results, and pages in the following items 1-5 are to \cite{hhg}}, the terminology of which we shall also use without repeating their definitions:
\begin{enumerate}
\item By our Lemma~\ref{lem:exmax}, the function $(\lambda, \bm{\delta}) \mapsto \Omega_{\lambda, \mathcal{F}}(\bm{\delta})$ is continuous on~$[0, 1] \times \mathbb{S}_K$. Hence, it is a Carath\'eodory Function (cf.~Definition 4.50); the unit interval is throughout equipped with its Borel sigma algebra. As a consequence of Part 3 of the ``Measurable Maximum Theorem'' 18.19, the nonempty-valued correspondence $\lambda \mapsto \argmax_{\bm{\delta} \in \mathbb{S}_K} \Omega_{\lambda, \mathcal{F}}(\bm{\delta})$ is measurable on~$[0, 1]$ (and hence weakly measurable by Lemma 18.2).
\item Due to Theorem 18.5 the distance function $$(\lambda, \bm{\gamma}) \mapsto d_1\left(\bm{\gamma},  \argmax_{\bm{\delta} \in \mathbb{S}_K} \Omega_{\lambda, \mathcal{F}}(\bm{\delta})\right)$$ associated with the nonempty-valued weakly measurable $\lambda \mapsto \argmax_{\bm{\delta} \in \mathbb{S}_K} \Omega_{\lambda, \mathcal{F}}(\bm{\delta})$ is a Carath\'eodory function.
\item Decomposing~$\mathbb{M}_{\mathcal{F}, \lambda}^{\rho}$ as 
\begin{equation*}
\big\{\bm{\gamma} \in \mathbb{S}_K : d_1\big( 
\bm{\gamma},
\argmax_{\bm{\delta} \in \mathbb{S}_K} \Omega_{\lambda, \mathcal{F}}(\bm{\delta})
\big) > \rho
\big \} \cup \big\{\bm{\gamma} \in \mathbb{S}_K : d_1\big( 
\bm{\gamma},
\argmax_{\bm{\delta} \in \mathbb{S}_K} \Omega_{\lambda, \mathcal{F}}(\bm{\delta})
\big) - \rho = 0
\big \}
\end{equation*}
and using Lemmas 18.7 and 18.8 together with the union part of 18.4 shows that $\lambda \mapsto \mathbb{M}_{\mathcal{F},  \lambda}^{\rho}$  is a measurable correspondence. 
\item The correspondence $\lambda \mapsto \check{\mathbb{M}}_{\mathcal{F}, \lambda}^{\rho}$ defined via $$\lambda \mapsto \begin{cases}
\mathbb{M}_{\mathcal{F},  \lambda}^{\rho} & \text{ if } 	\mathbb{M}_{\mathcal{F},  \lambda}^{\rho} \neq \emptyset \\
\mathbb{S}_K & \text{ else}.
\end{cases} $$
is weakly measurable, which is evident from~Definition 18.1, and has non-empty and compact values (for the latter compare the proof of Lemma~\ref{lem:mtune}).
\item We can write $$c_{\mathcal{F}}(\rho, \lambda) = \max_{\bm{\delta} \in \mathbb{S}_K} \Omega_{\lambda, \mathcal{F}}(\bm{\delta}) - \max_{\bm{\delta} \in \check{\mathbb{M}}_{\mathcal{F}, \lambda}^{\rho}} \Omega_{\lambda, \mathcal{F}}(\bm{\delta}).$$ Part 1 in the ``Measurable Maximum Theorem" 18.19 proves that~$\lambda \mapsto \max_{\bm{\delta} \in \mathbb{S}_K} \Omega_{\lambda, \mathcal{F}}(\bm{\delta})$ and~$\lambda \mapsto \max_{\bm{\delta} \in \check{\mathbb{M}}_{\mathcal{F}, \lambda}^{\rho}} \Omega_{\lambda, \mathcal{F}}(\bm{\delta})$ are measurable from which measurability of~$\lambda \mapsto c_{\mathcal{F}}(\rho, \lambda)$ follows.
\end{enumerate}

To establish the inequality in~\eqref{eqn:firstconsup}, we use Part 1 of Lemma~\ref{lem:mtune} with~$(\Theta, d) = (\mathbb{S}_K, d_1)$, $\Gamma = \Lambda$, $Q_n = \Omega_{\cdot, \hat{\mathcal{F}}_n}(\cdot)$, $Q = \Omega_{\cdot, \mathcal{F}}(\cdot)$,~$\hat{\theta}_n = \bm{\pi}_n^{\varepsilon_n, \hat{\lambda}_n}$, and~$\hat{\gamma}_n = \hat{\lambda}_n$. All assumptions in Part 1 of Lemma~\ref{lem:mtune} are satisfied, the continuity condition regarding~$Q$ was verified in Lemma~\ref{lem:exmax}. Fix~$\zeta > 0$, recall the definition of~$\mathbb{M}_{\mathcal{F},  \hat{\lambda}_n}^{\rho}$ from~\eqref{eqn:Mrhodef}, and note that the second condition in
\begin{equation}\label{eqn:implstart2}
\varepsilon_n + 6 \zeta \leq c_{\mathcal{F}}(\rho, \hat{\lambda}_{n}) ~~ \text{ and } ~~ \bm{\pi}_n^{\varepsilon_n, \hat{\lambda}_n} \in \mathbb{M}_{\mathcal{F},  \hat{\lambda}_n}^{\rho}
\end{equation}
implies via the first part of Lemma~\ref{lem:mtune} and the definition of~$\bm{\pi}_n^{\varepsilon_n, \hat{\lambda}_n}$ that
\begin{equation*}
2 \sup_{\bm{\delta} \in \mathbb{S}_K} |\Omega_{\hat{\lambda}_n, \hat{\mathcal{F}}_n}(\bm{\delta}) - \Omega_{\hat{\lambda}_n, \mathcal{F}}(\bm{\delta})| + \varepsilon_n \geq c_{\mathcal{F}}(\rho, \hat{\lambda}_n) \quad \text{ and } \quad c_{\mathcal{F}}(\rho, \hat{\lambda}_n) > 0,
\end{equation*}
so that an application of Lemma~\ref{lem:exmax} then implies
\begin{equation*}
4 \max_{z \in \mathcal{Z}}u_z(\mathcal{F}, \hat{\mathcal{F}}_n) + 2 \|\bm{p}_Z - \hat{\bm{p}}_Z\|_1 +  \varepsilon_n \geq c_{\mathcal{F}}(\rho, \hat{\lambda}_n) \quad \text{ and } \quad c_{\mathcal{F}}(\rho, \hat{\lambda}_n) > 0.
\end{equation*}
Hence, the two conditions in~\eqref{eqn:implstart2} together imply~$$\frac{2}{3} \max_{z \in \mathcal{Z}}u_z(\mathcal{F}, \hat{\mathcal{F}}_n) + \frac{1}{3}\|\bm{p}_Z - \hat{\bm{p}}_Z\|_1 \geq \zeta \quad  \text{ and } \quad c_{\mathcal{F}}(\rho, \hat{\lambda}_n) > 0.$$ We can thus upper bound
\begin{equation}
\begin{aligned}\label{eqn:upconssplit}
\mathbb{P}^* \left( \bm{\pi}_{n}^{\varepsilon_n, \hat{\lambda}_{n}}  \in \mathbb{M}_{\mathcal{F},  \hat{\lambda}_{n}}^{\rho}\right) \leq \mathbb{P} &\left(
\max_{z \in \mathcal{Z}} u_z(\hat{\mathcal{F}}_{n}, \mathcal{F})  \geq \zeta
\right) + \mathbb{P} \left(\|\hat{\bm{p}}_Z - \bm{p}_Z \|_1 \geq \zeta
\right) \\
& + \mathbb{P}^*\left( 0 <   c_{\mathcal{F}}(\rho, \hat{\lambda}_n) <\varepsilon_n + 6 \zeta \right),
\end{aligned}
\end{equation}
noting that the first two events in the upper bound are measurable (due to the specific form of~$u_z(\hat{\mathcal{F}}_{n}, \mathcal{F})$), and that the event in the third probability is measurable whenever~$\hat{\lambda}_n$ is so (as a composition of two measurable functions) from what has been shown above. Lemma~\ref{lem:uzaux} and a union bound shows that the first probability in the upper bound in~\eqref{eqn:upconssplit} is not smaller than
\begin{equation*}
2 |\mathcal{X}|
\left(K^2 + 1\right)  \sum_{z \in \mathcal{Z}} \max_{x, j}  e^{-\frac{nq_{x,z,j} \zeta^2}{8 |\mathcal{X}|^2 }},
\end{equation*}
Lemma~\ref{lem:inez} shows that 
$$\mathbb{P}\left( \|\hat{\bm{p}}_{Z} - \bm{p}_{Z}\|_1 \geq \zeta \right)  \leq  2 |\mathcal{Z}| e^{-\frac{n\zeta^2}{2 |\mathcal{Z}|^2}},$$ which establishes~\eqref{eqn:firstconsup}. The consistency statement after~\eqref{eqn:firstconsup} follows from~\eqref{eqn:firstconsup}. It remains to prove the last statement in the theorem. To this end, note that if~$\Lambda = \{\lambda^*_1, \hdots, \lambda^*_M\}$, for a fixed~$M \in \N$, we can trivially bound 
\begin{equation*}
\mathbb{P}^*\left( 0 <   c_{\mathcal{F}}(\rho, \hat{\lambda}_n) <\varepsilon_n + 6 \zeta_n \right) \leq \sum_{i = 1}^M \mathds{1}\left(  0 <   c_{\mathcal{F}}(\rho, \lambda_i^*) <\varepsilon_n + 6 \zeta_n \right),
\end{equation*}
which converges to~$0$ for any~$\zeta_n \to 0$ whenever~$\varepsilon_n \to 0$.
\end{proof}

\begin{proof}[Proof of Proposition~\ref{prop:conslamb}:]
We start with~\eqref{eqn:conslamb}. Suppose that~$$d_1\left( \bm{\pi}_n, \arg\max_{\bm{\delta} \in \mathbb{S}_K} \Omega_{  \hat{\lambda}_n, \mathcal{F}}(\bm{\delta})\right) \cps 0.$$
We first note that if $\arg\max_{\bm{\delta} \in \mathbb{S}_K} \Omega_{ \lambda^*, \mathcal{F}}(\bm{\delta}) = \mathbb{S}_K$, there is nothing to show, and we are done. Hence, we assume from now on that $\arg\max_{\bm{\delta} \in \mathbb{S}_K} \Omega_{ \lambda^*, \mathcal{F}}(\bm{\delta}) \neq \mathbb{S}_K$. Recall from~\eqref{eqn:Mrhodef} that

\begin{equation*}
\mathbb{M}_{\mathcal{F}, \lambda}^{\rho} := \big\{\bm{\gamma} \in \mathbb{S}_K : d_1\big( 
\bm{\gamma},
\argmax_{\bm{\delta} \in \mathbb{S}_K} \Omega_{\lambda, \mathcal{F}}(\bm{\delta})
\big) \geq \rho
\big \},
\end{equation*}
and fix~$\rho > 0$ small enough such that~$\mathbb{M}_{\mathcal{F}, \lambda^*}^{\rho} \neq \emptyset$ (such a~$\rho > 0$ exists, because the closed set~$\arg\max_{\bm{\delta} \in \mathbb{S}_K} \Omega_{ \lambda^*, \mathcal{F}}(\bm{\delta})$, cf.~Proposition~\ref{prop:exmax}, is a strict subset of~$\mathbb{S}_K$). Because the map~$$\bm{\gamma} \mapsto d_1\big( 
\bm{\gamma},
\argmax_{\bm{\delta} \in \mathbb{S}_K} \Omega_{\lambda, \mathcal{F}}(\bm{\delta})
\big)$$ is continuous (as the distance to a non-empty closed set, cf.~also Part 1 of Proposition~\ref{prop:exmax}), the set~$\mathbb{M}_{\mathcal{F}, \lambda^*}^{\rho}$ is closed, so that its complement in~$\mathbb{S}_K$ is an open neighborhood of~$\argmax_{\bm{\delta} \in \mathbb{S}_K} \Omega_{\lambda, \mathcal{F}}(\bm{\delta})$. Proposition~\ref{prop:exmax} established upper hemicontinuity, cf.~Definition 17.2 in~\cite{hhg}, of the correspondence~$$\lambda \mapsto 	\arg\max_{\bm{\delta} \in \mathbb{S}_K} \Omega_{  \lambda, \mathcal{F}}(\bm{\delta}),$$ so that there exists an~$\varepsilon = \varepsilon(\rho) > 0$ such that~$\Lambda_{\varepsilon} := [\lambda^* - \varepsilon, \lambda^* + \varepsilon] \cap [0, 1] \subseteq [0, 1]$ and such that
\begin{equation}\label{eqn:incluuhem}
\bigcup_{\lambda \in \Lambda_{\varepsilon}} \arg\max_{\bm{\delta} \in \mathbb{S}_K} \Omega_{  \lambda, \mathcal{F}}(\bm{\delta}) \subseteq \mathbb{S}_K \backslash \mathbb{M}_{\mathcal{F}, \lambda^*}^{\rho}.
\end{equation}
The (outer) probability of the event~$\hat{\lambda}_n \notin \Lambda_{\varepsilon}$ goes to~$0$ as~$n \to \infty$. Hence, we shall assume that~$\hat{\lambda}_n \in \Lambda_{\varepsilon}$ in the following. Given a non-empty and compact subset $A$ of a metric space $(V, d_V)$, we can associate to every~$v \in V$ an element $\Pi_A v \in A$, say, that satisfies $d_V(v, A) = \inf_{w \in A} d_V(v, w) = d_V(v, \Pi_A v)$. Denoting
$$\bm{\pi}_n^* := \Pi_{\arg\max_{\bm{\delta} \in \mathbb{S}_K} \Omega_{ \lambda^*, \mathcal{F}}(\bm{\delta})} \bm{\pi}_n, \quad \hat{\bm{\pi}}_n := \Pi_{\arg\max_{\bm{\delta} \in \mathbb{S}_K} \Omega_{ \hat{\lambda}_n, \mathcal{F}}(\bm{\delta})} \bm{\pi}_n,$$ as well as $$\hat{\bm{\pi}}_n^* := \Pi_{\arg\max_{\bm{\delta} \in \mathbb{S}_K} \Omega_{ \lambda^*, \mathcal{F}}(\bm{\delta})} \left[ \Pi_{\arg\max_{\bm{\delta} \in \mathbb{S}_K} \Omega_{ \hat{\lambda}_n, \mathcal{F}}(\bm{\delta})} \bm{\pi}_n\right] = \Pi_{\arg\max_{\bm{\delta} \in \mathbb{S}_K} \Omega_{ \lambda^*, \mathcal{F}}(\bm{\delta})} \hat{\bm{\pi}}_n,$$ we hence see that
\begin{align*}
d_1\left( \bm{\pi}_n, \arg\max_{\bm{\delta} \in \mathbb{S}_K} \Omega_{ \lambda^*, \mathcal{F}}(\bm{\delta}) \right) = d_1\left( \bm{\pi}_n, \bm{\pi}_n^* \right) \leq d_1\left( \bm{\pi}_n,  \hat{\bm{\pi}}_n^* \right) \leq d_1\left( \bm{\pi}_n, \hat{\bm{\pi}}_n  \right) + d_1\left( \hat{\bm{\pi}}_n , \hat{\bm{\pi}}_n^*  \right).
\end{align*}
The expression~$d_1( \bm{\pi}_n, \hat{\bm{\pi}}_n  )$ equals~$d_1( \bm{\pi}_n, \arg\max_{\bm{\delta} \in \mathbb{S}_K} \Omega_{  \hat{\lambda}_n, \mathcal{F}}(\bm{\delta}))$ by definition, and thus converges to~$0$ in (outer) probability by assumption. The expression~$d_1\left( \hat{\bm{\pi}}_n , \hat{\bm{\pi}}_n^*  \right)$ is bounded from above by~$\rho$ due to~$\hat{\lambda}_n \in \Lambda_{\varepsilon}$ and Equation~\eqref{eqn:incluuhem}. Because~$\rho > 0$ was arbitrary, this proves the first statement in the theorem.

To prove the second statement, note that the finiteness of~$\Lambda$ and the assumed convergence in outer probability implies~$\mathbb{P}^*(\hat{\lambda}_n \neq \hat{\lambda}_n^*) \to 0$ as~$n \to \infty$, which obviously proves the result.
\end{proof}

\begin{proof}[Proof of Theorem~\ref{thm:consitune}:]
We first note that by Assumption~\ref{as:MAIN} and Equations~\eqref{eqn:up1} and~\eqref{eqn:utouz}, we have
\begin{align*}
\max_{\lambda \in \Lambda} |\Delta_n(\lambda, \mathcal{F}) - \Delta_n(\lambda, \hat{\mathcal{F}}_n)| &\leq |\langle \bm{\pi}_{n}^{\varepsilon_n, 0}, \mathcal{F} \rangle - \langle \bm{\pi}_{n}^{\varepsilon_n, 0}, \hat{\mathcal{F}}_n \rangle| + \max_{\lambda \in \Lambda} |\langle \bm{\pi}_{n}^{\varepsilon_n, \lambda}, \mathcal{F} \rangle  - \langle \bm{\pi}_{n}^{\varepsilon_n, \lambda}, \hat{\mathcal{F}}_n\rangle | \\
&\leq 2u(\mathcal{F}, \hat{\mathcal{F}}_n) \leq 2 \max_{z \in \mathcal{Z}} u_z(\mathcal{F}, \tilde{\mathcal{F}}) + 2\|\bm{p}_Z - \tilde{\bm{p}}_Z\|_1.
\end{align*}
It hence follows from Lemmas~\ref{lem:uzaux} and~\ref{lem:inez} and a union bound that for every~$\rho > 0$ it holds that
\begin{equation*}
\mathbb{P}^*\left(\max_{\lambda \in \Lambda}|\Delta_n(\lambda, \mathcal{F}) - \Delta_n(\lambda, \hat{\mathcal{F}}_n)| \geq \rho \right) \leq  2 |\mathcal{X}|
\left(K^2 + 1\right) \sum_{z \in \mathcal{Z}}  \max_{x, j}  e^{-\frac{nq_{x,z,j} \rho^2}{128 |\mathcal{X}|^2 }} + 2 |\mathcal{Z}| e^{-\frac{n\rho^2}{32 |\mathcal{Z}|^2}}.
\end{equation*}
In particular, recalling that~$c_n = \sqrt{\log(n)/n}$, we obtain (recalling Assumption~\ref{as:posprob}) for every~$\alpha > 0$ that
\begin{equation}\label{eqn:Dcn0}
\mathbb{P}^*\left(\max_{\lambda \in \Lambda}|\Delta_n(\lambda, \mathcal{F}) - \Delta_n(\lambda, \hat{\mathcal{F}}_n)| \geq \alpha c_n \right) \to 0.
\end{equation}
To prove~\eqref{eqn:conscons}, just note that by definition it always holds that $$\Delta_n(\hat{\lambda}_n^*(\beta, \hat{\mathcal{F}}_n), \hat{\mathcal{F}}_n) \leq \beta(1 - c_n),$$ and that~$\hat{\lambda}_n^*(\beta, \hat{\mathcal{F}}_n) > \lambda_n^*(\beta, \mathcal{F})$ implies~$\Delta_n(\hat{\lambda}_n^*(\beta, \hat{\mathcal{F}}_n), \mathcal{F}) > \beta$. It hence follows that both sets of which outer probabilities are taken in~\eqref{eqn:conscons} are contained in the set defined by the relation~$\max_{\lambda \in \Lambda}|\Delta_n(\lambda, \mathcal{F}) - \Delta_n(\lambda, \hat{\mathcal{F}}_n)| \geq \beta c_n$. The outer probability of the latter event converges to~$0$ by~\eqref{eqn:Dcn0}. This verifies~\eqref{eqn:conscons}.

Next, we need to verify that under the additional condition that for some~$\alpha > 0$ it holds that $$\mathbb{P}^*\left( \beta - \Delta_n\left(\lambda_n^*(\beta, \mathcal{F}), \mathcal{F}\right) \leq [\alpha + \beta]c_n \right) \to 0,$$ we have
\begin{equation}\label{eqn:onlyneed}
\hat{\lambda}_n^*(\beta, \hat{\mathcal{F}}_n) - \lambda_n^*(\beta, \mathcal{F}) \cps 0;
\end{equation}
the convergence in~\eqref{eqn:consitune} then directly follows from the second part of Proposition~\ref{prop:conslamb}, noting that the consistency requirement to the left in~\eqref{eqn:conslamb} is satisfied as a consequence of Theorem~\ref{thm:consi}. To show~\eqref{eqn:onlyneed}, because~$\Lambda$ is finite, we need to verify that $$\mathbb{P}^*\left( \hat{\lambda}_n^*(\beta, \hat{\mathcal{F}}_n) \neq \lambda_n^*(\beta, \mathcal{F}) \right) \to 0.$$ Due to sub-additivity of outer probabilities and because we already know that~$\mathbb{P}^*(\hat{\lambda}_n^*(\beta, \hat{\mathcal{F}}_n) > \lambda_n^*(\beta, \mathcal{F})) \to 0$, it suffices to verify that $$\mathbb{P}^*\left( \hat{\lambda}_n^*(\beta, \hat{\mathcal{F}}_n) < \lambda_n^*(\beta, \mathcal{F}) \right) \to 0.$$ To this end, due to~\eqref{eqn:Dcn0}, it suffices to show that for an~$\alpha > 0$, which we also assume to satisfy~\eqref{eqn:addcond}, we have $$\mathbb{P}^*\left( \hat{\lambda}_n^*(\beta, \hat{\mathcal{F}}_n) < \lambda_n^*(\beta, \mathcal{F}) \text{ and } \max_{\lambda \in \Lambda}|\Delta_n(\lambda, \mathcal{F}) - \Delta_n(\lambda, \hat{\mathcal{F}}_n)| < \alpha c_n \right) \to 0.$$ Note that $\hat{\lambda}_n^*(\beta, \hat{\mathcal{F}}_n) < \lambda_n^*(\beta, \mathcal{F})$ implies $\Delta_n(\lambda_n^*(\beta, \mathcal{F}), \hat{\mathcal{F}}_n) > \beta (1-c_n)$, which together with~$$\max_{\lambda \in \Lambda}|\Delta_n(\lambda, \mathcal{F}) - \Delta_n(\lambda, \hat{\mathcal{F}}_n)| < \alpha c_n$$ gives $$\beta - \Delta_n(\lambda_n^*(\beta, \mathcal{F}), \mathcal{F}) \leq [\alpha + \beta]c_n,$$ the outer probability of which converges to~$0$ by assumption.
\end{proof}

\begin{lemma}\label{lem:bin2}
If the random variable~$B$ follows a Binomial distribution with parameters~$n \in \N$ and~$p \in (0, 1]$, then, for every~$\alpha \in (0, 1]$ and every~$m \in \N \cup \{0\}$, 
\begin{equation}\label{in:bin}
\mathbb{E}\left(B^{-\alpha m} \mathds{1}\left\{B > 0\right\}\right) \leq \left[\frac{2^{m(m+1)/2}[1-(1-p)^{n+m}]}{p^m \prod_{i = 1}^m (n+i) }\right]^\alpha \leq \frac{2^{\alpha m(m+1)/2}}{(pn)^{\alpha m}};
\end{equation}
where for~$m = 0$ we set~$\prod_{i = 1}^m (n+i) = 1$.
\end{lemma}

\begin{proof}
The second inequality in~\eqref{in:bin} is trivial. By Jensen's inequality we only need to establish the first inequality in~\eqref{in:bin} for the special case where~$\alpha = 1$. Furthermore, for~$m = 0$ that inequality is easily seen to be an equality. It therefore remains to verify the first inequality in~\eqref{in:bin} for~$\alpha = 1$ and for every~$m \in \N$,~$n \in \N$, and~$p \in (0, 1]$. Fix such a triple~$m$,~$n$, and~$p$. For every~$k \in \N$, we shall write~$B_k$ for a binomially distributed random variable with parameters~$k$ and~$p$, and note that
\begin{equation*}
\begin{aligned}
\mathbb{E}\left(B_n^{-m} \mathds{1}\left\{B_n > 0\right\}\right) &\leq 2^m \mathbb{E}\left((B_n+1)^{-m} \mathds{1}\left\{B_n > 0\right\}\right) \\ &= 2^m \sum_{i = 1}^n \frac{1}{(i+1)^m} {n \choose i} p^i (1-p)^{n-i}, 
\end{aligned}
\end{equation*}
which we can equivalently write (an analogous observation was used in~\cite{rempala}, cf.~the last display on their p.262) as 
\begin{equation*}
\begin{aligned}
&\frac{2^m}{p(n+1)} \sum_{i = 1}^n \frac{1}{(i+1)^{m-1}} {n + 1  \choose i + 1} p^{i+1} (1-p)^{n-i} \\  = ~~ &\frac{2^m}{p(n+1)} \sum_{i = 2}^{n+1} \frac{1}{i^{m-1}} {n + 1  \choose i} p^{i} (1-p)^{n + 1 - i} \\
\leq ~~ &\frac{2^m}{p(n+1)} \mathbb{E}\left(B_{n+1}^{-(m-1)} \mathds{1}\left\{B_{n+1} > 0\right\}\right).
\end{aligned}
\end{equation*}
Summarizing, we have shown that
\begin{equation*}
\mathbb{E}\left(B_n^{-m} \mathds{1}\left\{B_n > 0\right\}\right) \leq \frac{2^m}{p(n+1)} \mathbb{E}\left(B_{n+1}^{-(m-1)} \mathds{1}\left\{B_{n+1} > 0\right\}\right),
\end{equation*}
which we can iterate to obtain the first inequality in~\eqref{in:bin} from 
\begin{equation*}
\mathbb{E}\left( B_{n+m}^0 \mathds{1}\left\{B_{n+m} > 0\right\} \right) = 1 - (1-p)^{n+m}.
\end{equation*}
\end{proof}

\section{Proofs for Section~\ref{sec:contX}}\label{app:contX}

Throughout this section, all assumptions listed in Section~\ref{sec:contX} are maintained.

\begin{lemma}\label{lem:expf}
$
\mathbb{E}(f_{\bm{\delta}, c, z}(Y_{D_j, j}, X_j, Z_j, D_j)) = \langle \bm{\delta}, \mathcal{F} \rangle_z(c)
$ for every~$\bm{\delta} \in \Pi$,~$c\in \R$, and~$z \in \mathcal{Z}$.

\end{lemma} 

\begin{proof}[Proof of Lemma~\ref{lem:expf}:]
Fix~$\bm{\delta} \in \Pi$,~$c\in \R$, and~$z \in \mathcal{Z}$. We abbreviate~$\tilde{Y}_j = Y_{D_j, j}$ throughout. Note that $$0 \leq f_{\bm{\delta}, c, z}(\tilde{Y}_j, X_j, Z_j, D_j) \leq \max_{i = 1, \hdots, K} \frac{1}{p_Z(z) e_i(X_j, Z_j)}.$$ Hence, as a consequence of the maintained assumptions that~$p_Z(z) > 0$ and~$\mathbb{E}(e_i^{-1}(X_j, Z_j)) < \infty$ for every~$i = 1, \hdots, K$,~$f_{\bm{\delta}, c, z}(\tilde{Y}_j, X_j, Z_j, D_j)$ is integrable. Thus, by the law of iterated expectations, we can write
\begin{equation*}
\mathbb{E}(f_{\bm{\delta}, c, z}(\tilde{Y}_j, X_j, Z_j, D_j)) =  \sum_{i =1}^K \mathbb{E}\left[\mathbb{E}\left(\mathds{1}(D_j = i)f_{\bm{\delta}, c, z}(\tilde{Y}_j, X_j, Z_j, D_j) \mid X_j, Z_j\right)\right].
\end{equation*}
The inner expectation coincides (almost everywhere) with
\begin{equation*}
\frac{\delta_i(X_j)}{p_Z(z)e_i(X_j, z)} \times \mathds{1}(Z_j = z) \times \mathbb{E}(\mathds{1}(Y_{i,j} \leq c, D_j = i) \mid X_j, Z_j),
\end{equation*}
which, making use of the conditional independence assumption, reduces to
\begin{equation*}
\sum_{i = 1}^K \frac{\delta_i(X_j)}{p_Z(z)} \times \mathds{1}(Z_j = z)  \times F^i(c \mid X_j, Z_j),
\end{equation*}
so that (upon taking the outer expectation) we arrive (cf.~\eqref{eqn:langrang}) at
\begin{equation*}
\mathbb{E}(f_{\bm{\delta}, c, z}(Y_j, X_j, Z_j, D_j)) = \langle \bm{\delta}, \mathcal{F} \rangle_z(c)
\end{equation*}

\end{proof}

\begin{proof}[Proof of Theorem~\ref{thm:upregcont}:]
We first establish the bound for~$r(\bm{\pi}^{\varepsilon,   \hat{\lambda}_n}_n;   \hat{\lambda}_n, \mathcal{F})$: Using Lemma~\ref{lem:bdf} below, we can write 
$$r(\bm{\pi}^{\varepsilon,   \hat{\lambda}_n}_n; \hat{\lambda}_n, \mathcal{F}) = \sup_{\bm{\delta} \in \Pi} \Omega_{\hat{\lambda}_n, \mathcal{F}}(\bm{\delta}) - \sup_{\bm{\delta} \in \Pi} \hat{\Omega}_{\hat{\lambda}_n}\left(\bm{\delta} \right) + \sup_{\bm{\delta} \in \Pi} \hat{\Omega}_{\hat{\lambda}_n}\left(\bm{\delta}\right) - \Omega_{\hat{\lambda}_n, \mathcal{F}} \left(\bm{\pi}^{\varepsilon,   \hat{\lambda}_n}_n \right),$$ which, by the definition of the policy, is bounded from above by
$$
\sup_{\bm{\delta} \in \Pi} \Omega_{\hat{\lambda}_n, \mathcal{F}}(\bm{\delta}) - \sup_{\bm{\delta} \in \Pi} \hat{\Omega}_{\hat{\lambda}_n}\left(\bm{\delta} \right) + \hat{\Omega}_{\hat{\lambda}_n}\left(\bm{\pi}^{\varepsilon,   \hat{\lambda}_n}_n\right) - \Omega_{\hat{\lambda}_n, \mathcal{F}} \left(\bm{\pi}^{\varepsilon,   \hat{\lambda}_n}_n \right) + \varepsilon.
$$
This is bounded from above by
$
2 \sup_{\bm{\delta} \in \Pi} |\Omega_{\hat{\lambda}_n, \mathcal{F}}(\bm{\delta}) - \hat{\Omega}_{\hat{\lambda}_n}(\bm{\delta}) | + \varepsilon.
$ It is easy to verify that the inequalities in Assumption~\ref{as:MAIN} remain valid for~$F$ and~$\tilde{F}$ in the closure of~$\mathscr{D} \subseteq D_{cdf}([a,b])$ (equipped with the supremum metric); cf.~Remark~2.3 and Footnote 4 in~\cite{kpv1}. Making use of this observation and Assumption~\ref{as:MAIN}, together with the assumption~$\mathcal{F} \sqsubset_{\Pi} \mathscr{D}$ (implying also that~$\langle \bm{\delta}, \mathcal{F} \rangle = \sum_{z \in \mathcal{Z}} p_Z(z) \langle \bm{\delta}, \mathcal{F} \rangle_z$ is an element of the closure of~$\mathscr{D}$, because the latter is assumed to be convex), we can upper bound
\begin{equation}\label{eqn:contXrbdprf}
2 \sup_{\bm{\delta} \in \Pi} |\Omega_{\hat{\lambda}_n, \mathcal{F}}(\bm{\delta}) - \hat{\Omega}_{\hat{\lambda}_n}(\bm{\delta}) | + \varepsilon \leq 6 \max_{z \in \mathcal{Z}} \sup_{\bm{\delta} \in \Pi} \|\widehat{\langle \bm{\delta}, \mathcal{F} \rangle_z} - \langle \bm{\delta}, \mathcal{F} \rangle_z \|_\infty + \varepsilon.
\end{equation}
Fix~$\bm{\delta} \in \Pi$, $c \in \R$, and~$z \in \mathcal{Z}$, and note that (by the definition of~$\widehat{\langle \bm{\delta}, \mathcal{F} \rangle_z}$ and~$M_{a,b}$)~
\begin{equation}\label{eqn:contXrbdprf2}
|\widehat{\langle \bm{\delta}, \mathcal{F} \rangle_z}(c) - \langle \bm{\delta}, \mathcal{F} \rangle_z(c) | \leq | n^{-1} \sum_{j = 1}^n f_{\bm{\delta}, c, z}(Y_{D_j, j}, X_j, Z_j, D_j) - \langle \bm{\delta}, \mathcal{F} \rangle_z(c) | \leq \|P_n - P\|_{\mathcal{G}_{\Pi}}.
\end{equation} where the last inequality made use of Lemma~\ref{lem:expf}.

We now establish the bound for~$r(\tilde{\bm{\pi}}^{\varepsilon,   \hat{\lambda}_n}_n;   \hat{\lambda}_n, \mathcal{F})$: Clearly, we only need to consider the case where all denominators in~\eqref{eqn:paraprop} are nonzero. In this case, for later use, we first observe that
\begin{equation}
|f_{\bm{\delta}, c, z}(y, x, z^*, d) - \tilde{f}_{\bm{\delta}, c, z}(y, x, z^*, d)| \leq  \left|  \frac{1}{p_Z(z)e_d(x, z)} - \frac{1}{\hat{p}_Z(z)\hat{e}_d(x, z)} \right|,
\end{equation}
from which it follows that, for every~$\bm{\delta} \in \Pi$ and every~$z \in \mathcal{Z}$, we have 
\begin{equation}\label{eqn:propinp}
\left | \widehat{\langle \bm{\delta}, \mathcal{F} \rangle}_z - \widetilde{\langle \bm{\delta}, \mathcal{F} \rangle_z} \right | \leq n^{-1} \sum_{j = 1}^n \left|  \frac{1}{p_Z(z) e_{D_j}(X_j, z)} - \frac{1}{\hat{p}_Z(z)\hat{e}_{D_j}(X_j, z)} \right|.
\end{equation} 
Similarly as in the steps leading up to~\eqref{eqn:contXrbdprf} one obtains the bound
$$
r(\tilde{\bm{\pi}}^{\varepsilon,   \hat{\lambda}_n}_n; \hat{\lambda}_n, \mathcal{F}) \leq \|\hat{\bm{p}}_Z - \bm{p}_Z\|_1 + 
6 \max_{z \in \mathcal{Z}} \sup_{\bm{\delta} \in \Pi} \|\widetilde{\langle \bm{\delta}, \mathcal{F} \rangle_z} - \langle \bm{\delta}, \mathcal{F} \rangle_z \|_\infty + \varepsilon.
$$
Adding~$\pm \widehat{\langle \bm{\delta}, \mathcal{F} \rangle}_z$ inside the supremum norm, the triangle inequality together with Equations~\eqref{eqn:contXrbdprf2} and~\eqref{eqn:propinp} concludes the proof. 
\end{proof}

\begin{lemma}\label{lem:bdf}
The quantities~$\sup_{\lambda \in [0, 1]} \sup_{\bm{\delta} \in \Pi} \Omega_{\mathcal{F}, \lambda}(\bm{\delta})$ and $\sup_{\lambda \in [0, 1]} \sup_{\bm{\delta} \in \Pi} \hat{\Omega}_{\lambda}(\bm{\delta})$ are finite, regardless of whether the latter expression is based on~$\widehat{\langle \bm{\delta}, \mathcal{F} \rangle}$ and~$\widehat{\langle \bm{\delta}, \mathcal{F} \rangle_z}$ (defined in~\eqref{eqn:surro}) or on the quantities $\widetilde{\langle \bm{\delta}, \mathcal{F} \rangle}$ and $\widetilde{\langle \bm{\delta}, \mathcal{F} \rangle_z}$ (defined in~\eqref{eqn:surro2}).
\end{lemma}

\begin{proof}
Because~$\mathsf{S}$ is non-negative, we have, for every~$\lambda \in [0, 1]$ and every~$\bm{\delta} \in \Pi$, that~$$\Omega_{\mathcal{F}, \lambda}(\bm{\delta}) \leq |\mathsf{T}\left( \langle \bm{\delta}, \mathcal{F} \rangle \right)| + \max_{z \in \mathcal{Z}} \mathsf{S} \left(\langle \bm{\delta}, \mathcal{F} \rangle_z , \langle \bm{\delta}, \mathcal{F} \rangle \right);$$ and analogously $$\hat{\Omega}_{\lambda}(\bm{\delta})  \leq |\mathsf{T}(\widehat{\langle \bm{\delta}, \mathcal{F} \rangle})| + \max_{z \in \mathcal{Z}} \mathsf{S}(\widehat{\langle \bm{\delta}, \mathcal{F} \rangle_z}, \widehat{\langle \bm{\delta}, \mathcal{F} \rangle}),$$ with an analogous bound holding true when~$\hat{\Omega}_{\lambda}(\bm{\delta})$ is based on $\widetilde{\langle \bm{\delta}, \mathcal{F} \rangle_z}$ and $\widetilde{\langle \bm{\delta}, \mathcal{F} \rangle}$. Because $\langle \bm{\delta}, \mathcal{F} \rangle$, $\langle \bm{\delta}, \mathcal{F} \rangle_z$, $\widehat{\langle \bm{\delta}, \mathcal{F} \rangle}$, $\widehat{\langle \bm{\delta}, \mathcal{F} \rangle_z}$, $\widetilde{\langle \bm{\delta}, \mathcal{F} \rangle}$, and $\widetilde{\langle \bm{\delta}, \mathcal{F} \rangle_z}$ are  elements of $D_{cdf}([a,b])$, it suffices to verify that there exist real numbers~$C$ and~$C'$, say, such that for any $G \in D_{cdf}([a,b])$ and $G_z \in D_{cdf}([a,b])$, $z \in \mathcal{Z}$, it holds that $$|\mathsf{T}(G)| \leq C \quad \text{ and } \quad \mathsf{S}(G_z, G) \leq C'.$$ To this end, fix a $H \in \mathscr{D}$. By Assumption~\ref{as:MAIN},  $$|\mathsf{T}\left( G \right)| \leq |\mathsf{T}\left( H \right)| + \|H - G\|_{\infty} \leq |\mathsf{T}\left( H \right)| + 1,$$ and similarly (using that~$\mathsf{S}(H, H) = 0$) $$\mathsf{S}\left( G_z,  G \right) = \mathsf{S}\left( G_z,  G \right) - \mathsf{S}(H, H)\leq \|G_z - H\|_{\infty} + \|G - H\|_{\infty} \leq 2.$$
\end{proof}

\end{document}